%% file: dihiggs18.tex
\documentclass[11pt,a4paper]{article}
\usepackage[title]{appendix}

\usepackage[utf8]{inputenc}
\usepackage[affil-sl,auth-sc]{authblk}
\usepackage[top=2cm, bottom=2.5cm,left=2cm,right=2cm]{geometry}
\usepackage{lmodern} 
\usepackage{amssymb,amsmath,amsbsy}
\usepackage{bm}
\usepackage[colorlinks=true, linkcolor=blue, urlcolor=cyan,
  citecolor=cyan]{hyperref}
\usepackage{float}
\usepackage{caption}
\usepackage{wasysym}
\usepackage[nottoc]{tocbibind} 
\usepackage{array}
\usepackage{cite}
\usepackage{placeins}
\usepackage{adjustbox}
\usepackage{multirow}
\usepackage{color}
\usepackage{comment}

\definecolor{orange}{rgb}{0.9,0.2,0}
\definecolor{brown}{rgb}{0.7,0.3,0.2}
\definecolor{fuxia}{rgb}{1,0,1}
\definecolor{skyblue}{rgb}{0,0.1,0.9}
\definecolor{violetred}{rgb}{0.8,0.13,0.56}
\definecolor{deeppink}{rgb}{1.00,0.08,0.5}
\definecolor{pink}{rgb}{1.00,0.75,0.80}
\definecolor{orchid}{rgb}{0.85,0.44,0.84}
\definecolor{lightpink}{rgb}{1.00,0.71,0.76}
\definecolor{bluish}{rgb}{0,0.6,0.8}
\definecolor{lightgray}{rgb}{0.95,0.95,0.95}

\newcommand{\blue}[1]{\color{blue} #1 \color{black}}

\def\sfr{{\tt SmeftFR}~}

\def\bea{\begin{eqnarray}}
\def\eea{\end{eqnarray}}

\let\originalleft\left \let\originalright\right
\renewcommand{\left}{\mathopen{}\mathclose\bgroup\originalleft}
\renewcommand{\right}{\aftergroup\egroup\originalright}
 
\numberwithin{equation}{section}

\begin{document}

\title{ \bf Double Higgs Production via Vector Boson Fusion in SMEFT }

\author[1]{Athanasios Dedes\footnote{\tt adedes@uoi.gr}}

\author[2]{Janusz Rosiek\footnote{\tt Janusz.Rosiek@fuw.edu.pl}}

\author[3]{Micha\l{} Ryczkowski\footnote{\tt
  michaljakub.ryczkowski@unipd.it}}

\affil[1]{Department of Physics, University of Ioannina, GR 45110,
  Ioannina, Greece}

\affil[2]{Faculty of Physics, University of Warsaw, Pasteura 5, 02-093
  Warsaw, Poland}

\affil[3]{Dipartimento di Fisica e Astronomia ``Galileo Galilei”,
  Universita di Padova, Italy and Istituto Nazionale di Fisica
  Nucleare, Sezione di Padova, Padova, I-35131, Italy}

\date{September 18, 2025}

\maketitle

\begin{abstract}

While gluon fusion dominates Higgs pair production at the LHC, vector
boson fusion (VBF) offers a unique window into Beyond the Standard
Model (BSM) physics through its distinctive kinematic features and
direct sensitivity to Higgs-vector boson interactions.  We perform a
comprehensive analysis of double Higgs production via VBF in the
Standard Model Effective Field Theory (SMEFT), systematically
investigating how dimension-6 and dimension-8 bosonic operators --
particularly those involving field derivatives -- can enhance the
production rate.  We identify the most relevant Wilson coefficients
(WCs) affecting the trilinear Higgs coupling ($hhh$) and Higgs-vector
boson interactions ($hVV$, $hhVV$). Using constraints from global fits
and interpolating fit results for unconstrained WCs with Naive
Dimensional Analysis, we assess their effects on the $VV \to hh$ ($V =
W, Z$) scattering amplitudes and cross-sections. Our analysis includes
a study of EFT convergence and validity in models with scalar
extensions of the SM.  Numerical simulations for the planned
High-Luminosity LHC experiment (HL-LHC) in general reveal only a
modest and challenging to detect enhancement of the VBF di-Higgs
production rate over the SM prediction.  However, we show that in
optimistic scenarios, such a process could be observed at the HL-LHC.
In certain cases, when the enhancement is dominated by the dimension-6
or dimension-8 operators containing field derivatives (and thus
leading to stronger energy-dependent effects), this channel becomes
competitive with di-Higgs production via gluon fusion.  This work
highlights the role of VBF di-Higgs production as a complementary
channel for probing anomalous Higgs couplings and their impact on BSM
physics.
  
\end{abstract}

\newpage

{ \hypersetup{linkcolor=black} \tableofcontents }

\newpage


\section{Introduction}
\label{sec:intro}

The discovery of the Higgs boson at the Large Hadron Collider (LHC) in
2012~\cite{ATLAS:2012yve,CMS:2012qbp} marked a pivotal moment in
particle physics, confirming the mechanism of electroweak symmetry
breaking (EWSB) as described by the Standard Model
(SM)~\cite{Glashow,Weinberg:1967tq, Salam}. However, there are
couplings present in the SM have not yet been measured
experimentally. Among them, the trilinear Higgs self-coupling is
perhaps one of the most important, as it is vital for determining the
exact shape of the Higgs potential and for assessing the stability of
the Universe.
This coupling can be probed at the LHC through non-resonant Higgs
boson pair production (commonly termed as di-Higgs production), which
is one of the most promising directions in the search for the physics
beyond the Standard Model (BSM).  Existing observed and expected
$95\%$ CL experimental upper limits for inclusive gluon fusion (ggF)
and vector boson fusion (VBF) double Higgs production modes read (with
significant progress expected with the High-Luminosity LHC upgrade):
\begin{eqnarray}
\sigma(pp\rightarrow hh)^{\text{ggF}+\text{VBF}}_{\text{Obs. (Exp.)}}
& = & 2.9\,(2.4)\times \sigma^{SM}(pp\rightarrow hh),
~\text{ATLAS~\cite{ATLAS:2024ish}}\,, \nonumber\\
\sigma(pp\rightarrow hh)^{\text{ggF}+\text{VBF}}_{\text{Obs. (Exp.)}}
& = & 3.5\, (2.5)\times \sigma^{SM}(pp\rightarrow hh),
~\text{CMS~\cite{CMS:2024ymd}}\,.
\end{eqnarray}
These numbers indicate significant potential for new physics to affect
this process by enhancing its cross-section relative to the SM
prediction, leading lately to growing interest in studying
the process of double Higgs boson production.

\begin{figure}[htb!]
\centering
\includegraphics[width=0.9\linewidth]{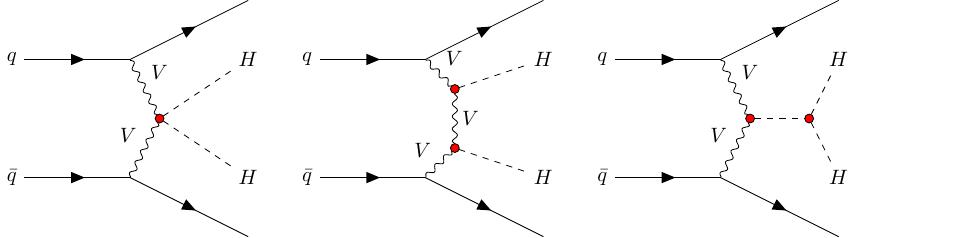}
\caption{\sl Diagrams contributing to the double Higgs production via
  vector boson fusion at tree-level and in unitary gauge (the
  $u$-channel is not shown).}
\label{fig:DiHiggs:diags:VBF}
\end{figure}

The two main production channels for Higgs boson pair production in
the SM are gluon fusion and vector boson fusion
processes~\cite{Baglio:2012np, Frederix:2014hta, Cepeda:2019klc,
  DiMicco:2019ngk}.  While gluon fusion $(gg\to
hh)$~\cite{Glover:1987nx,Plehn:1996wb, deFlorian:2013jea,
  Grigo:2015dia, deFlorian:2015moa, Borowka:2016ehy, Borowka:2016ypz,
  Grazzini:2018bsd, Bonciani:2018omm, Grober:2017uho, Davies:2018ood,
  Xu:2018eos, Baglio:2018lrj, Davies:2018qvx, Davies:2019dfy,
  Chen:2019lzz, Chen:2019fhs, Baglio:2020ini, Wang:2020nnr,
  Baglio:2020wgt, Davies:2022ram, Bellafronte:2022jmo, AH:2022elh,
  Davies:2023npk, Bagnaschi:2023rbx, Davies:2024znp, Heinrich:2024dnz,
  Bi:2023bnq, Jaskiewicz:2024xkd, Campbell:2024tqg, Davies:2025wke,
  Davies:2025ghl, Hu:2025aeo, Bonetti:2025vfd} dominates in the SM,
vector boson fusion $(qq \to qqVV \to qqhh)$~\cite{Ciccolini:2007ec,
  Figy:2008zd, Bolzoni:2010xr, Dolan:2013rja, Cacciari:2015jma,
  Dolan:2015zja, Dreyer:2016oyx, Cruz-Martinez:2018rod,
  Dreyer:2018qbw, Dreyer:2018rfu, Dreyer:2020urf, Dreyer:2020xaj,
  Jager:2025isz} constitutes a complementary channel with distinct
experimental signatures, including presence of forward jets and low
hadronic activity in the central region.  Moreover, although VBF
di-Higgs production is suppressed by an order of magnitude relative to
ggF channel, it is highly sensitive to anomalous Higgs couplings -
namely, the trilinear coupling $hhh$ and Higgs-gauge boson couplings
$hVV$ and $hhVV$ (where “$V$'' stands for either a $W$-boson or a
$Z$-boson).  Such couplings commonly arise in various BSM
theories. For reference, the leading-order tree-level diagrams
contributing to di-Higgs VBF production at LHC are shown in
Figure~\ref{fig:DiHiggs:diags:VBF}, with anomalous couplings marked by
red blobs.

The Standard Model Effective Field Theory
(SMEFT)~\cite{Weinberg:1980wa,Coleman:1969sm,Callan:1969sn,Buchmuller:1985jz,
  Grzadkowski:2010es} provides a model-independent framework to
parameterise BSM effects through higher-dimensional operators,
suppressed by a cutoff scale $\Lambda$.  These operators modify Higgs
and gauge boson interactions, leading to deviations in cross-sections
and kinematic distributions.  In the context of double Higgs VBF
production, dimension-6 and dimension-8 operators can, in principle,
significantly alter the scattering amplitudes, offering a window into
new physics entering the trilinear Higgs coupling that may be
inaccessible in single-Higgs processes.

Over the years, numerous studies have investigated double Higgs
production in the context of various BSM scenarios~\cite{Cao:2013si,
  Ellwanger:2013ova, Hespel:2014sla, Chen:2014ask,
  Cacciapaglia:2017gzh, Grober:2017gut, Cheung:2020xij,
  Cappati:2022skp}.  A number of recent analyses have employed EFT
approach, including both SMEFT and Higge Effective Field Theory
(HEFT)~\cite{Feruglio:1992wf, Buchalla:2012qq, Alonso:2012px,
  Brivio:2013pma, Brivio:2016fzo}, to study single, double, and multi
Higgs production. Some focused on the ggF
channel~\cite{Goertz:2014qta, Azatov:2015oxa, Grober:2015cwa,
  Maltoni:2016yxb, Deutschmann:2017qum, Buchalla:2018yce,
  Heinrich:2022idm, Alasfar:2023xpc, Assi:2024zap}, while others
examined modifications to the trilinear Higgs coupling~\cite{
  DiVita:2017eyz, DiLuzio:2017tfn, Falkowski:2019tft, Chang:2019vez,
  Bhattiprolu:2024tsq, Li:2024iio, Hoeve:2025yup}, or concentrated on
the VBF channel~\cite{ Bishara:2016kjn, Gomez-Ambrosio:2022qsi,
  Domenech:2022uud, Anisha:2024ljc, Mahmud:2024iyn, Delgado:2023ynh,
  Anisha:2024ryj, Braun:2025hvr}.

In this work, we investigate double Higgs production via VBF in the
SMEFT framework, examining the impact of higher-dimensional operators
on the production rate and differential distributions, including terms
$\propto \frac{1}{\Lambda^2}$ (LO in the EFT expansion) and $\propto
\frac{1}{\Lambda^4}$ (NLO in the EFT expansion).  We identify the most
sensitive operators and assess the prospects for detecting BSM effects
at current (LHC) and future (HL-LHC) collider experiments.  Our
analysis highlights the importance of VBF di-Higgs production as a
unique probe of Higgs interactions and the Higgs sector, and
underlines the potential of SMEFT as a tool for uncovering indirect
signs of new physics.  Finally, utilising SMEFT as a tool should be
treated with care, as we also illustrate in this study.

The paper is organised as follows. Section~\ref{sec:DiHiggs:SMEFT}
identifies the dimension-6 and dimension-8 SMEFT operators relevant
for this analysis, and outlines our assumptions leading to bounds on
maximal values of their Wilson coefficients used further in numerical
analysis.  In Section~\ref{sec:DiHiggs:UVScalar} we discuss the EFT
convergence and accuracy, illustrating it with examples of the three
BSM models involving heavy scalar singlets and triplets, by comparing
the amplitudes calculated in renormalisable UV-theories with the
corresponding ones obtained in the decoupling limit.  Analytical
formulae for $VV\to hh$ helicity amplitudes and cross-sections are
presented in Section~\ref{sec:DiHiggs:Ampls} and
Appendices~\ref{App:DiHiggs:WWHH:Ampl}
and~\ref{App:DiHiggs:ZZHH:Ampl}. Finally,
Section~\ref{sec:DiHiggs:Num} contains the results of our numerical
simulations, directly addressing the question of maximal potential
enhancement of VBF di-Higgs production in SMEFT for LHC and HL-LHC
experiments. Section~\ref{sec:summary} concludes the paper.

\section{Contributing operators and bounds on Wilson coefficients}
\label{sec:DiHiggs:SMEFT}

\subsection{Relevant operators}
\label{subsec:DiHiggs:Operators}

To affect the process of double Higgs production via VBF, the SMEFT
operators can be inserted into three types of vertices: $hhh$, $hVV$,
and $hhVV$ (red dots on diagrams in
Figure~\ref{fig:DiHiggs:diags:VBF}, $h$ indicates the Higgs and $V$
indicates the gauge boson).  The relevant classes of the SMEFT
operators that can impact this process are all summarised in
Table~\ref{tab:DiHiggs:Classes}.
\begin{table}[htb!]
\centering
\begin{tabular}{|c|c|c|}
\hline & \multicolumn{2}{c|}{Operator class} \\ \hline Vertex &
Dimension-6 & Dimension-8 \\ \hline
$hhh$ & $\varphi^6$, $\varphi^4 D^2$ & $\varphi^8$, $\varphi^6 D^2$\\
\hline
$hVV$ & $\varphi^4 D^2$, $X^2\varphi^2$ & $\varphi^6 D^2$,
$X^2\varphi^4$, $X\varphi^4D^2$\\
\hline
$hhVV$ & $\varphi^4 D^2$, $X^2\varphi^2$ & $\varphi^4 D^4$, $\varphi^6
D^2$, $X^2\varphi^4$, $X^2\varphi^2D^2$
\\
\hline
\end{tabular}
\caption{\sl Classes of SMEFT dimension-6 and dimension-8 operators
  affecting vertices relevant for double Higgs boson production via
  VBF.}
\label{tab:DiHiggs:Classes}
\end{table}

Table~\ref{tab:DiHiggs:Operators} presents the specific operators that
may affect one or both of the two VBF processes of interest: $WW
\rightarrow hh$ and $ZZ \rightarrow hh$.  All these operators have
been included in the corresponding matrix elements presented in
Section~\ref{sec:DiHiggs:Ampls} and analytical results are given in
Appendices \ref{App:DiHiggs:WWHH:Ampl} and
\ref{App:DiHiggs:ZZHH:Ampl}.

\begin{table}[htb!]
\centering
\begin{tabular}{|c|c|c|c|} 
\hline \multicolumn{2}{|c|}{Dimension-6} &
\multicolumn{2}{c|}{Dimension-8} \\ \hline
\multicolumn{2}{|c|}{$\varphi^6$} & \multicolumn{2}{c|}{$\varphi^8$}
\\
\hline $Q_\varphi$ & $(\varphi^\dagger\varphi)^3$ & $Q_{\varphi^8}$ &
$(\varphi^\dagger \varphi)^4$ \\
\hline \multicolumn{2}{|c|}{$\varphi^4 D^2$} &
\multicolumn{2}{c|}{$\varphi^4 D^4$} \\ \hline
$Q_{\varphi\Box}$ & $(\varphi^\dagger
\varphi)\raisebox{-.5mm}{$\Box$}(\varphi^\dagger \varphi)$ &
$Q_{\varphi^4 D^4}^{(1)}$ & $(D_{\mu} \varphi^{\dagger} D_{\nu}
\varphi) (D^{\nu} \varphi^{\dagger} D^{\mu} \varphi)$ \\
$Q_{\varphi D}$ & $\left(\varphi^\dagger D^\mu\varphi\right)^{\ast}
\left(\varphi^\dagger D_\mu\varphi\right)$ & $Q_{\varphi^4 D^4}^{(2)}$
& $(D_\mu \varphi^\dagger D_\nu \varphi) (D^\nu \varphi^\dagger D^\mu
\varphi)$\\
& & $Q_{\varphi^4 D^4}^{(3)}$ & $(D_\mu \varphi^\dagger D^\mu \varphi)
(D_\nu \varphi^\dagger D^\nu \varphi)$\\
\cline{3-4} &&\multicolumn{2}{c|}{$\varphi^6 D^2$} \\
\cline{3-4} & & $Q_{\varphi^{6} D^{2}}$ & $(\varphi^{\dagger} \varphi)
(\varphi^\dagger D_\mu\varphi)^{\ast} (\varphi^\dagger D^\mu\varphi)$
\\
& & $Q_{\varphi^6 \square}$ & $(\varphi^\dagger \varphi)^{2} \square
(\varphi^\dagger \varphi)$\\
\hline \multicolumn{2}{|c|}{$X^2 \varphi^2$} &
\multicolumn{2}{c|}{$X^2 \varphi^4$} \\ \hline
$Q_{\varphi W}$ & $\varphi^\dagger \varphi\, W^I_{\mu\nu} W^{I\mu\nu}$
& $Q_{W^2\varphi^4}^{(1)}$ & $(\varphi^\dag \varphi)^2 W^I_{\mu\nu}
W^{I\mu\nu}$ \\
$Q_{\varphi B}$ & $\varphi^\dagger \varphi\, B_{\mu\nu} B^{\mu\nu}$ &
$Q_{B^2\varphi^4}^{(1)}$ & $(\varphi^\dag \varphi)^2 B_{\mu\nu}
B^{\mu\nu}$ \\
\cline{3-4} $Q_{\varphi WB}$ & $ \varphi^\dagger \tau^I \varphi\,
W^I_{\mu\nu} B^{\mu\nu}$ & \multicolumn{2}{c|}{$X^2 \varphi^2 D^2$}
\\ \cline{3-4}
& & $Q_{W^2\varphi^2D^2}^{(1)}$ & $(D^{\mu} \varphi^{\dag} D^{\nu}
\varphi) W_{\mu\rho}^I W_{\nu}^{I \rho}$\\
& & $Q_{W^2\varphi^2D^2}^{(2)}$ & $(D^\mu \varphi^\dagger \tau^I D^\nu
\varphi) W_{\mu\rho}^I W_{\nu}^{\rho}$\\
& & $Q_{B^2\varphi^2D^2}^{(1)}$ & $(D^{\mu} \varphi^{\dag} D^{\nu}
\varphi) B_{\mu\rho} B_{\nu}^{\rho}$\\
& & $Q_{B^2\varphi^2D^2}^{(2)}$ & $(D^\mu \varphi^\dagger D_\mu
\varphi) B_{\nu\rho} B^{\nu\rho}$\\
\cline{3-4} && \multicolumn{2}{|c|}{$X\varphi^4 D^2$} \\ \cline{3-4}
& & $Q_{W\varphi^4D^2}^{(1)}$ & $(\varphi^{\dag} \varphi) (D^{\mu}
\varphi^{\dag} \tau^I D^{\nu} \varphi) W_{\mu\nu}^I$\\
& & $Q_{B\varphi^4D^2}^{(1)}$ & $(\varphi^{\dag} \varphi) (D^{\mu}
\varphi^{\dag} D^{\nu} \varphi) B_{\mu\nu}$\\
\hline
\end{tabular}
\caption{\sl Chosen bosonic SMEFT dimension-6 and dimension-8
  operators affecting the $VV\rightarrow hh$ process, and
  consequently, double Higgs boson production via VBF.  Operators have
  been categorised based on their field structure: $X$ denotes the
  field strength tensor, $D$ the covariant derivative and $\varphi$
  the Higgs doublet.}
\label{tab:DiHiggs:Operators}
\end{table}

Some important comments are in place here.  As we are interested in
the most general features of the VBF part of the process, we neglect
the impact of the SMEFT operators on the external quark lines and on
further decays of the Higgs boson pair into the final states.  At the
dimension-6 level, the primary potential contributions may originate
from the dipole operators $Q_{uW}^{11}, Q_{dW}^{11}$ that are
chirality-flipping, as well as the chiral operators $Q_{\varphi
  ud}^{11}$ and $[Q_{\varphi q}^{(3)}]^{11}$.  Especially the latter
could amplify the longitudinal $W$-boson scattering, and as it was
shown in ref.~\cite{Dedes:2020xmo} for $WW\to WW$ scattering, these
operators result in distinct kinematic distributions, possibly leading
to the same effects in $WW\to hh$.  However, they are strongly
constrained by the $Z$-pole precision observables, e.g.  for
$[Q_{\varphi q}^{(3)}]^{11}$, a recent electroweak
analysis~\cite{Bellafronte:2023amz} shows a tight bound on its
corresponding WC, of less than $0.02 \,\frac{1}{\mathrm{TeV}^2}$.
Despite that, the concern about such fermionic operators affecting
studied process remains, especially at the dimension-8 level.
Effects of a subset of such operators have been recently studied in
Ref.~\cite{Assi:2024zap} in the case of single Higgs production.
Additionally, we neglect all CP-violating operators. We also do not
consider the effects of the Renormalisation Group Equation (RGE)
running of the WCs. Although it has been shown in the literature (see,
e.g.,~\cite{Maltoni:2016yxb, Aoude:2022aro, DiNoi:2023onw,
  Maltoni:2024dpn, DiNoi:2024ajj}) that RGE effects are often
non-negligible, we have verified that, in our case, they can change
the values of the dimension-6 WCs by about $\mathcal{O}(20\%)$. This
may slightly modify our results, but with no significant impact on the
final conclusions.

\subsection{Bounds on Wilson coefficients}
\label{subsec:DiHiggs:Assumptions}

To estimate the possible impact of SMEFT operators on double Higgs
production via VBF, one must assign numerical values to the Wilson
coefficients.  For dimension-6 WCs, one can assume the maximum values
that current fits to experimental data allow, as detailed in
Refs.~\cite{Ellis:2020unq, Ethier:2021bye, Celada:2024mcf}.  These
values are displayed in Table~\ref{tab:fits}.
%
\begin{table}[htb!]
\centering
\begin{tabular}{|c|c|c|c|c|}
\hline WC & \multicolumn{2}{c|}{95\% CL bounds,
  $\mathcal{O}(\Lambda^{-2})$} & \multicolumn{2}{c|}{95\% CI bounds,
  $\mathcal{O}(\Lambda^{-4})$} \\
\cline{2-5} & Individual & Marginalised & Individual & Marginalised \\
\hline
$C_{\varphi B}$ & [-0.005, 0.062] & [-0.310, 0.573] & [-0.005, 0.002]
$\cup$ [0.085, 0.092] & [-0.103, 0.152] \\
$C_{\varphi W}$ & [-0.018, 0.006] & [-0.273, 0.707] & [-0.017, 0.006]
$\cup$ [0.282, 0.305] & [-0.067, 0.338] \\
$C_{\varphi WB}$ & [0.007, 0.003] & [-0.525, 0.504] & [-0.007, 0.003]
& [-0.190, 0.263] \\
$C_{\varphi \Box}$ & [0.416, 1.193] & [-1.715, 1.879] & [-0.429,
  1.141] & [-1.856, 1.199] \\
$C_{\varphi D}$ & [-0.027, -0.003] & [-1.063, 1.149] & [-0.027,
  -0.003] & [-0.513, 0.483] \\
\hline
\end{tabular}
\caption{\sl The 95\% CL bounds on the EFT coefficients from linear,
  $\mathcal{O}(\Lambda^{-2})$, and quadratic,
  $\mathcal{O}(\Lambda^{-4})$ EFT fits taken
  from~\cite{Celada:2024mcf}.  $\Lambda = 1$~TeV is assumed.  Results
  from individual and marginalised fits are presented separately.}
\label{tab:fits}
\end{table}

The maximum allowed values for $C_\varphi$ (not shown in
Table~\ref{tab:fits}) can be obtained from the current allowed 95\%
confidence level intervals for the trilinear Higgs coupling modifier
$\kappa_{\lambda_3} = \lambda_3 / \lambda_3^{SM}$. Bounds on
$\kappa_{\lambda_3}$ derived from combined single- and double Higgs
production searches read (with other couplings set to their SM
predictions)\footnote{There is also new CMS measurement constraining
$\kappa_{\lambda_3}$ to similar range, $\kappa_{\lambda_3} \in
[-1.39,\, 7.02]$, based on di-Higgs production
only~\cite{CMS:2024ymd}.}:
\begin{align}
\kappa_{\lambda_3} &\in [-0.4,\,
  6.3],\text{ATLAS~\cite{ATLAS:2022jtk}}\;, \nonumber \\
\kappa_{\lambda_3} &\in [-1.2,\, 7.5],\text{CMS~\cite{CMS:2024awa}} \;.
\label{eq:kappa:exp}
\end{align}
However, within linear field redefinitions in SMEFT, the trilinear
Higgs coupling $hhh$ is a momentum-dependent quantity through the WCs
$C_{\varphi D}$ and $C_{\varphi\Box}$ and related WCs associated with
dimension-8 operators.  Under the assumption that all other WCs are
negligibly small compared to $C_\varphi$ and $C_{\varphi^8}$, this
limit can be translated into a bound on the following combination in
the $\{G_F,m_W,m_Z\}$-input parameter scheme~\cite{Dedes:2023zws,
  Alasfar:2023xpc},
\begin{equation}
\kappa_{\lambda_3} = 1 - \frac{ C_\varphi}{(G_F m_h^2) (G_F \Lambda^2)} 
- \frac{\sqrt{2}\, C_{\varphi^8}}{(G_F m_h^2) (G_F\Lambda^2)^2}\;,
\label{eq:kappal3}
\end{equation}
where $\lambda_3^{SM}=\sqrt{2} G_F m_h^2$.  Assuming $C_{\varphi^8}=0$
and $\Lambda=1$ TeV, one can translate the bound on
$\kappa_{\lambda_3}$ from Eq.~\eqref{eq:kappa:exp} into the direct
bound on $C_\varphi$. Using the less stringent constraints from CMS
one gets,
\begin{equation}
-\frac{13.8}{\text{TeV}^2} \lesssim C_\varphi \lesssim
\frac{4.7}{\text{TeV}^2}\;.
\label{eq:Cphi:num}
\end{equation}
While our study adopts conservative bounds on $C_\varphi$, we note
that recent trilinear coupling analysis~\cite{Maura:2025rcv,
  terHoeve:2025hfx} suggests that somewhat stronger constraints may
apply.

In the case of dimension-8 Wilson coefficients, there have been some
attempts to impose constraints based on available experimental data
and theoretical considerations, such as unitarity
bounds~\cite{Cappati:2022skp,Bresciani:2025toe}.  However, these
constraints are significantly weaker than the values adopted in this
work following assumptions, which we list below.  Consequently, we do
not include them in our analysis.  Instead, one can roughly estimate
their maximal values using the approach of~\cite{Contino:2016jqw},
based on the study of the validity of the EFT expansion, which leads
to a particular power-counting of WCs, also known as Naive Dimensional
Analysis (NDA),
\begin{equation}
  \frac{C_i}{\Lambda^{D-4}}=\frac{g_*^{n_i-2}\times
    c_i}{\Lambda^{D-4}}\;,
\label{eq:C8:num}
\end{equation}
where $D$ is the dimension of SMEFT operator, $n_i$ is the number of
fields involved, $g_*\leq 4\pi$ is a generic UV coupling, and $c_i$ is
an additional suppression factor related, e.g., to additional
symmetries such as shift symmetry or custodial symmetry.  As a result,
we get,
\begin{equation}
\begin{aligned}
       C_{\varphi^8} &= g_*^6\, c_{\varphi^8}\;, \quad
       C_{\varphi^6\Box} = g_*^4\, c_{\varphi^6\Box}\;, \quad
       C_{\varphi^6 D^2} = g_*^4\, c_{\varphi^6 D^2}\;, \quad
       C_{\varphi^4 D^4}^{(i)} = g_*^2 \, c_{\varphi^4 D^4}^{(i)}
       \;,\\
    C_{V^2 \varphi^4}^{(i)} &= g_*^4 \, c_{V^2 \varphi^4}^{(1)} \;,
    \qquad C_{V^2 \varphi^2 D^2}^{(i)} = g_*^2 \, c_{V^2 \varphi^2
      D^2}^{(1)} \;, \qquad C_{V \varphi^4 D^2}^{(i)} = g_*^3 \, c_{V
      \varphi^4 D^2}^{(1)} \;\;,
\end{aligned}
\label{eq:C8:num:1}
\end{equation}
with $V=W$ or $B$.

Further assumption of the maximal possible value of $g_*=4\pi$ and no
additional suppression factors ($c_i=1$) is in principle allowed in
strongly coupled UV scenarios, but may lead to problems with EFT
expansion convergence.  Let us consider as an example the Higgs boson
mass in SMEFT.  At the tree level, it receives contributions from all
higher order operators of the form $Q_{\varphi^n}
(\varphi^\dagger\varphi)^{n+3}$.  After EWSB, assuming the NDA form of
\eqref{eq:C8:num} for $C_{\varphi^n}$, one
has~\cite{Trifyllis:2022yee},
\begin{eqnarray}
m_h^2 = \frac{1}{Z_h^2}\left[\lambda v^2 - g_*^2 v^2 \sum_{n=0}^\infty
  (n+2)(n+3) c_{\varphi^n} \, \left(\frac{g_*^2
    v^2}{2\Lambda^2}\right)^{n+1}\right] \;, \label{eq:mh2}
\end{eqnarray}
where the constant $Z_h$ is required to ensure canonical normalisation
of the Higgs field in the kinetic terms, and is dependent on WCs of
the form $\varphi^{4+2n}D^2\;, n=0,1,2,\dots$ in
Table~\ref{tab:DiHiggs:Operators}.  Assuming that all $c_{\varphi^n}$
are of order ${\cal O}(1)$ (or at least not growing faster than $n$ to
some fixed power), such a power series converges if,
\begin{eqnarray}
  \frac{ g_*^2 v^2}{2\Lambda^2} < 1 \;.
  \label{eq:maxnda}
\end{eqnarray}
Incidentally, Eq.~\eqref{eq:mh2} displays a certain fine-tuning for
keeping $m_h^2\ll \Lambda^2$, when the upper bound of
\eqref{eq:maxnda} is nearly saturated.  Similar constraints on WCs
containing Higgs field derivatives follow from the requirements of
consistency of the SMEFT Lagrangian.  E.g.  $Z_h$ factor at
dimension-6 level has the form~\cite{Dedes:2017zog},
\begin{equation}
    Z_h = 1 + \frac{1}{4} C_{\varphi D} \, \frac{v^2}{M^2} -
    C_{\varphi\Box} \, \frac{v^2}{M^2}\,.
\end{equation}
This has to be greater than zero (otherwise a Higgs tachyon in SMEFT
appears), leading to the condition,
\begin{equation}
  \frac{|4\, C_{\varphi\Box} - C_{\varphi D}|v^2}{\Lambda^2} <
  \sqrt{2} \;,
\end{equation}
or equivalently, using NDA for WCs through Eq.~\eqref{eq:C8:num:1}, to
a bound,
\begin{eqnarray}
  \frac{ g_*^2 v^2}{\Lambda^2} \lesssim {\cal O}(1)\;.
  \label{eq:maxnda1}
\end{eqnarray}
Consequently, in the context of NDA estimates for the upper bounds on
the magnitudes of Wilson coefficient, the assumption $g_*\sim 4\pi$ is
generally invalid when the scale $\Lambda$ is below approximately 3.1
TeV, where to ensure the convergence of the EFT series, the effective
coupling $g_*$ should satisfy a tighter constraint $g_* \lesssim
\frac{\Lambda}{v}$.

Additionally, assuming the NDA scaling of Eq.~\eqref{eq:C8:num}, the
bounds on dimension-6 Wilson coefficients obtained from fits to the
experimental data set a natural limit on UV coupling $g_*$.  The most
stringent limit arises from constraints on $C_\varphi$ in
Eq.~\eqref{eq:Cphi:num}, leading to,
\begin{eqnarray}
  g_* \lesssim \mathrm{min}\left( |C_{\varphi}^{\mathrm max}|^{1/4} \,
  \sqrt{\frac{\Lambda}{\text{1 TeV}}}, \, 4\pi\right) \, ,
  \label{eq:maxgstar}
\end{eqnarray}
where, to be conservative, we use a maximal allowed value of
$C_{\varphi}^{\mathrm max}=13.8$ derived from Eq.~\eqref{eq:Cphi:num}.
Likewise, applying the NDA estimation to the Wilson coefficient
$C_{\varphi \Box}$ yields the following bound,
\begin{eqnarray}
  g_* \lesssim \mathrm{min}\left( \sqrt{|C_{\varphi\Box}^{\mathrm
      max}|} \, \frac{\Lambda}{\text{1 TeV}}, \, 4\pi\right) \, ,
  \label{eq:maxgstar1}
\end{eqnarray}
with $|C_{\varphi\Box}^{\mathrm max}|\approx 2$ following from
Table~\ref{tab:fits}.  For our estimates of possible enhancements to
double Higgs production via the VBF channel in the SMEFT, we use the
$g_*$ limits displayed above as the maximal NDA value for WCs that
currently have no experimental constraints.  As operators with and
without derivatives may arise from distinct underlying physics
processes, we have chosen to employ different bounding schemes for
each category, using the constraint~(\ref{eq:maxgstar}) for
$C_{\varphi^8}$ and~\eqref{eq:maxgstar1} for the remaining WCs not yet
included in fits to experimental data.

Indeed, the above-mentioned ``educated-NDA'' procedure works quite
well for keeping the EFT convergence under control.  To illustrate
this point, if we take the scaling of $C_{\varphi^8}$ from
Eq.~\eqref{eq:C8:num:1} and impose the bound from
Eq.~\eqref{eq:maxgstar}, we arrive at ($\Lambda$ expressed in TeV):
\begin{eqnarray}
  \frac{C_{\varphi^8}}{\Lambda^4} \sim \frac{g_*^6}{\Lambda^4} \sim
  \frac{10^{3/2}}{\Lambda}\;.
\end{eqnarray}
This demonstrates that $\frac{C_{\varphi^8}}{\Lambda^4}$ decreases as
$\Lambda$ grows and remains bounded.  Critically, assuming a large
coupling $g_* = 4\pi$ even at a relatively high scale like $\Lambda =
10 \, \text{TeV}$ would lead to a drastically different result,
$\frac{C_{\varphi^8}}{\Lambda^4} = 125 \pi$, which is orders of
magnitude larger and likely inconsistent with Eq.~\eqref{eq:kappal3}
and the convergence of EFT expansion.

In summary, to check the role of VBF processes in di-Higgs production,
we employ the following bounds.  For WCs corresponding to a
non-derivative operator $Q_{\varphi^8}$, we use the experimental CMS
bound~\cite{CMS:2024awa} on $\kappa_{\lambda_3}$ and the resulting
limits on $C_\varphi$ in Eq.~\eqref{eq:Cphi:num} together with the NDA
scaling of \eqref{eq:C8:num} for $c_{\varphi^8}=1$. We then obtain,
\begin{equation}
\begin{aligned}
  -\mathrm{min}\left( 1.8 \, \sqrt{\frac{\Lambda}{\text{1 TeV}}}, \,
  4\pi\right) \, \lesssim & \, g_*^\prime \lesssim \mathrm{min}\left(
  1.4 \, \sqrt{\frac{\Lambda}{\text{1 TeV}}}, \, 4\pi\right) \,, \\
  -\frac{31.6}{\text{TeV}^4} & \lesssim
  \frac{C_{\varphi^8}}{\Lambda^4} \lesssim \frac{7.4}{\text{TeV}^4}\;.
  \label{eq:maxgstar:prime}
\end{aligned} 
\end{equation}
However, as can be seen from the analytical expressions of the
helicity amplitudes $VV\to hh$ given in
Appendices~\ref{App:DiHiggs:WWHH:Ampl}
and~\ref{App:DiHiggs:ZZHH:Ampl}, the dominant contribution to di-Higgs
production both in $WW$ or $ZZ$ scattering is proportional to the same
combination of $C_{\varphi}$ and $C_{\varphi^8}$ as the one entering
$\kappa_{\lambda_3}$ in Eq.~\eqref{eq:kappal3}. For this reason,
individual estimates of both WCs do not significantly affect numerical
results and remain bounded primarily by already known experimental
limits on $\kappa_{\lambda_3}$, as we discuss further in
Section~\ref{sec:DiHiggs:Num}.

For derivative operators, again being conservative, we assume maximal
values slightly larger than those derived from global fits for
\textit{marginalised} and \textit{linear} case presented in
Table~\ref{tab:fits},
\begin{equation}
  \left|C_{\varphi \Box}\right| \lesssim \frac{2}{\text{TeV}^2}, \quad
  \left|C_{\varphi D}\right|\lesssim \frac{1.5}{\text{TeV}^2}, \quad
  \left|C_{\varphi \left(W,B,WB\right)}\right|\lesssim
  \frac{1}{\text{TeV}^2}\;,
\label{eq:WC:D6:num}
\end{equation}
which, after making use of \eqref{eq:maxgstar1} result in bounds,
\begin{equation}
\begin{aligned}
  g_*& \lesssim \text{max}\left(\sqrt{2}\frac{\Lambda}{1\text{
      TeV}},\, 4\pi\right)\;,\\
  \frac{|C_{\varphi^6 D^2}|}{\Lambda^4}& =
  \frac{|C_{V^2\varphi^4}|}{\Lambda^4}= \frac{g_*^4}{\Lambda^4}
  \lesssim \frac{4}{(1\text{ TeV})^4}\;,\\
  \frac{|C_{\varphi^4 D^4}|}{\Lambda^4}& = \frac{|C_{V^2\varphi^2
      D^2}|}{\Lambda^4} = \frac{g_*^2}{\Lambda^4} \lesssim
  \frac{2}{(1\text{ TeV})^2}\:\frac{1}{\Lambda^2}\;,\\
  \frac{|C_{W\varphi^4 D^2}|}{\Lambda^4}& = \frac{|C_{V\varphi^4
      D^2}|}{\Lambda^4} = \frac{g_*^3}{\Lambda^4} \lesssim
  \frac{2\sqrt{2}}{(1\text{ TeV})^3}\:\frac{1}{\Lambda}\;.
    \label{eq:maxgstar:unprime}
\end{aligned}
\end{equation}
The bounds from Eqs.~\eqref{eq:maxgstar:prime} and
\eqref{eq:maxgstar:unprime} will serve as the primary input values for
the subsequent analysis.

\section{Wilson coefficients and EFT convergence in  scalar extensions
  of the SM}
\label{sec:DiHiggs:UVScalar}

Several questions arise when discussing the maximal enhancement of the
double Higgs boson production within SMEFT.  First, are the NDA
estimates realistic, i.e. can one explicitly construct UV models which
after decoupling of heavy particles lead to large values of WCs?
Second, if WCs are large, can they affect the convergence of the EFT
expansion in inverse powers of heavy mass scale? Third, how reliable
is the SMEFT approximation for the growing energy of the process?

We address these questions by considering examples of a few simple
extensions of the SM.  These UV-completions are models with a heavy
real scalar singlet field $S$ or two types of heavy scalar triplets
$T_1$, $T_2$.  They provide us with a useful theoretical laboratory
for studying and testing the VBF di-Higgs production in the SMEFT.

\subsection{Real scalar model}

The matching onto the SMEFT of a heavy scalar singlet model has been
worked out in several studies, with dimension-6 operators at
tree-level~\cite{Egana-Ugrinovic:2015vgy,Dawson:2021jcl} and
one-loop~\cite{Haisch:2020ahr,Cohen:2020fcu} as well as dimension-8
operators~\cite{Ellis:2023zim}.  By incorporating all renormalisable
interactions between the heavy singlet $S$ and the SM fields, we have
\begin{equation}
  \mathcal{L}_{S} = \mathcal{L}_{SM} + \frac{1}{2} (\partial_\mu S)^2
  -\frac{1}{2} M^2 S^2 - \alpha \, M \, S \varphi^\dagger \varphi -
  \frac{1}{2}\kappa \, |\varphi|^2 S^2 - \frac{1}{3!} \beta \, M \,
  S^3 -\frac{1}{4!} \lambda_S\, S^4\;,
\end{equation}
where $\mathcal{L}_{SM}$ is the SM-Lagrangian in the notation of
Refs.~\cite{Grzadkowski:2010es,Dedes:2017zog} and $\varphi$ is the SM
Higgs doublet. Besides the heavy singlet mass $M$, all other
parameters are dimensionless. By promoting the parameters of the
theory into auxiliary fields, a background $Z_2$-symmetry under which
$\varphi\to \varphi$ and $S\to -S$, assigns charges $(-1)$ to $\alpha$
and $\beta$ and $(+1)$ to all other couplings.  The Wilson
coefficients in front of the $Z_2$-invariant EFT expansions in terms
of $v^2/M^2$ (or $s/M^2$ when calculating momentum-dependent terms in
scattering amplitudes) should also have an $Z_2$-invariant polynomial
form,
\begin{equation}
  C_i = f(\alpha^{n_1} \beta^{n_2},...;,\kappa,\lambda_S)\;, \quad
  n_1+n_2=\mathrm{even}\;,
\end{equation}

\begin{figure}[t]
\centering
\includegraphics[width=0.8\linewidth]{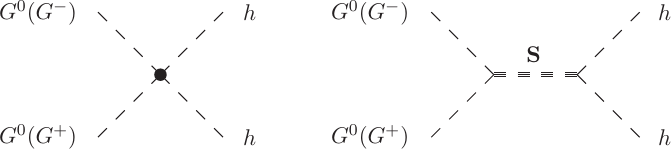}
\caption{\sl Dominant diagrams contributing to the di-Higgs production
  in the real scalar SM extension.  The black dot indicates the
  quartic Higgs coupling $\lambda$. The fields $G^0,G^\pm$ are the
  Goldstone bosons in the $R_\xi$-gauge we are using and $h$ is the
  Higgs field.}
\label{fig:uvscat}
\end{figure}

Alternatively to the standard matching procedure, one can try to
identify SMEFT WCs by comparing the corresponding terms of the
amplitudes in both UV and effective theory in the energy region $v^2
\ll s \ll M^2$.  Relevant to the di-Higgs production processes, we can
use longitudinally polarised vector bosons $W_LW_L\to hh$ and
$Z_LZ_L\to hh$ scattering, although the method works as well for other
choices (e.g.  $W_L W_L \to W_L W_L$ and $W_L W_L \to Z_L Z_L $).  Let
us compare the leading (not suppressed by additional $v^2/M^2$ powers)
contributions to coefficients of $s$ powers in the amplitude.  In the
high $s\gg v^2$ regime, the Goldstone Boson Equivalence
Theorem~\cite{Cornwall:1973tb, Vayonakis:1976vz, Chanowitz:1985hj}
simplifies the calculation since at high energy in SMEFT we only have
to calculate $G^+G^-\to hh$ and $G^0G^0\to hh$ and match to the same
amplitudes in full theory at $s\ll M^2$.  The tree level longitudinal
amplitudes in the UV model for both processes are given by the same
expression, including contributions from the SM contact 4-scalar
interaction and the single $s$-channel exchange of the heavy scalar
(see Fig~\ref{fig:uvscat}),
\begin{equation}
  \mathcal{M}_{00}^{(G^+G^-\to hh)}(s)=\mathcal{M}_{00}^{(G^0G^0\to
    hh)}(s) = -\lambda + \frac{(-i \alpha M)^2}{s-M^2} \overset{s\ll
    M^2}{=} -\lambda + \alpha^2 \, \left [ 1+ \frac{s}{M^2} +
    \frac{s^2}{M^4} \dots + \right ]\;.
    \label{eq:sfull}
\end{equation}
where $\lambda$ is the quartic Higgs coupling in $\mathcal{L}_{SM}$:
$-\frac{\lambda}{2} (\varphi^\dagger\varphi)^2$.

It is crucial to note the same prefactor $\alpha^2$ for coefficients
of $(s/M^2)^n\;, n=0,1,2,..$ powers.  On the other hand, the SMEFT
amplitudes at large-$s$, but always below the singlet mass squared,
give (in dimension-6 basis of \cite{Grzadkowski:2010es} and
dimension-8 basis of \cite{Murphy:2020rsh}; the explicit expressions
for the amplitudes are presented in
Appendices~\ref{App:DiHiggs:WWHH:Ampl}
and~\ref{App:DiHiggs:ZZHH:Ampl}):
\begin{align} 
  \mathcal{M}_{00}^{(G^+G^-\to hh)}(v^2\ll s\ll M^2) = -\lambda_{eff}
  &-\frac{1}{2} (4 C^{\varphi\Box}-C^{\varphi D})\frac{s}{M^2}
  \nonumber \\[2mm]
  &+ \frac{1}{8} \biggl [(C^{(1)}_{\varphi^4D^4} +
    C^{(2)}_{\varphi^4D^4})(1+\cos^2\theta) + 4\,
    C^{(3)}_{\varphi^4D^4}\biggr ]\frac{s^2}{M^4} \;,\label{eq:gpeft}
  \\[2mm]
  \mathcal{M}_{00}^{(G^0G^0\to hh)}(v^2\ll s\ll M^2) = -\lambda_{eff}
  &- (2 C^{\varphi\Box}+C^{\varphi D})\frac{s}{M^2} \nonumber \\[2mm]
  &+ \frac{1}{4} \biggl [2 (C^{(1)}_{\varphi^4D^4} +
    C^{(3)}_{\varphi^4D^4}) - C^{(2)}_{\varphi^4D^4}
    (1-\cos^2\theta)\biggr ]\frac{s^2}{M^4}\;, \label{eq:g0eft}
\end{align}
where $\theta$ is the scattering angle.  By comparing the $s^0,
s^1,s^2$ pieces of Eqs.~\eqref{eq:sfull}, \eqref{eq:gpeft} and
\eqref{eq:g0eft} we get
\begin{align}
  & \lambda_{eff} = \lambda - \alpha^2 \;, \label{eq:leff}\\
  & C_{\varphi\Box} = - \frac{\alpha^2}{2}\;, \quad C_{\varphi
    D}=0\;, \label{eq:boxalpha} \\
  & C_{\varphi^4 D^4}^{(3)} = 2 \alpha^2\;, \quad C_{\varphi^4
    D^4}^{(1)} = C_{\varphi^4 D^4}^{(2)}=0 \;, \label{eq:phialpha}
\end{align}
where $\lambda_{eff}$ corresponds to low-energy quartic SM Higgs
coupling.

These results are in agreement with the
literature~\cite{Egana-Ugrinovic:2015vgy, Haisch:2020ahr,
  Cohen:2020fcu, Dawson:2021jcl, Ellis:2023zim}.  Note also that the
dimension-8 WCs are positive semi-definite according to the
corresponding positivity bound~\cite{Remmen:2019cyz}.  Furthermore,
$C_\varphi$ can be found by comparing the corresponding amplitudes for
$\varphi^6$ processes in full and effective theory.  It
is~\cite{Cohen:2020fcu}
\begin{equation}
  C_\varphi = -\frac{\alpha^2}{2} \biggl (\kappa -
  \frac{\alpha\beta}{3} \biggr )\;.
\end{equation}
The above procedure suggests the matching conditions to be taken at
$v^2\ll s=\Lambda_0^2\ll M^2$, with $\Lambda_0$ could naturally be
chosen at the scale defined by the unitarity violation of SMEFT.  As
usual, WCs at low scale can be then calculated using the RGEs
evolution.

From the UV model's perspective, $\alpha$ is a free parameter,
eventually constrained by perturbativity of couplings.  In such case,
the UV loop-expansion parameter should be limited to
\begin{equation}
  \frac{\alpha^2}{(4\pi)^2} \lesssim 1\;,
\end{equation}
so that $\alpha \lesssim 4\pi$ and maximal value of $C_{\varphi\Box}$
can saturate the NDA limit.  Therefore, the SM extension with a real
scalar provides a simple, explicit example of a model that can really
produce large values for WCs in the low-energy limit\footnote{
However, one should note that some WCs may be constrained by
additional conditions, such as the properties of the Higgs potential.
Higher order operators contribute to it, ultimately affecting its
stability and positions of minima.  We do not study these issues in
detail here, as our goal is simply to provide an example of
consistent UV models which may produce large WCs.}.

Actually, it is possible to get even larger WCs from perturbative UV
models.  If instead of one heavy real scalar we add to the SM $n$
copies of such fields with couplings $\alpha_1, \alpha_2, \ldots$ and
masses $M_1 \leq M_2 \leq \ldots$ (in addition we assume for
simplicity that heavy scalars interact feebly with each other and
mainly couple to the SM Higgs-doublet), the formula for
$C_{\varphi\Box}$ takes the form,
\begin{eqnarray}
C_{\varphi\Box} = - \frac{1}{2 M_1^2} \sum_{i=1}^n\alpha_i^2 M_i^2 \;.
\end{eqnarray}
Obviously by increasing the number of heavy fields with similar masses
one can make $C_{\varphi\Box}$ considerable, while keeping all the UV
couplings $\alpha_i$ being perturbative.

To address the question of EFT convergence, let us note that
comparison of Eq.~(\ref{eq:sfull}) with Eqs.~\eqref{eq:gpeft} and
\eqref{eq:g0eft} shows that WCs of {\it all} higher order operators
contributing to the dominant terms in the amplitude are proportional
to $\alpha^2$ and not to the higher powers of $\alpha$.  Consequently,
they do not grow uncontrollably with increasing mass dimension and do
not endanger the EFT series convergence.  This remark is also in
agreement with NDA as on dimensional ground all such couplings should
be proportional to $g_*^2$.  From Eq.~(\ref{eq:sfull}), and, at tree
level in $\hbar$-expansion, it is clear that EFT series always
converges if we are below the heavy mass threshold, $s< M^2$,
independently of WC values.  Other WCs which may contribute to
sub-dominant terms in the amplitude are proportional to higher powers
of UV parameters and thus can be larger, but they are always
multiplied by powers of $(v/M)$ and therefore should not affect EFT
convergence as long as condition~\eqref{eq:maxnda} is fulfilled.

\subsection{Triplet scalar models}

As can be seen from Eq.~\eqref{eq:sfull}, the dominant contribution to
$W_LW_L\to hh$ and $Z_LZ_L\to hh$ in real scalar model has always
opposite sign to SM term $-\lambda_{eff}$ ($\lambda_{eff}$ should be
positive for Higgs potential to be stable and to have proper minima).
Thus, in such a model both processes are suppressed compared to the
SM, unless the $\alpha$-parameter is huge and BSM terms dominate the
SM amplitude.  This may not be the case for scalar triplet extensions.

Let us consider models containing heavy scalar triplets, first neutral
$T_1(1,3,0)$ and then hypercharged $T_2(1,3,1)$ (numbers is
parenthesis denote as usual $SU(3)_c\times SU(2)_L \times U(1)_Y$
quantum numbers).  In the first case, the interesting part of the
Lagrangian is
\begin{equation}
  \mathcal{L}_{T_1} = \mathcal{L}_{SM} + (D_\mu T_1)^\dagger (D^\mu
  T_1) - \alpha_{T_1}\: M\, \varphi^\dagger \sigma^a T_1^a \varphi +
  \dots
\end{equation}
where $\sigma^a$ are the Pauli matrices. By exploiting the same
$Z_2$-symmetry arguments as for the real singlet scalar, we deduce
that WCs will only depend on $\alpha_{T_1}^2$.

In this case, the amplitudes $\mathcal{M}_{00}^{G^0G^0\to hh}$ and
$\mathcal{M}_{00}^{G^+G^-\to hh}$ are different, at high energy the
first one comes from the $s-$channel only with neutral component of
the triplet mediator $T_1^0$ plus the contact SM-like graph in
Figure~\ref{fig:uvscat}, while the second one also contains
contributions from the $t$ and $u$ channels with charged mediators
exchange (not shown in Figure~\ref{fig:uvscat}).  The amplitudes are
\begin{equation}
  \mathcal{M}_{00}^{G^0G^0\to hh}(s) = -\lambda + \frac{(i
    \alpha_{T_1} M)^2}{s-M^2}= -\lambda+\alpha_{T_1}^2 \biggl [1 +
    \frac{s}{M^2} + \frac{s^2}{M^4} + \dots \biggr ]\;,
    \label{eq:MT00}
\end{equation}
and
\begin{align}
  \mathcal{M}_{00}^{G^+G^-\to hh}(s) &= -\lambda + \alpha_{T_1}^2\:
  M^2\, \biggl [\frac{1}{s-M^2} - \frac{1}{t-M^2} - \frac{1}{u-M^2}
    \biggr ] \nonumber \\[2mm]
&= -\lambda + \alpha_{T_1}^2 - 2 \alpha_{T_1}^2 \frac{s}{M^2} -
  \frac{\alpha_{T_1}^2}{2} \frac{s^2}{M^4} (1-\cos^2\theta) + \dots\;.
\label{eq:MTpm}
\end{align}
By performing the UV and EFT matching procedure we described earlier
for the amplitudes in the real scalar scenario, and using the leading
terms of amplitudes given in Appendices~\ref{App:DiHiggs:WWHH:Ampl}
and~\ref{App:DiHiggs:ZZHH:Ampl}, we arrive at,
\begin{align}
  \lambda_{eff} &= \lambda -\alpha_{T_1}^2 \;, \\[2mm]
  C_{\varphi\Box} &= \frac{1}{2} \alpha_{T_1}^2\;, \quad C_{\varphi D}
  = - 2 \alpha_{T_1}^2 \;, \\[2mm]
  C_{\varphi^4 D^4}^{(1)} &= 4 \alpha_{T_1}^2\;, \quad C_{\varphi^4
    D^4}^{(2)} =0\;, \quad C_{\varphi^4 D^4}^{(3)}=-2 \alpha_{T_1}^2
  \;.  \label{eq:CT0}
\end{align}
These results are consistent with Ref.~\cite{Ellis:2023zim} derived
with functional matching.  In such a model, $C_{\varphi\Box}$ is
positive but non-vanishing negative $C_{\varphi D}$ is generated, so
that effectively $Z_LZ_L\to hh$ cross-section is still suppressed
compared to SM but $W_LW_L\to hh$ becomes enhanced.  Nevertheless,
both effects cannot be large due to constraints on the $C_{\varphi D}$
which enters alone in the electroweak $\rho$-parameter
as~\cite{Dedes:2017zog}
\begin{equation}
  \rho -1 = \frac{1}{2}\frac{v^2}{\Lambda^2} \, C_{\varphi D}\;.
\end{equation}
The $\rho$-parameter deviation from unity is constrained by the
electroweak precision observables to be at one per mile level.  Hence,
$|\alpha_{T_1}| \lesssim 0.01$ and the leading neutral scalar triplet
contribution affecting di-Higgs production at high energies is
strongly constrained.

The situation is different in charged heavy triplet model
$T_2(1,3,1)$, with the Lagrangian
\begin{equation}
  \mathcal{L}_{T_2} = \mathcal{L}_{SM} + (D_\mu T_2)^\dagger (D^\mu
  T_2) - \alpha_{T_2}\: M\, \widetilde{\varphi}^\dagger \sigma^a T_2^a
  \varphi \ + \ \mathrm{h.c.} \ + \dots \;.
\end{equation}
In this case, high energy scattering amplitudes are given by
\begin{align}
  \mathcal{M}_{00}^{G^0G^0\to hh}(s) &= -\lambda \ +\ 2\,
  \alpha_{T_2}^2\, M^2\, \biggl[
    \frac{1}{s-M^2}-\frac{1}{t-M^2}-\frac{1}{u-M^2} \biggr] \nonumber
  \\[2mm]
  &=-\lambda + 2 \alpha_{T_2}^2 - 4\, \alpha_{T_2}^2\, \frac{s}{M^2} -
  \alpha_{T_2}^2 (1-\cos^2\theta) \frac{s^2}{M^4} + \dots \;,
  \label{eq:MT100}
\end{align}
\begin{align}
  \mathcal{M}_{00}^{G^+G^-\to hh}(s) &= -\lambda \ - \alpha_{T_2}^2 \,
  M^2\, \biggl [ \frac{1}{t-M^2} + \frac{1}{u-M^2} \biggr ] \nonumber
  \\[2mm]
  &= -\lambda + 2\, \alpha_{T_2}^2 - \alpha_{T_2}^2\, \frac{s}{M^2} +
  \frac{\alpha_{T_2}^2}{2}\, (1+\cos^2\theta)\, \frac{s^2}{M^4} \ +
  \ \dots
  \label{eq:MT1pm}
\end{align}
and matching to SMEFT as before provides
\begin{align}
  \lambda_{eff} &= \lambda -2\,\alpha_{T_2}^2 \;, \\[2mm]
  C_{\varphi\Box} &= \alpha_{T_2}^2\;, \quad C_{\varphi D} = 2
  \alpha_{T_2}^2 \;, \\[2mm]
  C_{\varphi^4 D^4}^{(1)} &= 0 \;, \quad C_{\varphi^4 D^4}^{(2)} = 4
  \alpha_{T_2}^2\;, \quad C_{\varphi^4 D^4}^{(3)}=0 \;.
  \label{eq:CT1}
\end{align}
Comparing to the singlet and the neutral triplet scenarios, both
$C_{\varphi\Box}$ and $C_{\varphi D}$ are now positive and both
$Z_LZ_L\to hh$ and $W_LW_L\to hh$ channels of di-Higgs production are
enhanced in this case.  In addition, adding both types of triplets
(with identical masses) to the SM, one has
\begin{align}
  \lambda_{eff} &= \lambda -\,\alpha_{T_1}^2 -2\,\alpha_{T_2}^2 \;,
  \nonumber\\[2mm]
  C_{\varphi\Box} &= \frac{1}{2}\alpha_{T_2}^2 + \alpha_{T_2}^2\;,
  \\[2mm]
  C_{\varphi D} &= -2 (\alpha_{T_1}^2-\alpha_{T_2}^2)\;,
  \nonumber\\[2mm]
  C_{\varphi^4 D^4}^{(1)} &= 4 \alpha_{T_1}^2 \;, \quad C_{\varphi^4
    D^4}^{(2)} = 4 \alpha_{T_2}^2\;, \quad C_{\varphi^4 D^4}^{(3)}=-2
  \alpha_{T_1}^2 \;.
  \label{eq:CT12}
\end{align}
and in case of approximately (or exactly, due to some underlying
symmetry) equal couplings, $\alpha_{T_1}\approx\alpha_{T_2}$, it is
possible to get enhancement of the double Higgs production avoiding
the constraints from the $\rho$-parameter. One should note that RGEs
evolution can mix various WCs after decoupling and in principle lead,
at lower energies, to problems with $\rho$-parameter constraints even
if they were initially fulfilled. However, we have checked that in
SMEFT for considered low $\Lambda$ scales such effect is small and
does not affect our conclusions.

In summary, we demonstrated with explicit examples that it is possible
to find renormalisable UV-models which produce at low energies large
WCs in SMEFT, that may increase or decrease the di-Higgs production
rates through VBF and are consistent with EFT expansion convergence.

\subsection{EFT accuracy for large $\sqrt{s}$}

\begin{figure}[tb]
  \centering
  \begin{tabular}{rr}
    \includegraphics[width=0.44\linewidth]{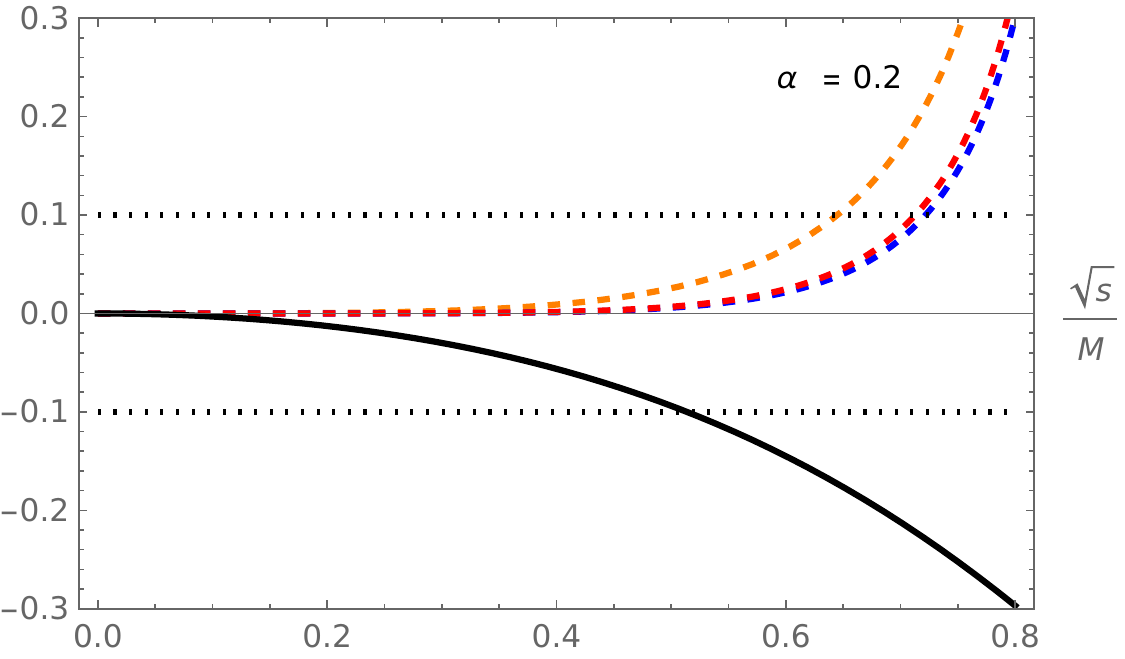} &
    \includegraphics[width=0.44\linewidth]{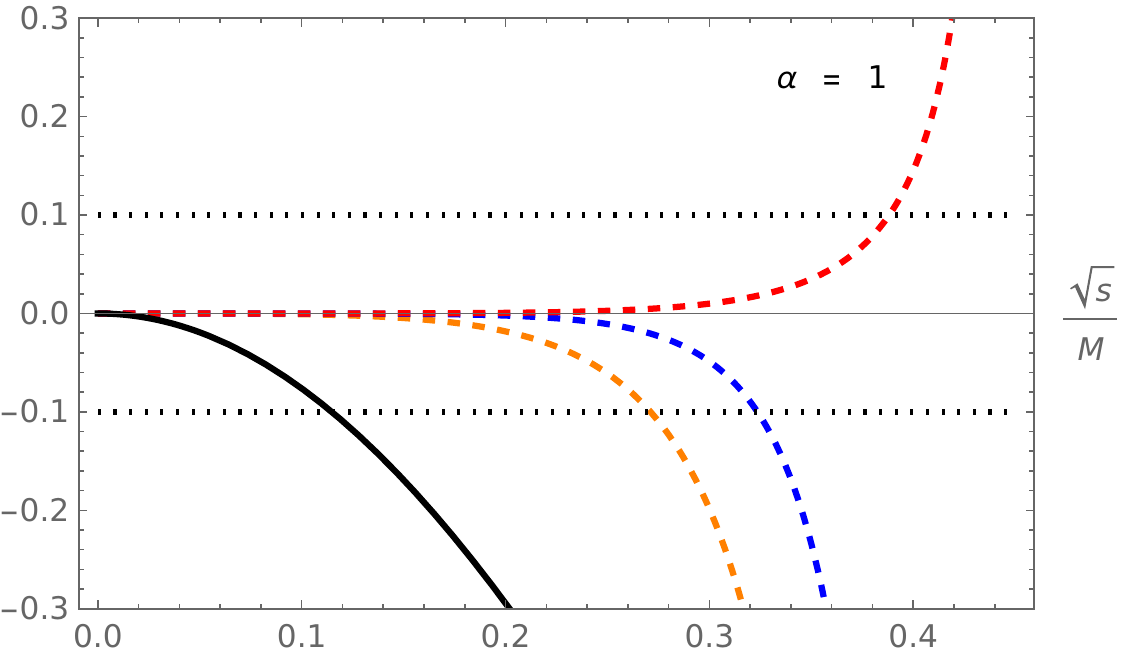} \\
    \includegraphics[width=0.44\linewidth]{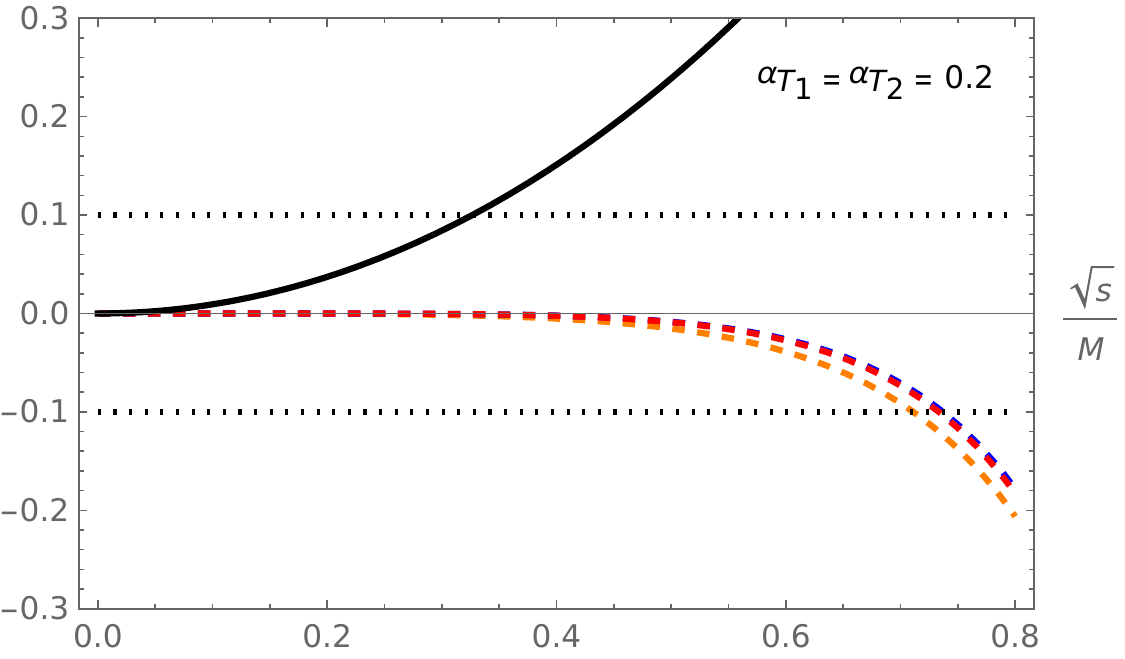} &
    \includegraphics[width=0.44\linewidth]{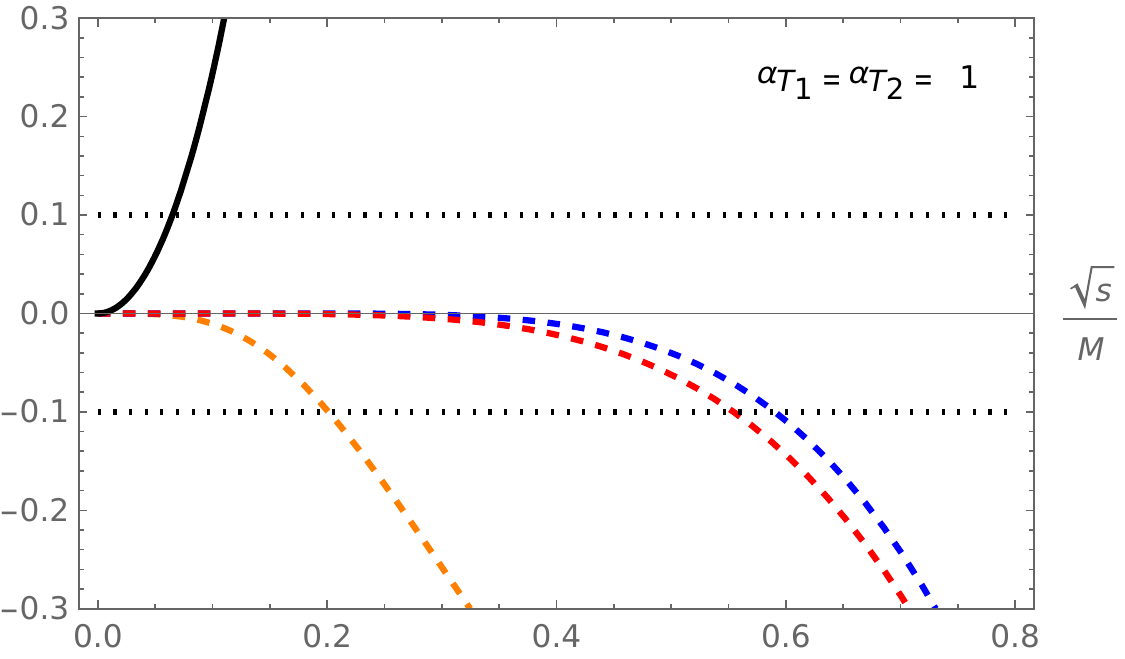} \\
  \end{tabular}
\caption{\sl The plots compare the relative accuracy of the $W_LW_L
  \to hh$ high energy scattering cross-section as predicted by the
  SMEFT and underlying UV models.  The dashed lines represent
  $\Delta_{EFT}$ from Eq.~\eqref{eq:deltaeft} as a function of
  $\sqrt{s}/M$. The orange curve shows the EFT result with dimension-6
  terms in the amplitude alone, with higher-order terms truncated in
  the cross-section. The blue and red curves include dimension-8 terms
  in the amplitude, either truncated or fully kept in its square,
  respectively.  The dotted lines mark the $\pm 10\%$ range of
  relative accuracy.  The solid black line shows the deviation of the
  cross-section ratios from unity in the scalar models relative to the
  SM.  The upper panels present the findings for a real scalar model,
  and the lower panels show the findings for a model extended with
  both neutral and charged triplets.
    \label{fig:eftvsuv} }
\end{figure}

The issue related to the EFT convergence is how well the EFT
approximation reproduces the full UV-model, depending on the value of
$\sqrt{s}$.  For a rough estimate, in Figure~\ref{fig:eftvsuv} we
plotted the relative accuracy of EFT approximation defined as
\begin{equation}
  \Delta_{EFT} =
  \left|\frac{\mathcal{M}_{00~\mathrm{(EFT)}}^{(G^+G^-\to
      hh)}}{\mathcal{M}_{00~\mathrm{(UV)}}^{(G^+G^-\to hh)}}\right|^2
  - 1\;,
  \label{eq:deltaeft}
\end{equation}
with $\mathcal{M}_{00~\mathrm{(EFT)}}^{(G^+G^-\to hh)}$ and
$\mathcal{M}_{00~\mathrm{(UV)}}^{(G^+G^-\to hh)}$ given by expressions
for the real scalar or triplet models given above.
Figure~\ref{fig:eftvsuv} presents results for two scenarios: the upper
panels correspond to the real scalar model, and the lower panels to a
Standard Model extension incorporating both neutral and charged
triplets with degenerate masses and identical couplings, denoted by
$\alpha_{T_1}=\alpha_{T_2}$.  We also assumed that $\lambda_{eff}$ is
predominantly determined by the Higgs mass value, so that
$\lambda_{eff} \approx 0.26$.

Figure~\ref{fig:eftvsuv} indicates that for small couplings $\alpha
(\alpha_{T_i})=0.2$, the EFT remains a reliable tool for all
considered UV-models up to at least $\sqrt{s}/M\sim 0.7$, provided
dimension-8 terms are accounted for.  With large couplings, around
$\mathcal{O}(1)$, the EFT approximation maintains good agreement with
UV-models in scenarios with enhanced cross-sections.  However, its
validity is reduced to $\sqrt{s}/M\sim 0.4$ when SM-BSM interference
is negative, potentially leading to a complete cancellation of the
amplitude.  This effect is exacerbated by the smallness of
UV-amplitude in the denominator of Eq.~\eqref{eq:deltaeft}, increasing
the discrepancy.  Considering the limited experimental interest in
suppressed cross-sections, the EFT offers a safe approach for studying
enhanced di-Higgs VBF production at the LHC, even provided that the
new physics scale is not much larger than the highest investigated
invariant mass of the Higgs pair.  Based on the preceding analysis, we
adopt the bound on the ratio of the maximal center-of-mass energy to
the new physics (NP) scale to ensure EFT convergence:
\begin{equation}
     \frac{\sqrt{s_{max}}}{\Lambda}\leq x\approx 1\;.
     \label{eq:validity}
\end{equation}

Although the value $x \approx 0.7$ is theoretically preferable for
inequality \eqref{eq:validity}, we adopt the more relaxed value
$x=1$. This choice is justified by our numerical analysis in Section
\ref{sec:DiHiggs:Num}, which shows that even $x=1$ severely limits
potential process enhancements. Consequently, the stricter bound $x
\approx 0.7$ would yield even less favorable predictions.

It is also important to note that retaining only linear terms in
dimension-6 Wilson coefficients yields considerably less accurate
results.  Furthermore, Figure~\ref{fig:eftvsuv} demonstrates that when
the discrepancy between EFT and UV predictions becomes significant for
stronger couplings. In some cases, EFT more accurately reproduces the
full result if the amplitude is truncated at dimension-8 terms, even
when formally higher-order terms (up to dimension-12) arising from
squaring the amplitude are retained in the cross-section.  More
details are provided in the following subsection.

Finally, constraints on the maximum $\sqrt{s}$ for consistent EFT
application can also arise from perturbative unitarity requirements
(see, e.g. Refs.~\cite{Corbett:2014ora, Kalinowski:2018oxd,
  Kilian:2018bhs, Almeida:2020ylr, Dedes:2020xmo, Logan:2022uus,
  Mahmud:2025wye, Bresciani:2025toe}).  However, for center-of-mass
energies achievable at the LHC, unitarity violation typically does not
impose stringent bounds.  For instance, for WCs of operators involving
Higgs doublet derivatives, it leads to a condition of the form
$C\frac{s}{\Lambda^2}\lesssim {\cal O}(4\pi)$.  This condition is
generally satisfied if the relevant dimension-6 WCs remain within the
bounds derived from fits to experimental data, as summarised in
Eqs.~\eqref{eq:WC:D6:num} and \eqref{eq:maxgstar:unprime}.  Similarly,
our checks indicate that unitarity constraints from $2\to 3$ and $2\to
4$ processes, as discussed in Ref.~\cite{Mahmud:2025wye}, also do not
provide meaningful constraints for $\sqrt{s} \lesssim 2$ TeV.
Therefore, we do not consider such constraints in our
analysis\footnote{We also note that the existing literature on
unitarity constraints in SMEFT generally does not consistently account
for dimension-6$^2$ contributions and thus cannot be directly applied
to our analysis, which includes such terms.}.

\subsection{A note on truncation: amplitude vs cross-section}
\label{sec:truncations}

\begin{figure}[tb]
  \centering
  \begin{tabular}{rr}
    \includegraphics[width=0.44\linewidth]{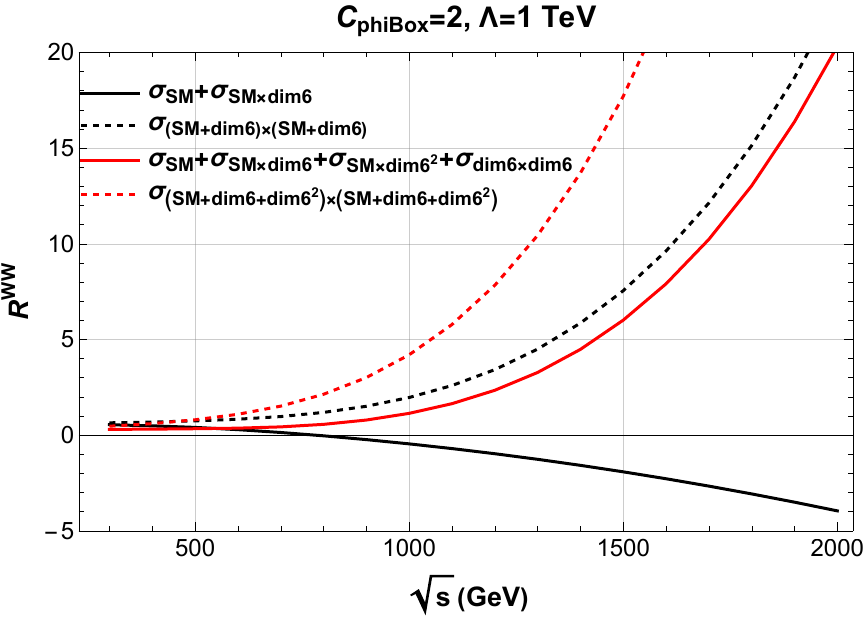} &
    \includegraphics[width=0.44\linewidth]{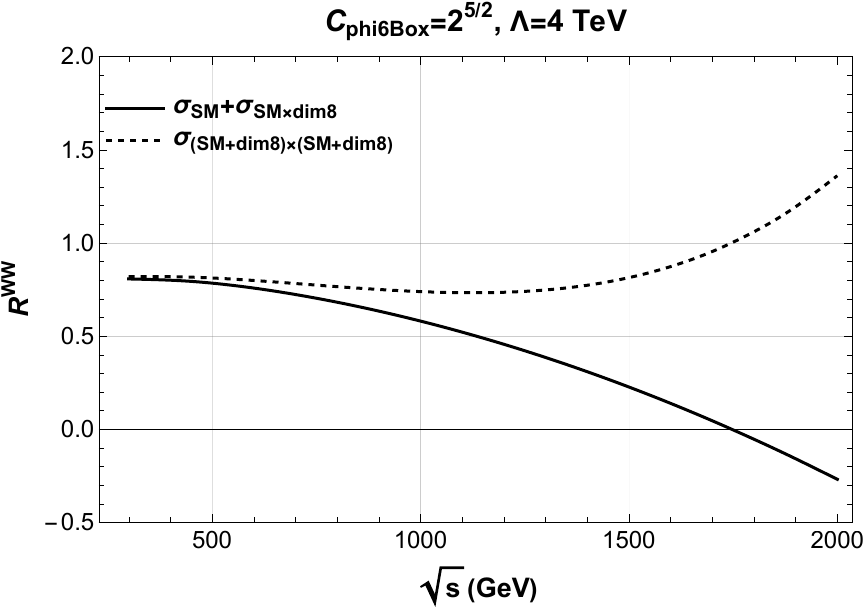} \\
  \end{tabular}
\caption{\sl Comparison of truncation schemes using the ratio $R^{WW}$
  between angular-integrated cross-sections in the SMEFT and the SM
  [\textit{c.f.} Eq.~\eqref{eq:RVV}] for $C_{\varphi \Box}$
  dimension-6 WC and $C_{\varphi^6\Box}$ dimension-8 WC (with
  numerical values displayed in the corresponding plots). Solid lines
  indicate a cross-section truncated to terms of the same maximal
  dimension as included in amplitude, dashed lines indicate a
  cross-section calculated by squaring amplitudes without any
  truncation.}
\label{fig:truncation} 
\end{figure}

The general structure of the EFT amplitude includes terms at the
leading order $\propto\frac{1} {\Lambda^2}$ (LO) and next-to-leading
order $\propto\frac{1}{\Lambda^4}$ (NLO) in the EFT expansion,
\begin{equation}
    \mathcal{M}=\mathcal{M}_{\text{SM}} +
    \frac{1}{\Lambda^2}\mathcal{M}_{\text{dim6}} +
    \frac{1}{\Lambda^4}\mathcal{M}_{\text{dim6}^2} +
    \frac{1}{\Lambda^4}\mathcal{M}_{\text{dim8}},
    \label{eq:M:trunc}
\end{equation}
where $\mathcal{M}_{\text{dim6}}$ contains single insertions of
dimension-6 WCs, $\mathcal{M}_{\text{dim6}^2}$ their double
insertions, and $\mathcal{M}_{\text{dim8}}$ single insertions of
dimension-8 WCs.  At the level of cross-section, this leads to
corrections of order as high as $\propto\frac{1}{\Lambda^8}$,
\begin{equation}
    \sigma\propto \left|\mathcal{M}\right|^2\equiv
    \sigma_{(\text{SM}+\text{dim6}+\text{dim6}^2+
      \text{dim8})\times(\text{SM}+\text{dim6}+\text{dim6}^2+
      \text{dim8})}\supset \frac{1}{\Lambda^8}
    \left|\mathcal{M}_{\text{dim6}^2}+\mathcal{M}_{\text{dim8}}\right|^2.
    \label{eq:sigma:trunc}
\end{equation}
In this work, we truncate only the amplitude in the EFT expansion.
This means we calculate the amplitude up to a specific order but
retain all higher-order terms, as those in Eq.~\eqref{eq:sigma:trunc},
that arise when squaring it to obtain the cross-section.  Truncation
at the cross-section level would omit terms like those presented in
Eq.~\eqref{eq:sigma:trunc}, which are formally of dimension-12.  While
Figure~\ref{fig:eftvsuv} shows that amplitude truncation does not
always guarantee better accuracy than cross-section truncation at a
given order, a significant drawback of cross-section truncation is its
potential to yield negative cross-section values.  This behaviour is
illustrated in Figure~\ref{fig:truncation}.

As a concrete example, let's discuss the issue of truncation within
the real scalar singlet model of this section.  At a very high energy,
the amplitude \eqref{eq:sfull} governs the dynamics of the
longitudinal vector bosons scattering $V_LV_L\to hh$ with $V=Z,W^\pm$,
\begin{equation}
  \mathcal{M}_{00}^{V_LV_L\to hh}(s) \ = -\lambda_{eff} \ +
  \ \alpha^2\, \biggl (\frac{s}{M^2} + \frac{s^2}{M^4} + \dots\biggr
  )\;,
    \label{eq:M00}
\end{equation}
where $\lambda_{eff} = \lambda - \alpha^2$.  Now we square
\eqref{eq:M00}, to find
\begin{align}
  |\mathcal{M}_{00}^{V_LV_L\to hh}(s)|^2 \ &=
  \overbrace{\underbrace{\lambda_{eff}^2 \ - \ 2\, \lambda_{eff} \,
      \alpha^2 \frac{s}{M^2}}_{\sigma_{1/\Lambda^{2}}} \ +\ \alpha^4\:
    \frac{s^2}{M^4}}^{|\mathcal{M}_{1/\Lambda^{2}}|^2}
  \ \ \ \overbrace{\underbrace{-2 \,\lambda_{eff}\, \alpha^2
      \,\frac{s^2}{M^4}}_{\sigma_{1/\Lambda^{4}}} \ +\ 2\, \alpha^2\,
    \frac{s^3}{M^6} \ +\ \alpha^4\,
    \frac{s^4}{M^8}}^{|\mathcal{M}_{1/\Lambda^{4}}|^2} \;.
\end{align}
Clearly, truncating the cross-section at order
$O(\sigma_{1/\Lambda^2})$ or $O(\sigma_{1/\Lambda^4})$ can lead to
negative values.  In the strong coupling regime $(\alpha \gtrsim 1)$,
this readily occurs for energies above $\sqrt{s} > \frac{M}{\alpha}
\sqrt{\frac{\lambda_{eff}}{2}}$.  Conversely, truncating the amplitude
in such cases allows the non-interference term to dominate, preventing
negative cross-sections.  Therefore, unless the coupling $\alpha$ is
weak, comparable to $\lambda_{eff}$, truncating the full amplitude is
the correct approach.  Issues related to truncation and EFT validity
have been discussed in Refs.~\cite{Heinrich:2022idm, Brivio:2022pyi,
  Allwicher:2024mzw}, where similar conclusions were reached.

\section{Helicity amplitudes and cross-sections}
\label{sec:DiHiggs:Ampls}

Double Higgs boson production via VBF is a complicated multi-particle
process in the SM, which becomes even more involved in SMEFT.  For
this reason, before performing the exact numerical calculations of the
cross-section for the $pp\to hhjj$ process, it is instructive to study
the anatomy of individual on-shell sub-amplitudes contributing to the
full amplitude.  More specifically, we study the on-shell scattering
of opposite-sign $W^\pm$ bosons or a pair of neutral $Z$-bosons into a
pair of Higgs bosons $h$, namely:
\begin{equation}
V (p_1,\lambda_1)+ V (p_2,\lambda_1) \rightarrow h(p_3) + h(p_4),
\end{equation}
where $V=W^\pm$ or $Z$, $p_i$ are the four-momenta of the initial and
final states, and $\lambda_i$ represent the polarizations of the
initial heavy vector bosons (two transverse — $\pm$ and longitudinal —
$0$).
The helicity amplitude for the $VV\rightarrow hh$ process can be
expressed as:
\begin{equation}
  \mathcal{M}^{VV}_{\lambda_1 \lambda_2}\equiv\mathcal{M}_{\lambda_1
    \lambda_2}^{VVHH}(\theta, s, G_F, M_W, M_Z, M_H, C_i)\;,
\end{equation}
where $\theta\in [0,\pi]$ is the scattering angle between the incoming
and outgoing particles in the Center of Mass frame.

Not all the 9 possible helicity configurations are independent.  As we
neglect CP-violating WCs, the following relations hold:
\begin{equation}
  \mathcal{M}^{VV}_{++}=\mathcal{M}^{VV}_{--},\quad
  \mathcal{M}^{VV}_{+-}=\mathcal{M}^{VV}_{-+},\quad
  \mathcal{M}^{VV}_{0\pm}=\mathcal{M}^{VV}_{\pm 0},
\end{equation}
leaving a total of 4 independent helicity structures.  The
differential cross-section for a given helicity is given by,
\begin{equation}
\biggl ( \frac{d \sigma^{VV}}{d \Omega} \biggr )_{\lambda_1 \lambda_2
} = \frac{1}{2!} \, \frac{1}{9} \, \frac{1}{64 \pi^2 s} \:
|\mathcal{M}^{VV}_{\lambda_1 \lambda_2}(\theta, s, G_F, M_W, M_Z, M_H,
C_i) |^2 \;,
\end{equation} 
and the helicity cross-section,
\begin{equation}
\sigma(s)^{VV}_{\lambda_1 \lambda_2 } = \int_0^{2\pi} d\varphi
\int_{\theta_{cut}}^{\pi -\theta_{cut}} d\theta \sin\theta \: \biggl (
\frac{d \sigma^{VV}}{d \Omega} \biggr )_{\lambda_1 \lambda_2 } \;.
\end{equation}
By summing over all relevant helicities, the total cross-section is:
\begin{equation}
    \sigma(s)^{VV}=\sigma(s)^{VV}_{00}+2\sigma(s)^{VV}_{++} +
    2\sigma(s)^{VV}_{+-}+4\sigma(s)^{VV}_{0+}\;.
\end{equation}
It is important to note that, as discussed in
Sec.~\ref{sec:DiHiggs:UVScalar}, although we consistently truncate all
amplitudes to dimension-8 level, we {\em do not} truncate expressions
for cross-sections in the same way, keeping instead all terms arising
from squaring the amplitudes.  This means that we include in cross
section calculations also the subset of terms up to dimension-12 (such
subset is gauge-invariant\footnote{ We have analytically and
explicitly checked (with the help of {\tt
  SmeftFR}-code~\cite{Dedes:2023zws}) that the amplitude $VV\to hh$ is
gauge invariant, i.e., gauge-fixing $\xi$-parameter independent, at
tree level for dim6 and dim6$^2$ contributions alone.  We expect the
same holds for the dim8 amplitude terms, although we have not verified
the latter cases.}).
For large values of WCs it can lead to numerically significant
differences comparing to cross-sections truncated to dimension-8 terms
only.

Explicit formulae for the leading order helicity amplitudes are
provided in Appendix~\ref{App:DiHiggs:WWHH:Ampl} for the
$WW\rightarrow HH$ process and in Appendix~\ref{App:DiHiggs:ZZHH:Ampl}
for the $ZZ\rightarrow HH$ process.  All amplitudes and corresponding
numerical results in this section were generated using the following
chain of programs:
\begin{center}
  {\tt SmeftFR~\cite{Dedes:2023zws} $\to$ FeynArts~\cite{Hahn:2000kx}
    $\to$ FeynCalc~\cite{Shtabovenko:2020gxv}}
\end{center}

The new feature of this analysis is the {\em consistent} inclusion of
$O(1/\Lambda^4)$ dimension-6$^2$ terms.  Neglecting such terms e.g. in
normalisation factors $Z_h, Z_W, \ldots$ between weak eigenbasis and
physical fields (see Refs.~\cite{Dedes:2017zog,Dedes:2023zws}), leads
to error by a factor of ${\cal O}(10)$ in the amplitudes when the
dimension-6$^2$ terms are relevant.  Even more important is the fact
that, failure of a consistent truncation in $Z$-normalisation factors
here would lead to gauge-fixing parameter dependent results and
therefore meaningless $S$-matrix elements.  This omission may turn out
to be important in setting bounds to Wilson coefficients or
predictions for physical observables in SMEFT.

Inspecting the formulae for helicity amplitudes displayed in
Appendices~\ref{App:DiHiggs:WWHH:Ampl} and~\ref{App:DiHiggs:ZZHH:Ampl}
one can see that contributions from some dimension-6 operators
($Q_{\varphi\Box}, Q_{\varphi D}, Q_{\varphi B}, Q_{\varphi W},
Q_{\varphi WB}$) may grow linearly with $s$.  Including
dimension-6$^2$ terms does not change this conclusion, the new
contributions have the form $C^2 s/(G_F \Lambda^4)$.  However,
corrections to amplitudes due to some dimension-8 operators
($Q_{\varphi^4 D^4}^{(1)}, Q_{\varphi^4 D^4}^{(2)}, Q_{\varphi^4
  D^4}^{(3)}, Q_{B^2\varphi^2 D^2}^{(1)}, Q_{W^2\varphi^2 D^2}^{(1)},
Q_{W^2\varphi^2 D^2}^{(2)} $) grow faster with energy, like $s^2$.

\begin{figure}[htb!]
  \centering
  \begin{tabular}{rr}
    \includegraphics[width=0.4\linewidth]{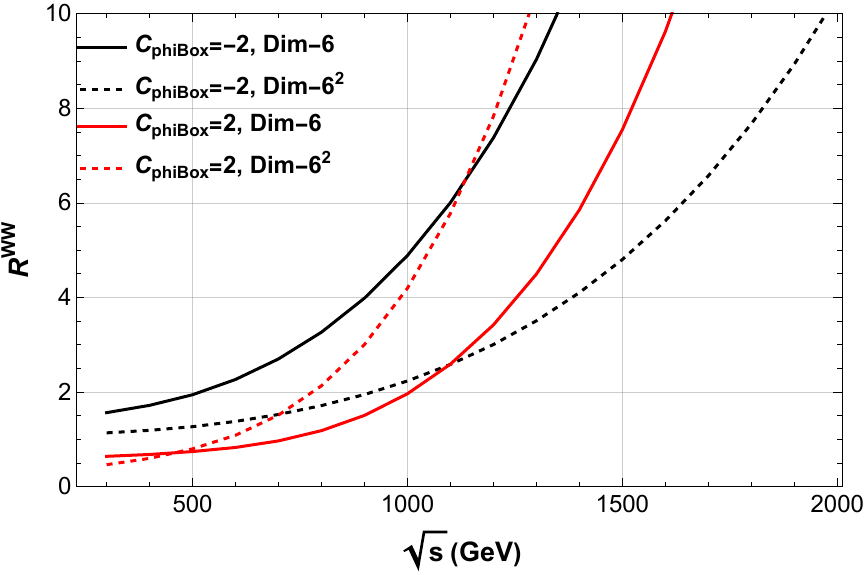} &
    \includegraphics[width=0.4\linewidth]{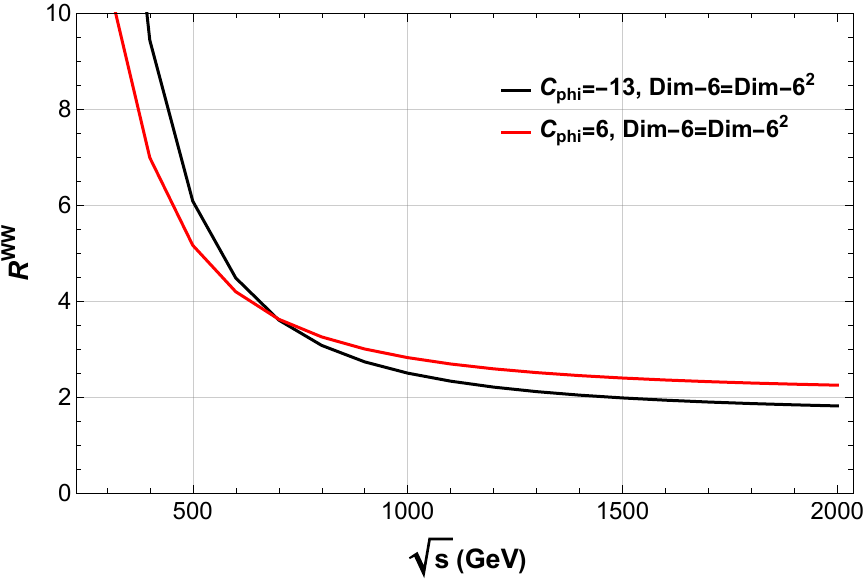} \\ 
    \includegraphics[width=0.4\linewidth]{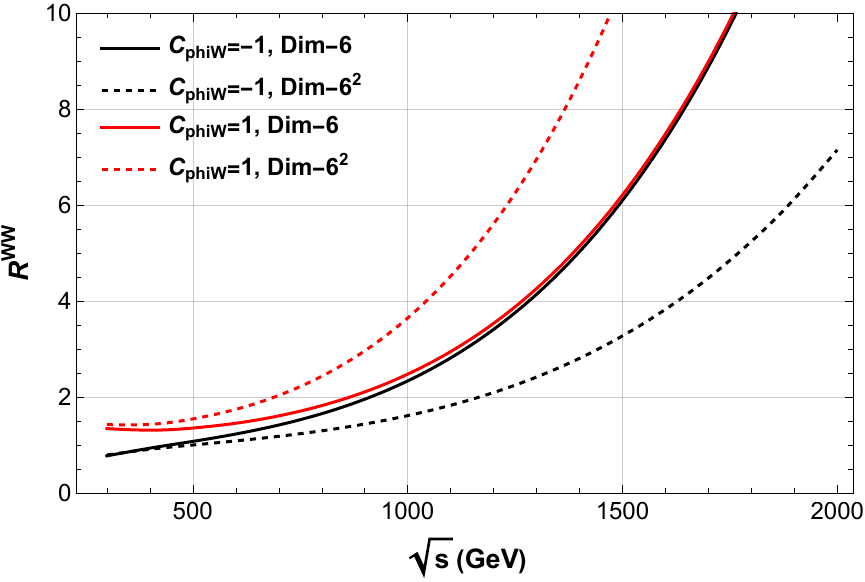} &
    \includegraphics[width=0.4\linewidth]{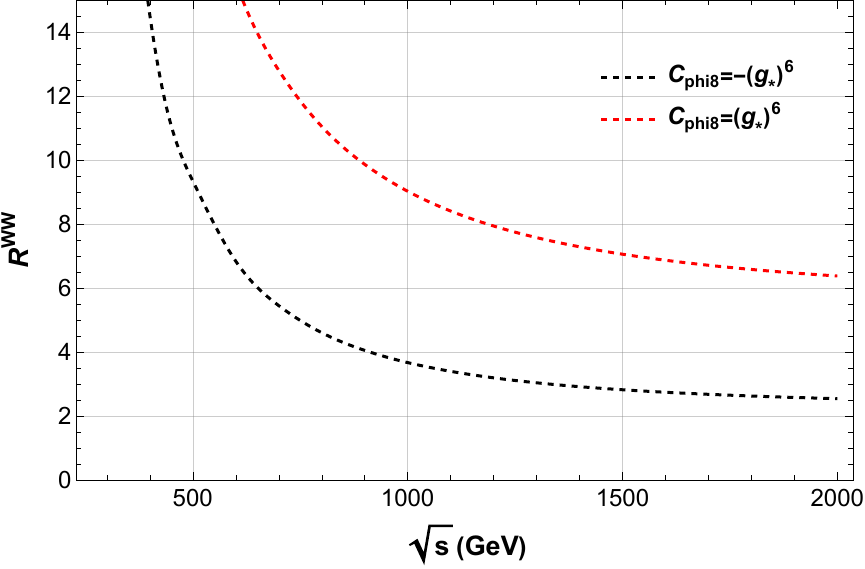  } \\
    \includegraphics[width=0.4\linewidth]{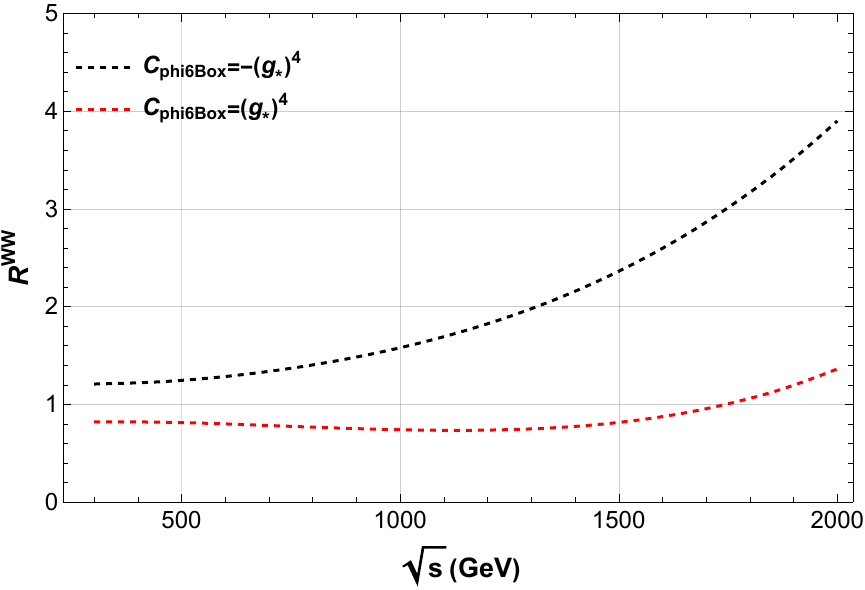} &
    \includegraphics[width=0.4\linewidth]{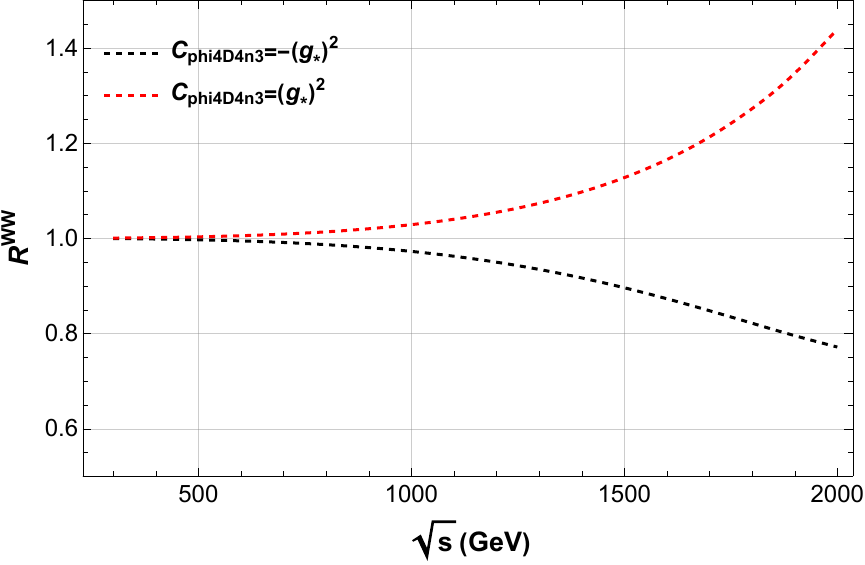  } \\
    \includegraphics[width=0.42\linewidth]{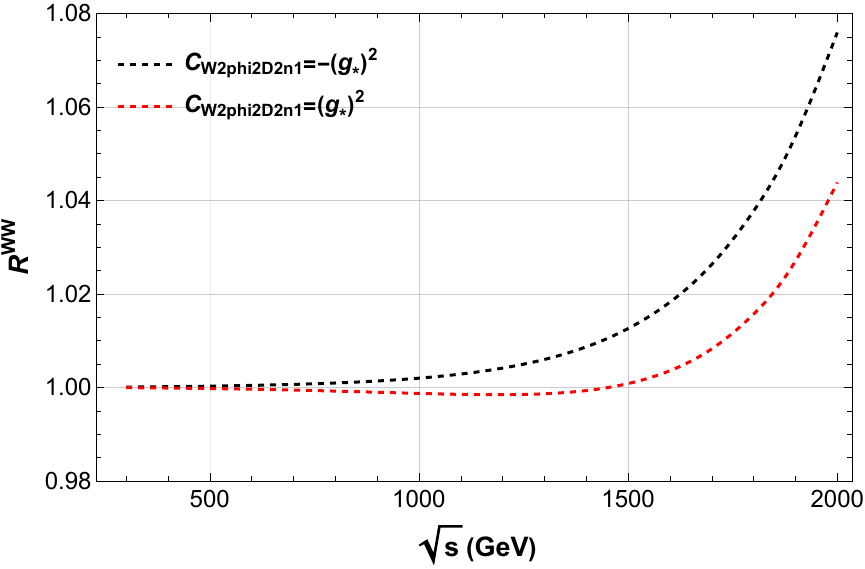} &
    \includegraphics[width=0.4\linewidth]{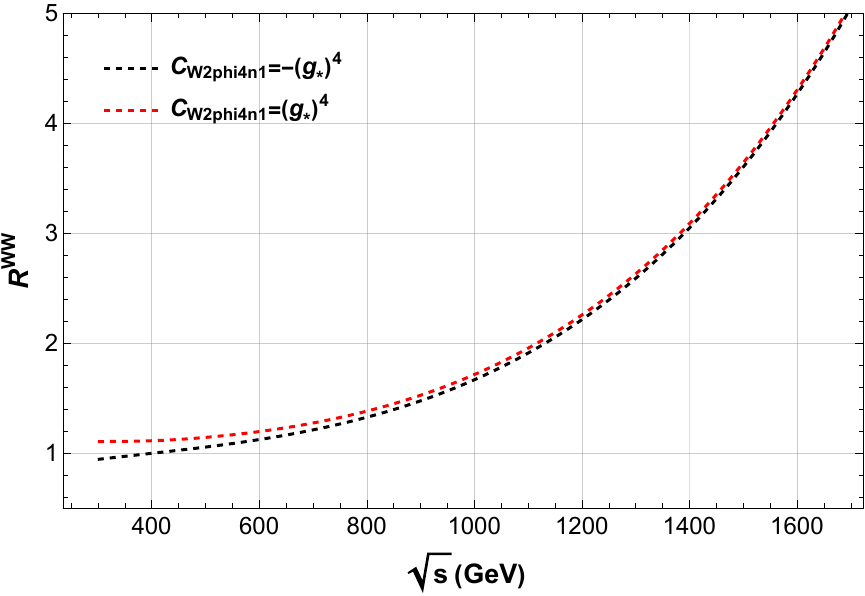  }
  \end{tabular}
\caption{\sl Ratio $R^{WW}$ between angular-integrated cross-sections
  in the SMEFT and the SM for chosen dimension-6 WCs $C_{\varphi
    \Box}$, $C_{\varphi}$, $C_{\varphi W}$, including dimension-6
  (solid lines) and dimension-6$^2$ (dashed lines) terms, and for
  dimension-8 WCs $C_{\varphi^8}$, $C_{\varphi^6\Box}$, $C_{\varphi^4
    D^4}^{(3)}$, $C_{W^2\varphi^2 D^2 }^{(1)}$, $C_{W^2\varphi^4
  }^{(1)}$.  Values of dimension-6 WCs are given on corresponding
  plots and $\Lambda=1$ TeV is assumed.  For dimension-8 WCs, NDA
  scaling is assumed with numerical values of
  $g_*=\sqrt{2}\frac{\Lambda}{1\text{ TeV}}$ and $\Lambda=4$ TeV.}
    \label{fig:CSWWHH}
\end{figure}

In Figure~\ref{fig:CSWWHH} we illustrate possible enhancement of the
$ZZ\to HH$ cross-sections in SMEFT due to effects growing with energy.
We present results in terms of the ratio to the SM cross-section,
defined as:
\begin{equation}
R^{VV}=\frac{\sigma_{SMEFT}(VV\to HH)}{\sigma_{SM}(VV\to HH)}\;,\quad
V=(W^\pm,\, Z)\;.
\label{eq:RVV}
\end{equation} 
We show only the results for the $W^+W^-\rightarrow HH$ process, as
for all operators which contribute to $ZZ$ scattering their
$s$-dependence is qualitatively similar to the $W^+W^-$ case.  For all
plots and numerical results, we assume one non-vanishing WC at a time
and its sign is chosen to maximise the cross-section.

Plots in Figure~\ref{fig:CSWWHH} are chosen to illustrate several
important points.  First, one can see that for dimension-6 WCs values
close to the upper bounds dictated by fits to experimental data (see
Table~\ref{tab:fits}), dimension-6$^2$ terms become dominant for
$\sqrt{s}\gtrsim 600~\mathrm{GeV}$ (excluding $C_\varphi$, for which
there are no double insertions at the level of the amplitude, hence
the LO and NLO results are identical).
Moreover, one can expect significant enhancement of the $VV\rightarrow
HH$ cross-section from all the classes of dimension-6 WCs, with
specific examples presented in Figure~\ref{fig:CSWWHH}: $\varphi^6$:
$C_\varphi$, $\varphi^4D^2$: $C_{\varphi\Box}$ and $X^2\varphi^2$:
$C_{\varphi W}$.  For dimension-8 WCs, the most promising classes are:
$\varphi^8$: $C_{\varphi^8}$, $\varphi^6D^2$: $C_{\varphi^6\Box}$ and
$X^2\varphi^4$: $C_{W^2\varphi^4}^{(1)}$.  The relative magnitude of
enhancement from various WCs is primarily dependent on their maximal
allowed values given by numerical fits or NDA.  The $s$-dependence
can, for some operators, further significantly increase the $R^{VV}$
ratio, but one should remember that in the high-$s$ range, the number
of produced events is lower and statistical errors of measurements
larger, so this tail effect may be more difficult to be observed.
However, even if not relevant for the HL-LHC, it might become more of
significance for future planned experiments such as FCC-hh.

\section{Maximal enhancement of the double Higgs production via VBF in SMEFT}
\label{sec:DiHiggs:Num}

We now turn our attention towards finally answering the question:
\textit{What is the maximum enhancement in SMEFT of the double Higgs 
boson production rate through vector boson fusion at the HL-LHC?}

The answer requires an analysis of the full process $pp\to hh +
\mathrm{jets}$, rather than just the $VV\to hh$ sub-process which we
discussed in detail in the previous sections, to analytically
understand the most important effects of the higher-dimensional SMEFT
operators.  To this end, we performed a numerical Monte Carlo analysis
of this process for the LHC and its upgrade HL-LHC.  The results were
obtained using {\tt MadGraph 5 - v3\_4\_1}~\cite{Alwall:2011uj} event
generator utilising {\tt UFO}~\cite{Degrande:2011ua} model files
generated with {\tt SmeftFR v3.02}~\cite{Dedes:2023zws}.  The process
of interest is:
\begin{center}
{\tt p p > h h j j} 
\end{center}
where {\tt p} represents the initial protons, {\tt h} the Higgs
bosons, and {\tt j} the final-state jets.  We utilised the
VBF-specific cuts on kinematic variables that are based on the ATLAS
recommendations (following~\cite{Kalinowski:2018oxd, Bishara:2016kjn,
  Dedes:2020xmo}) on $M_{jj}$ - invariant mass of a pair of jets,
$\Delta \eta_{jj}$ - the rapidity separation of jets, $\eta_{j}$ - the
rapidity of individual jets, $\Delta p_{Tj}$ - the transverse momentum
of jets, and $M_{hh}$ - the invariant mass of the pair of Higgs
bosons:
\begin{equation}
M_{jj}\geq 700\text{ GeV},\quad \Delta \eta_{jj}\geq 5,\quad
|\eta_{j}|\leq 4.5,\quad p_{Tj}\geq 25\text{ GeV},\quad M_{hh}\geq
400\text{ GeV}.
\end{equation}
It should be noted that a more detailed study including Higgs decays
to final states, may require more refined and final-state-specific
cuts on kinematic variables.

Moreover, we assumed all fermion masses — apart from the top quark —
to be equal to zero, and the flavour mixing matrices to be equal to
identity.  The numerical values of input parameters are in line with
the default \sfr v3 “$G_F$'' default input scheme and read:
\begin{equation}
\begin{aligned}
G_F & = 1.1638\times 10^{-5}\text{ GeV}^{-2}\;, \quad
M_Z=91.1876\text{ GeV}\;, \quad M_W=80.379\text{ GeV}\;,\\
M_H&=125.35~\text{GeV}\;,\quad M_t=172.76~\text{GeV}\;.
\end{aligned}
\end{equation}
We run the simulations for the LHC and HL-LHC experiments, in the rest
of the paper assuming the energy $\sqrt{s}=13$ TeV and integrated
luminosity of $\mathcal{L}=30$ fb$^{-1}$ for the LHC and $\sqrt{s}=14$
TeV, $\mathcal{L}=3000$ fb$^{-1}$ for the HL-LHC.

As the maximal values of Wilson coefficients allowed by current
experimental data (eventually extrapolated using the NDA assumption)
we adopt their numerical values listed in
Section~\ref{subsec:DiHiggs:Assumptions}. As some of them depend on
the new physics scale $\Lambda$, we employ the EFT validity criterion
of Eq.~\eqref{eq:validity} to determine its minimal value, limited by
the condition that less than $1\%$ of all generated Higgs pairs have
invariant mass higher than $\Lambda$ (see Sec.~\ref{sec:dim8numerics}
for more details).

We start with a study assuming one non-zero WC at a time (with an
exception to $C_{\varphi}$ and $C_{\varphi^8}$ affecting
$\kappa_{\lambda_3}$, see discussion in
Section~\ref{subsec:KappaLambda:Num}).  For each run, we calculate the
cross-section and the number-of-event enhancements understood as:
\begin{equation}
  \Delta\sigma = \frac{\sigma^{SMEFT}}{\sigma^{SM}},\quad \Delta N =
  N^{SMEFT} - N^{SM}\;.
  \label{eq:DiHiggs:DeltaN}
\end{equation}
Whenever necessary, we further distinguish the contributions from
various subclasses of terms, e.g.  $\Delta\sigma_{D6},
\Delta\sigma_{D6^2}, \Delta\sigma_{D8}$ denote cross-section
enhancement considering, respectively, dimension-6 terms only,
dimension-6 and dimension-6$^2$ terms or dimension-8 terms only (and
similarly for event excesses $\Delta N_{D6}, \Delta N_{D6^2}, \Delta
N_{D8}$). We use such a self-explanatory notation in all following
Tables and Figures.

We display the results for both positive and negative ends of the
allowed WC ranges.  Moreover, we present kinematic distributions of
Higgs pair invariant mass $M_{hh}$ to illustrate where the BSM effects
can be easiest to identify.

\subsection{WCs affecting Higgs self-couplings}
\label{subsec:KappaLambda:Num}

\begin{figure}[htb!]
  \centering
  \begin{tabular}{cc}
    \includegraphics[width=0.475\linewidth]{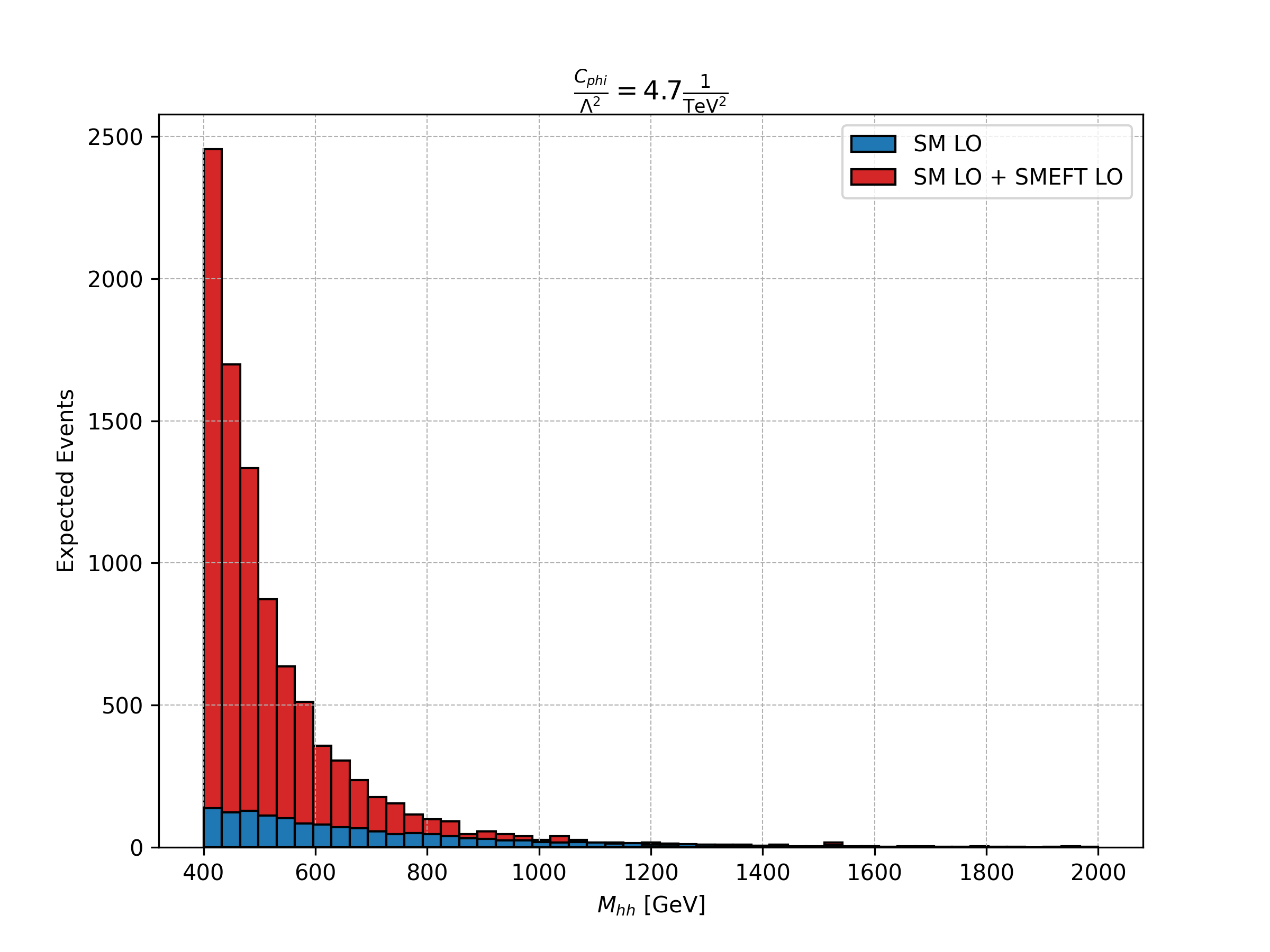} &
    \includegraphics[width=0.475\linewidth]{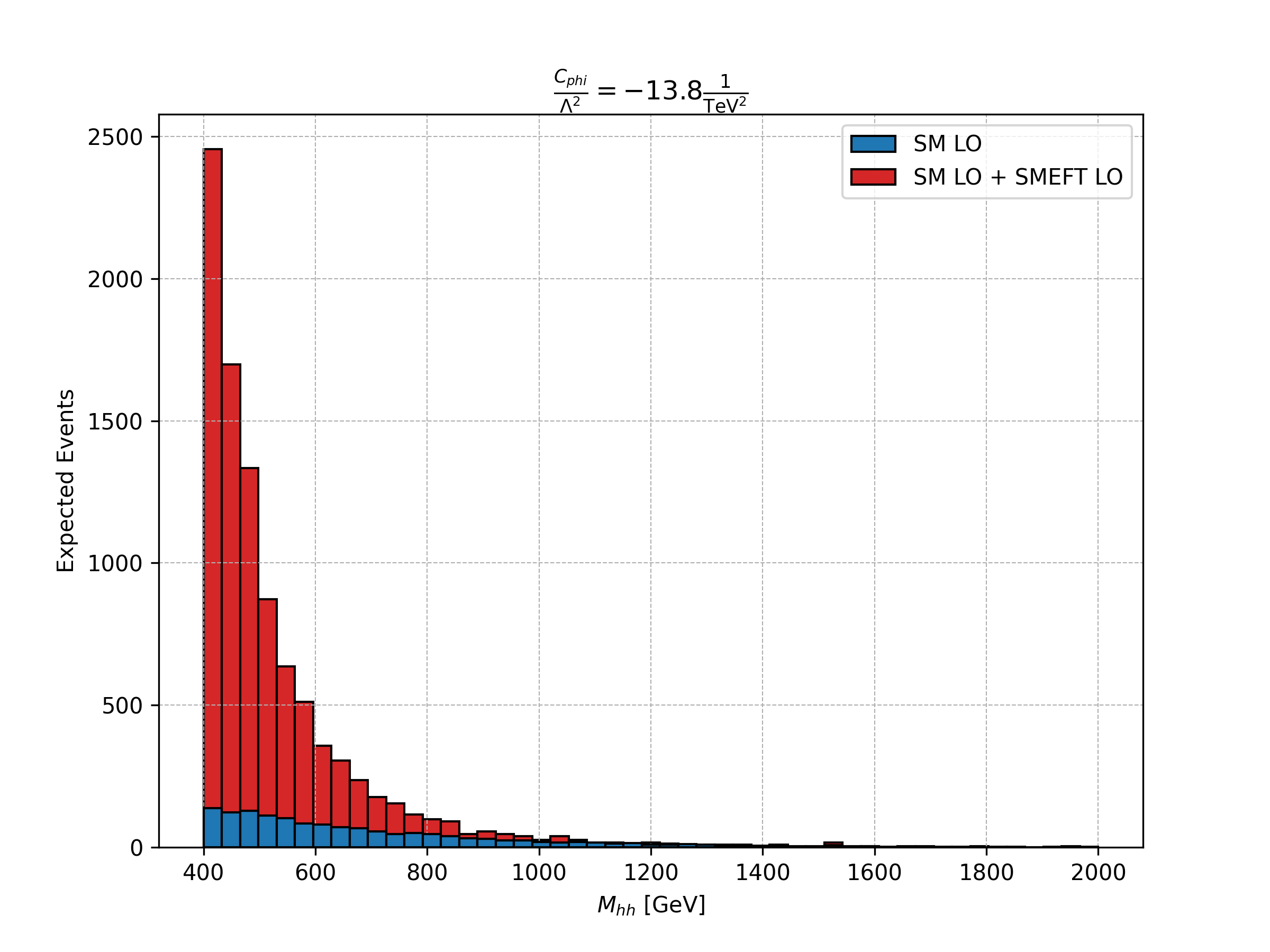}
  \end{tabular}
  \caption{\sl Higgs pair invariant mass $M_{hh}$ distribution for
    $C_\varphi$ modifying $\kappa_{\lambda_3}$.  }
    \label{fig:Dim6:Histos:KappaLambda}
\end{figure}

We start by analysing the enhancement associated with the triple Higgs
coupling modifier $\kappa_{\lambda_3}$.  As discussed in
Section~\ref{subsec:DiHiggs:Assumptions}, experimental constraints on
$\kappa_{\lambda_3}$ can be translated into bounds on two Wilson
coefficients that primarily affect this coupling: $C_\varphi$ and
$C_{\varphi^8}$.  However, as can be seen from the formulae in
Appendices~\ref{App:DiHiggs:WWHH:Ampl}
and~\ref{App:DiHiggs:ZZHH:Ampl}, the dominant contributions from both
WCs affect mainly longitudinal scattering amplitudes
$\mathcal{M}^{WW(ZZ)}_{00}$ and enter into relevant expression in the
same linear combination as into $\kappa_{\lambda_3}$ (note also that
there are no $C_\varphi^2$ contributions, distinguishing LO and NLO
orders of EFT expansion).  Therefore, bounds on di-Higgs production
are in this case directly given by current limits on
$\kappa_{\lambda_3}$ [see Eq.~\eqref{eq:kappal3}], independently of
individual values of $C_\varphi$ and $C_{\varphi^8}$. For
illustration, we present them assuming $C_\varphi \neq 0$ and
$C_{\varphi^8} = 0$ using limits of Eq.\eqref{eq:Cphi:num}. The
numerical results of the impact of these Wilson coefficients on
di-Higgs production are shown in Table~\ref{Tab:pphhjj:KappaLambda},
while the invariant mass distributions are displayed in
Figure~\ref{fig:Dim6:Histos:KappaLambda}.  As can be seen from
Table~\ref{Tab:pphhjj:KappaLambda}, the enhancement can be
substantial, especially in the negative sign case.  The distributions
in Figure~\ref{fig:Dim6:Histos:KappaLambda} show, consistent with
Figure~\ref{fig:CSWWHH}, that the most significant excess of events is
expected at low energies, with $M_{hh} = \sqrt{s} \lesssim 1000$ GeV.

It should be noted that while in this instance the enhancement can
reach factor $\mathcal{O}(20)$, gluon fusion would also cause a
similar excess in the dominant production channel, resulting in a much
earlier measurement.  For this reason, we consider other Wilson
coefficients, endemic for VBF double Higgs production, to be more
interesting, as discussed in the following sections.

\begin{table}[htb!]
  \begin{center}
    \renewcommand{\arraystretch}{1} 
    \begin{tabular}{cc}
    \begin{tabular}{|c|c||c|c|}
    \hline
\multicolumn{4}{|c|}{\textbf{LHC}}\\
    \hline
\multicolumn{2}{|c||}{WC} & $\Delta\sigma$ & $\Delta N$ \\
    \hline
\multirow{2}{*}{$\frac{C_\varphi}{1\text{ TeV}^{2}}$} &
    $4.7$ & 6.0 & 81 \\
    \cline{2-4} & $-13.8$ & 22.2 & 336 \\
    \hline
    \end{tabular}
    &
    \begin{tabular}{|c|c||c|c|}
    \hline
    \multicolumn{4}{|c|}{\textbf{HL-LHC}}\\
    \hline
\multicolumn{2}{|c||}{WC} & $\Delta\sigma$ & $\Delta N$ \\
    \hline
    \multirow{2}{*}{$\frac{C_\varphi}{1\text{ TeV}^{2}}$} & $4.7$ &
    5.9 & 9550 \\
    \cline{2-4} & $-13.8$ & 21.5 & 39784 \\
    \hline
    \end{tabular}
  \end{tabular}
\end{center}
  \caption{\sl Maximal enhancement of the di-Higgs production via VBF
    for $C_\varphi$ modifying $\kappa_{\lambda_3}$.  }
  \label{Tab:pphhjj:KappaLambda}
\end{table}

\subsection{WCs affecting Higgs-vector boson interactions}

\subsubsection*{Dimension-6 operators}

In the case of dimension-6 Wilson coefficients, we assume numerical
values based on the results of global fits cited in
Table~\ref{tab:fits}.  We use wider, “marginalised'' bounds, as we
first aim to examine their individual effects, keeping in mind that,
in realistic scenarios, they never appear in isolation.  We illustrate
the magnitude of difference in results when only single insertions of
dimension-6 WCs in the amplitudes are allowed (i.e. linear order),
with the case in which we allow double insertions (quadratic order).
Moreover, as argued in Section~\ref{sec:truncations}, we do not
truncate higher order EFT terms at the level of the cross-section as
the consistent and correct choice.

\begin{figure}[htb!]
  \centering
  \begin{tabular}{cc}
    \includegraphics[width=0.47\linewidth]{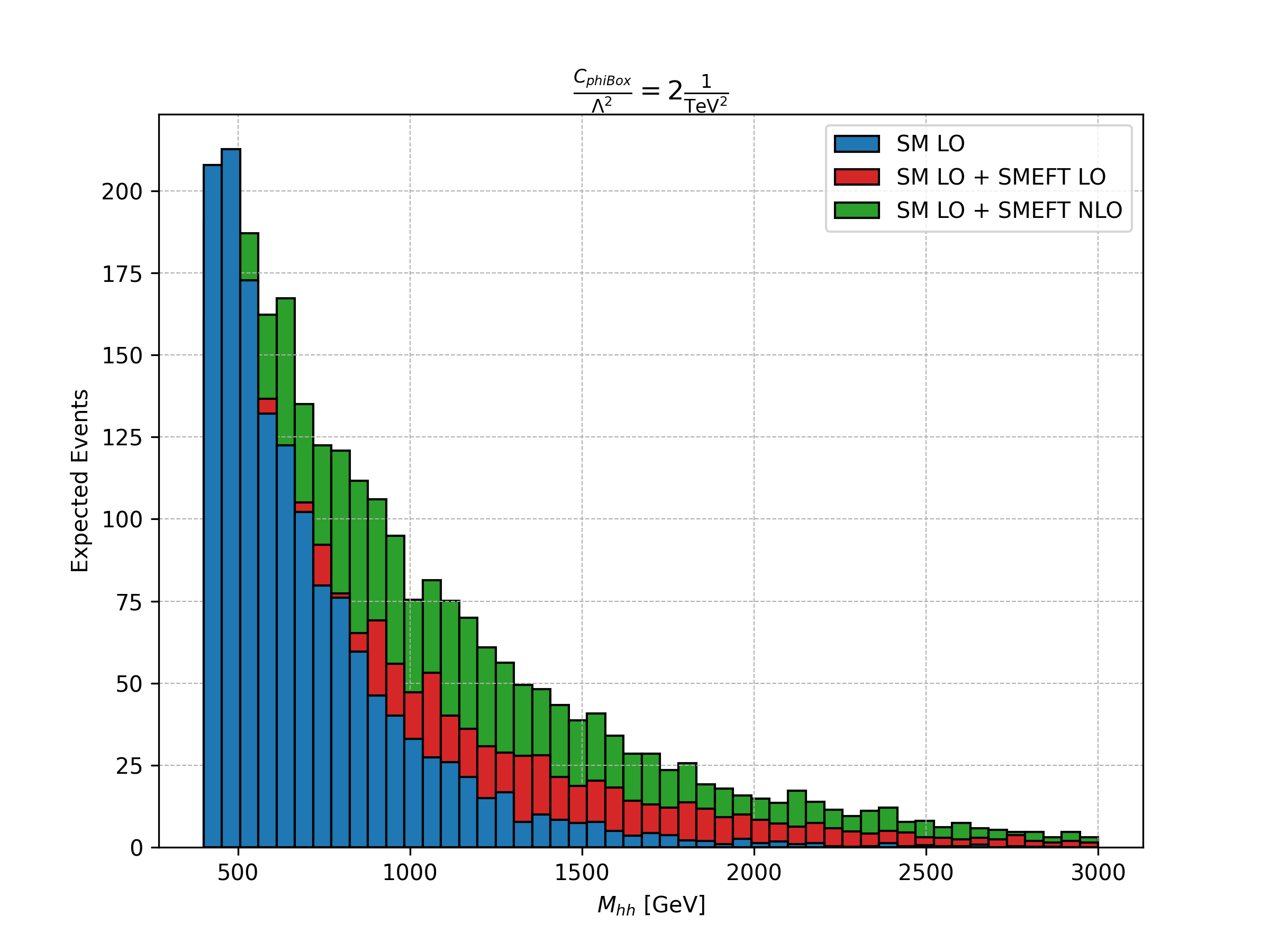} &
    \includegraphics[width=0.47\linewidth]{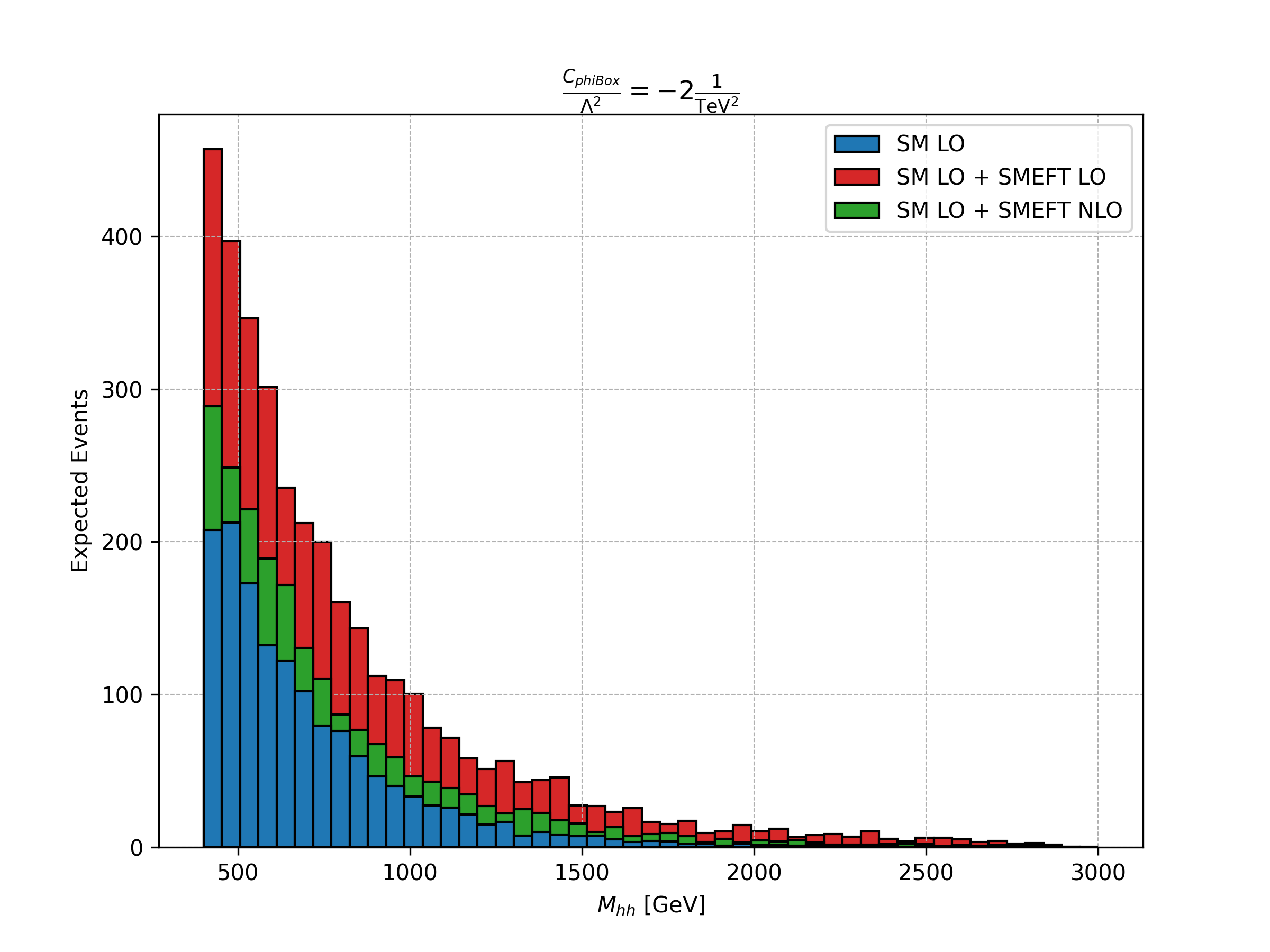} \\
    \includegraphics[width=0.47\linewidth]{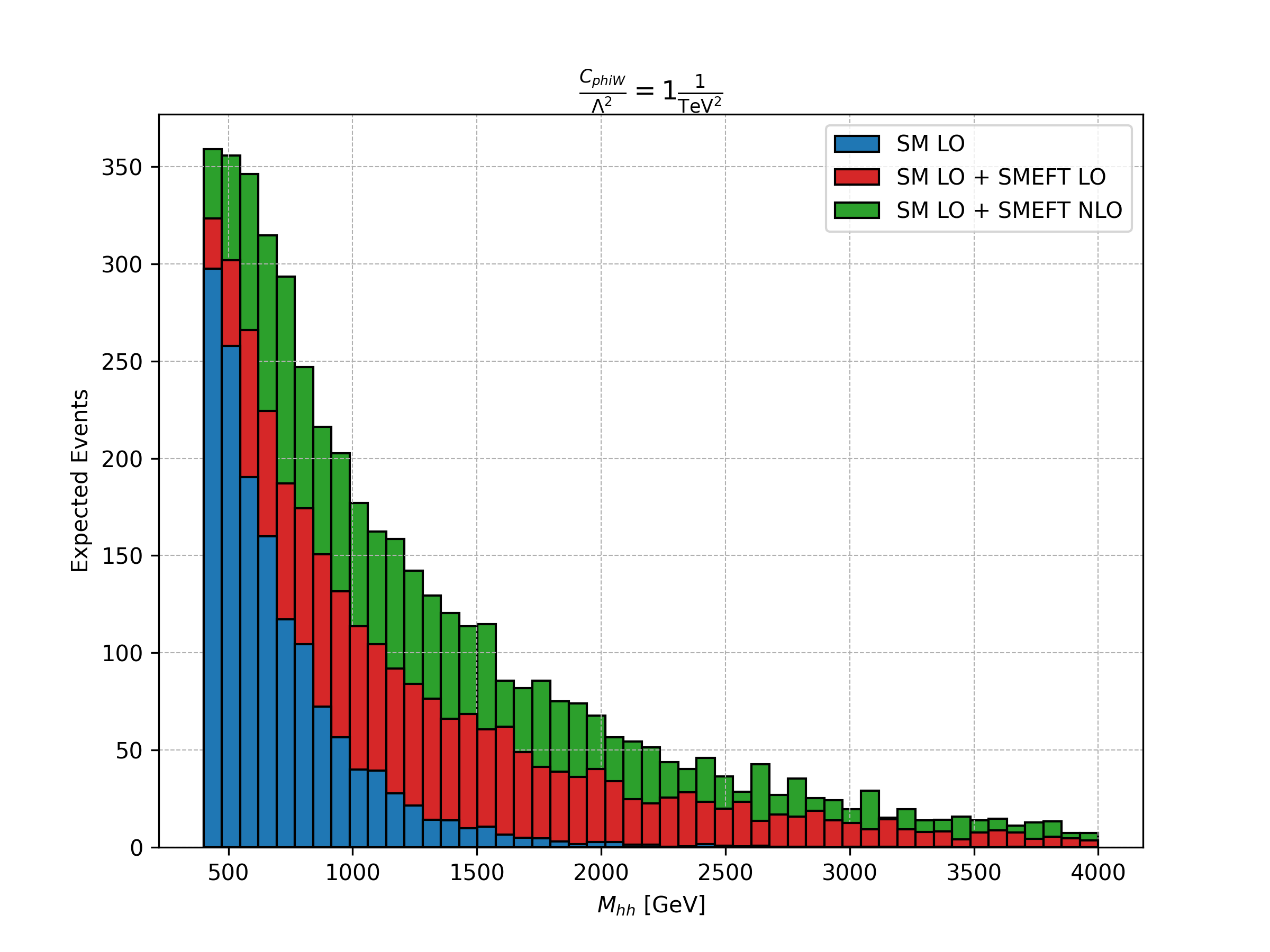} &
    \includegraphics[width=0.47\linewidth]{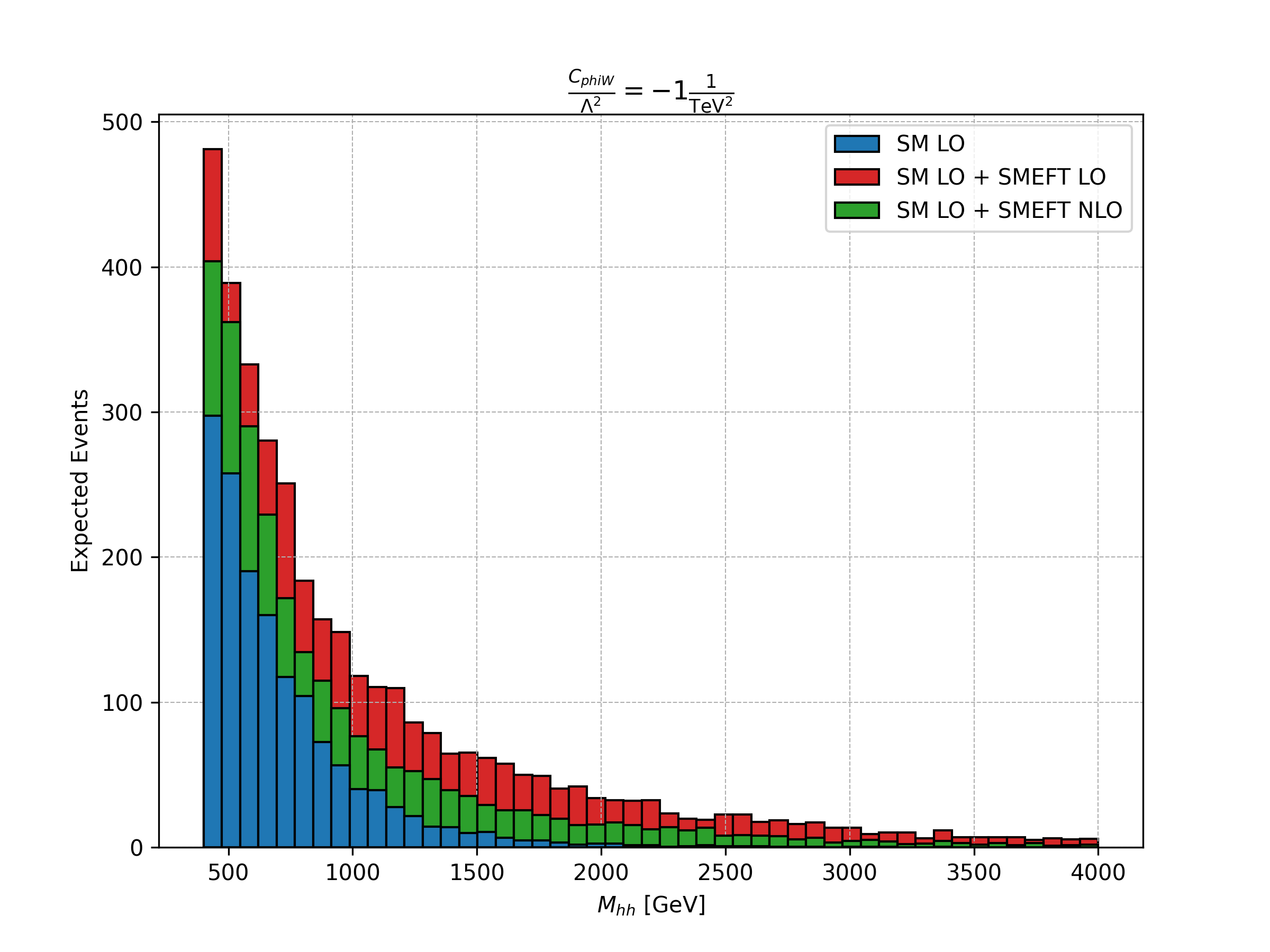  }
  \end{tabular}
  \caption{\sl Higgs pair invariant mass $M_{hh}$ distribution for
    $C_{\varphi\Box}$ and $C_{\varphi W}$.  }
  \label{fig:Dim6:Histos}
\end{figure}

Our results are presented in Table~\ref{Tab:pphhjj:Dim6} and
illustrated in Figure~\ref{fig:Dim6:Histos} for the $C_{\varphi\Box}$
and $C_{\varphi W}$ WCs, which correspond to the strongest
enhancements.  These results indicate that the strongest enhancement
of di-Higgs VBF production due to dimension-6 Wilson coefficients at
the HL-LHC reaches at most 2.7 for $C_{\varphi W}$, and is
significantly lower for all remaining WCs.  Furthermore, they
highlight the importance of NLO (dimension-6$^2$) contributions in
comparison to LO (dimension-6) ones.  For example, in the case of
$C_{\varphi\Box} = -2$, the LO contribution is much larger than the
NLO one, consistent with the behaviour shown in the plots in
Figure~\ref{fig:CSWWHH}.  This is further illustrated by the Higgs
invariant mass histograms in Figure~\ref{fig:Dim6:Histos}.  The reason
for this difference is the negative interference between the SM, LO,
and NLO contributions in the negative-sign case, which can be examined
in Eq.~\eqref{eq:DiHiggs:WWHH:Dim6}.  The histograms in
Figure~\ref{fig:Dim6:Histos} also demonstrate the energy growth of the
cross-section, as can be seen in Figure~\ref{fig:CSWWHH}.  Finally,
from these results, one can deduce the EFT validity scale in line with
the requirement of Eq.~\eqref{eq:validity}: $\Lambda \gtrsim 3$ TeV
for $C_{\varphi\Box}$ and $\Lambda \gtrsim 4$ TeV for $C_{\varphi W}$.

\begin{table}[tb!]
  \begin{center}
    \begin{adjustbox}{width=\textwidth, max width=\textwidth,
        max totalheight=\textheight, keepaspectratio}
  \begin{tabular}{cc}
    \begin{tabular}{|c|c||c|c||c|c|}
    \hline
    \multicolumn{6}{|c|}{\textbf{LHC}}\\
    \hline
    WC & $\frac{C}{1\text{ TeV}^2}$ & $\Delta\sigma_{D6}$ & $\Delta
    N_{D6}$ & $\Delta\sigma_{D6^2}$ & $\Delta N_{D6^2}$ \\
    \hline
    \multirow{2}{*}{$C_{\varphi \Box}$} & $2$ & 1.04 & 1 & 1.53 & 9
    \\
    \cline{2-6} & $-2$ & 2.41 & 23 & 1.41 & 7 \\
    \hline
    \multirow{2}{*}{$C_{\varphi D}$} & $1.5$ & 1.08 & 1 & 1.07 & 1
    \\
    \cline{2-6} & $-1.5$ & 1.00 & 0 & 1.02 & 0 \\
    \hline
    \multirow{2}{*}{$C_{\varphi W}$} & $1$ & 1.74 & 12 & 2.49 & 24
    \\
    \cline{2-6} & $-1$ & 2.20 & 19 & 1.62 & 10 \\
    \hline
    \multirow{2}{*}{$C_{\varphi WB}$} & $1$ & 1.27 & 4 & 1.14 & 2
    \\
    \cline{2-6} & $-1$ & 1.23 & 4 & 1.41 & 7 \\
    \hline
    \multirow{2}{*}{$C_{\varphi B}$} & $1$ & 1.39 & 6 & 1.69 & 11 \\
    \cline{2-6} & $-1$ & 1.38 & 6 & 1.17 & 3 \\
    \hline
    \end{tabular}
    &
    \begin{tabular}{|c|c||c|c||c|c|}
    \hline
    \multicolumn{6}{|c|}{\textbf{HL-LHC}}\\
    \hline
    WC & $\frac{C}{1\text{ TeV}^2}$ & $\Delta\sigma_{D6}$ & $\Delta
    N_{D6}$ & $\Delta\sigma_{D6^2}$ & $\Delta N_{D6^2}$ \\
    \hline
    \multirow{2}{*}{$C_{\varphi \Box}$} & $2$ & 1.07 & 134 & 1.59 &
    1144 \\
    \cline{2-6} & $-2$ & 2.42 & 2770 & 1.41 & 800 \\
    \hline
    \multirow{2}{*}{$C_{\varphi D}$} & $1.5$ & 1.07 & 140 & 1.07 & 131
    \\
    \cline{2-6} & $-1.5$ & 1.01 & 23 & 1.02 & 38 \\
    \hline
    \multirow{2}{*}{$C_{\varphi W}$} & $1$ & 1.85 & 1645 & 2.72 & 3340
    \\
    \cline{2-6} & $-1$ & 2.31 & 1504 & 1.67 & 1303 \\
    \hline
    \multirow{2}{*}{$C_{\varphi WB}$} & $1$ & 1.28 & 540 & 1.14 & 275
    \\
    \cline{2-6} & $-1$ & 1.25 & 481 & 1.44 & 848 \\
    \hline
    \multirow{2}{*}{$C_{\varphi B}$} & $1$ & 1.40 & 773 & 1.72 & 1396
    \\
    \cline{2-6} & $-1$ & 1.39 & 756 & 1.17 & 336 \\
    \hline
    \end{tabular}
  \end{tabular}
  \end{adjustbox}
  \end{center}
  \caption{\sl Maximal enhancement of VBF di-Higgs production process
    for dimension-6 WCs.  }
  \label{Tab:pphhjj:Dim6}
\end{table}

\subsubsection*{Dimension-8 operators}
\label{sec:dim8numerics}

In the case of the dimension-8 Wilson coefficients considered in this
paper, we assume their numerical values in line with
Eq.~\eqref{eq:maxgstar:unprime}, leading to coefficients $C_{\varphi^4
  D^4}$, $C_{V^2 \varphi^2 D^2}$, and $C_{V\varphi^4 D^2}$ scaling
with the new physics scale $\Lambda$.  To estimate their maximal
effect on di-Higgs production, we must first find a minimal scale
$\Lambda$ consistent with the validity criterion of
Eq.~\eqref{eq:validity}.

We do that by generating several event samples for $\Lambda = 1, 2, 3,
4, 5$ TeV, while assigning corresponding numerical values to the
Wilson coefficients, assuming their scaling in
Eq.~\eqref{eq:maxgstar:unprime}.  For each sample, we impose a cut on
the Higgs pair invariant mass, assuming $M_{hh} < \sqrt{s_\text{max}}
= \Lambda$, and define the “minimal'' value of $\Lambda$ as the one
for which the number of events excluded by this cut is of the order of
$\mathcal{O}(1\%)$.  We emphasise once more that a more stringent
scrutiny of our assumptions would only result in a further reduction
of the event excess.
This procedure yields maximal absolute values for dimension-8 WCs
ranging from $0.13/\mathrm{TeV}^4$ to $4/\mathrm{TeV}^4$, with EFT
validity scale $\Lambda$ generally exceeding 2-4 TeV.  Then, at NLO,
the most significant enhancements of di-Higgs VBF production at the
HL-LHC, compared to Standard Model predictions, are at most a factor
of $1.6$ for $C_{W^2 \varphi^4}^{(1)}$ and $1.3$ for $C_{\varphi^6
  \Box}$.  Enhancements for other WCs are considerably smaller.  These
increases are even less pronounced than those from dimension-6 WCs,
making them unlikely to produce any observable effects given the
expected HL-LHC sensitivity.

\subsection{Interference effects of multiple WCs}

If one assumes that several WCs do not vanish at the same time, at the
NLO level, quadratic contributions from dimension-6$^2$ WCs may produce
interesting interference effects, rising the possibility of additional
di-Higgs production enhancement.


\begin{table}[htb!]
  \begin{center}
  \begin{tabular}{cc}
    \begin{tabular}{|c|c|c|c|}
    \hline
    \multicolumn{4}{|c|}{\textbf{BP1}}\\
    \hline
     \multicolumn{2}{|c|}{\textbf{WC}} & $C_{\varphi}$& $C_{\varphi\Box}$  \\
    \hline
    \multirow{4}{*}{$\frac{C}{1\text{ TeV}^{2}}$} & A & 4.7 & 2\\
    \cline{2-4}
    & B & 4.7 & -2\\
    \cline{2-4}
    & C & -13.8 & 2\\
    \cline{2-4}
    & D & -13.8 & -2\\
    \hline
\end{tabular} 
&
\begin{tabular}{|c|c|c|c|c|c|}
    \hline
    \multicolumn{6}{|c|}{\textbf{BP2}}\\
    \hline
     \multicolumn{2}{|c|}{\textbf{WC}} & $C_{\varphi}$& $C_{\varphi^8}$& $C_{\varphi\Box}$& $C_{\varphi^6\Box}$  \\
    \hline
    \multirow{4}{*}{$\frac{C}{1\text{ TeV}^{(2,4)}}$} & A & 3.8 & 7.4 & 2 & 4\\
    \cline{2-6}
    & B & 3.8 & 7.4 & -2 & -4\\
    \cline{2-6}
    & C & -10 & -13.6 & 2 & 4 \\
    \cline{2-6}
    & D & -10 & -13.6 & -2 & -4 \\
    \hline
\end{tabular} \\ \\
\begin{tabular}{|c|c|c|c|}
    \hline
    \multicolumn{4}{|c|}{\textbf{BP3}}\\
    \hline
     \multicolumn{2}{|c|}{\textbf{WC}} & $C_{\varphi\Box}$& $C_{\varphi W}$  \\
    \hline
    \multirow{4}{*}{$\frac{C}{1\text{ TeV}^{(2)}}$} & A & 2 & 1 \\
    \cline{2-4}
    & B & 2 & -1 \\
    \cline{2-4}
    & C & -2  & 1\\
    \cline{2-4}
    & D & -2  & -1 \\
    \hline
\end{tabular}
&
\begin{tabular}{|c|c|c|c|c|c|}
    \hline
    \multicolumn{6}{|c|}{\textbf{BP4}}\\
    \hline
     \multicolumn{2}{|c|}{\textbf{WC}} & $C_{\varphi\Box}$& $C_{\varphi^6\Box}$& $C_{\varphi W}$& $C^{(1)}_{W^2 \varphi^4 }$  \\
    \hline
    \multirow{4}{*}{$\frac{C}{1\text{ TeV}^{(2,4)}}$} & A & 2 & 4 & 1 & 4\\
    \cline{2-6}
    & B & 2 & 4 & -1 & -4\\
    \cline{2-6}
    & C & -2 & -4 & 1 & 4\\
    \cline{2-6}
    & D & -2 & -4 & -1 & -4\\
    \hline
\end{tabular}
\end{tabular}
  \end{center}
  \caption{\sl Four benchmark scenarios used to evaluate possible
    enhancement of the VBF di-Higgs production process. New physics
    scale $\Lambda=4$ TeV is assumed. \textbf{Letters A, B, C and D
      label sub-scenarios that take into account different signs
      configuration.}}
  \label{Tab:pphhjj:Benchmarks}
\end{table}
  
To assess the impact of several selected non-zero Wilson coefficients
on VBF double Higgs production, we collect the results for chosen
benchmark points defined in Table~\ref{Tab:pphhjj:Benchmarks}.  The
relations between WCs in points \textbf{BP1} and \textbf{BP2} follow
the structure generated in the scalar singlet model discussed in
Section~\ref{sec:DiHiggs:UVScalar} (see also e.g.
\cite{Ellis:2023zim}). In such model, the leading contributions arise
from the $C_\varphi$, $C_{\varphi^8}$, $C_{\varphi \Box}$ and
$C_{\varphi^6 \Box}$ WCs. \textbf{BP1} includes only dimension-6 WCs,
whereas \textbf{BP2} also considers dimension-8 operators\footnote{At
the amplitude level, dimension-8 WCs add linearly to other
terms. However, as we do not truncate higher-order terms in cross
sections, also dimension-8 terms may lead to non-trivial interference
with other WCs; thus we include them in our benchmark points.}. Points
\textbf{BP3} and \textbf{BP4} include $C_{\varphi \Box}$,
$C_{\varphi^6 \Box}$, $C_{\varphi W}$ and $C_{W^2\varphi^4}^{(1)}$ WCs
(all potentially loop-generated here) and are chosen to illustrate the
possible enhancement with $\kappa_{\lambda_3} \simeq 1$, since
$C_\varphi$ and $C_{\varphi^8}$ are set to zero.

\begin{table}[htb!]
  \begin{center}
    \begin{adjustbox}{width=\textwidth,max width=\textwidth,max totalheight=\textheight,keepaspectratio}
\begin{tabular}{cc}
  \begin{tabular}{|c||c|c|c||c|c|}
    \hline
    \multicolumn{6}{|c|}{\textbf{LHC}}\\
    \hline
    \multicolumn{2}{|c|}{Scenario} & $\Delta\sigma_{D6}$ & $\Delta
    N_{D6}$ & $\Delta\sigma_{D6^2+D8}$ & $\Delta N_{D6^2+D8}$ \\
    \hline
    \multirow{4}{*}{BP1}
    &A & 3.95 & 48 & 6.36 & 87 \\ \cline{2-6}
    &B & 9.63 & 140 & 3.73 & 44 \\ \cline{2-6}
    &C & 28.17 & 440 & 65.64 & 1046 \\ \cline{2-6}
    &D & 16.97 & 259 & 3.98 & 48 \\ \hline\hline
    \multirow{4}{*}{BP2}
    &A & 2.90 & 31 & 5.17 & 68 \\ \cline{2-6}
    &B & 7.77 & 110 & 5.19 & 68 \\ \cline{2-6}
    &C & 15.30 & 231 & 46.48 & 736 \\ \cline{2-6}
    &D & 7.63 & 107 & 3.16 & 35 \\ \hline\hline
    \multirow{4}{*}{BP3}
    &A & 1.96 & 15 & 4.73 & 60 \\ \cline{2-6}
    &B & 2.06 & 17 & 2.76 & 29 \\ \cline{2-6}
    &C & 2.99 & 32 & 1.89 & 14 \\ \cline{2-6}
    &D & 3.77 & 45 & 1.64 & 10 \\ \hline\hline
    \multirow{4}{*}{BP4}
    &A & 1.96 & 15 & 8.00 & 113 \\ \cline{2-6}
    &B & 2.06 & 17 & 5.10 & 66 \\ \cline{2-6}
    &C & 2.99 & 32 & 3.80 & 45 \\ \cline{2-6}
    &D & 3.77 & 45 & 2.90 & 31 \\ \hline
    \end{tabular}
    &
    \begin{tabular}{|c||c|c||c|c|c|}
    \hline
    \multicolumn{6}{|c|}{\textbf{HL-LHC}}\\
    \hline
    \multicolumn{2}{|c|}{Scenario} & $\Delta\sigma_{D6}$ & $\Delta N_{D6}$ & $\Delta\sigma_{D6^2+D8}$ & $\Delta N_{D6^2+D8}$ \\
    \hline
    \multirow{4}{*}{BP1}
    &A & 3.87 & 5581 & 6.23 & 10177 \\ \cline{2-6}
    &B & 9.48 & 16498 & 3.69 & 5230 \\ \cline{2-6}
    &C & 27.69 & 51934 & 65.02 & 124564 \\ \cline{2-6}
    &D & 16.57 & 30304 & 3.91 & 5656 \\ \hline\hline
    \multirow{4}{*}{BP2}
    &A & 2.87 & 3634 & 5.13 & 8035 \\ \cline{2-6}
    &B & 7.66 & 12955 & 5.13 & 8029 \\ \cline{2-6}
    &C & 15.13 & 27490 & 46.08 & 87724 \\ \cline{2-6}
    &D & 7.53 & 12700 & 3.12 & 4132 \\ \hline\hline
    \multirow{4}{*}{BP3}
    &A & 2.10 & 2134 & 5.21 & 8197 \\ \cline{2-6}
    &B & 2.21 & 2356 & 2.98 & 3856 \\ \cline{2-6}
    &C & 3.12 & 4129 & 1.98 & 1912 \\ \cline{2-6}
    &D & 3.91 & 5653 & 1.66 & 1282 \\ \hline\hline
    \multirow{4}{*}{BP4}
    &A & 2.10 & 2134 & 8.87 & 15316 \\ \cline{2-6}
    &B & 2.21 & 2356 & 5.62 & 8983 \\ \cline{2-6}
    &C & 3.12 & 4129 & 4.13 & 6097 \\ \cline{2-6}
    &D & 3.91 & 5653 & 3.02 & 3934 \\ \hline
    \end{tabular}
  \end{tabular}
  \end{adjustbox}
  \end{center}
  \caption{Enhancement of VBF di-Higgs production for benchmark
    scenarios BP1–BP4. $\Delta\sigma_{D6}$ and
    $\Delta\sigma_{D6^2+D8}$ represent cross-section enhancements at
    linear and full quadratic order (i.e., including in the amplitude
    dimension-6 terms only or also dimension-6$^2$ and dimension-8
    contributions), respectively. $\Delta N_{D6}$ and $\Delta
    N_{D6^2+D8}$ indicate the respective event yield enhancements.  }
  \label{Tab:pphhjj:manyWCs}
\end{table}

\begin{figure}[htb!]
  \centering
    \includegraphics[width=0.4\linewidth]{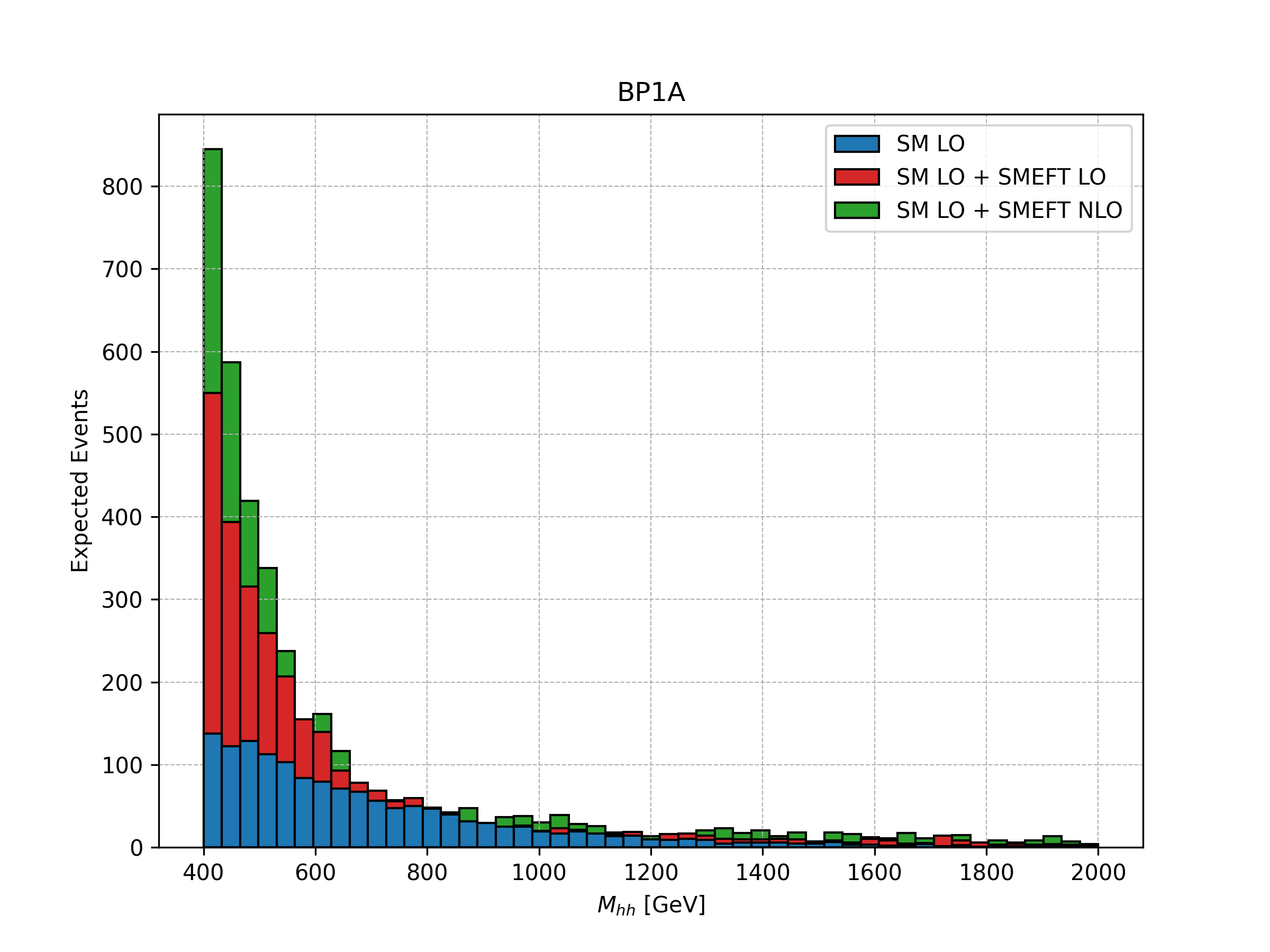}
    \includegraphics[width=0.4\linewidth]{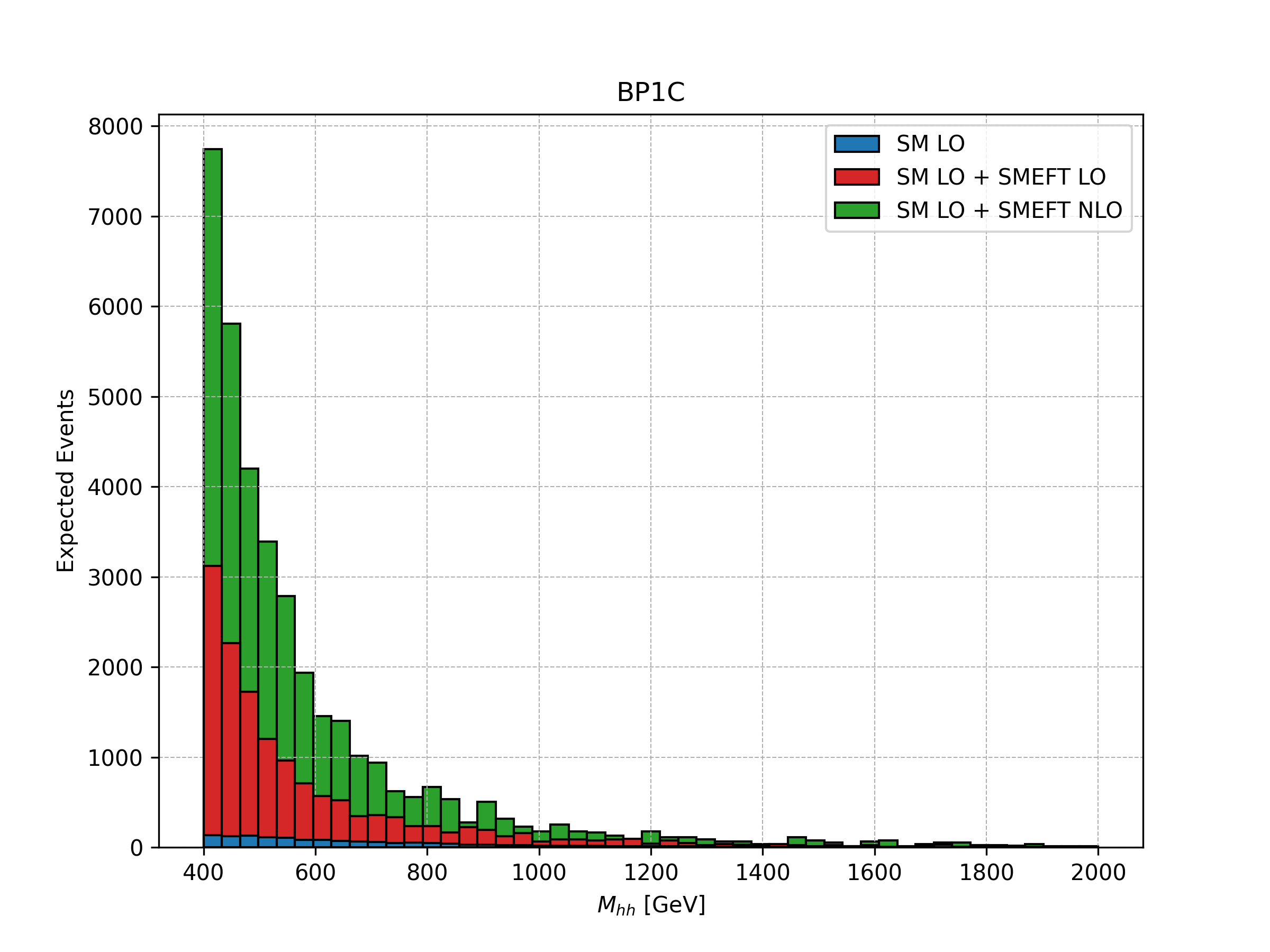}
    \includegraphics[width=0.4\linewidth]{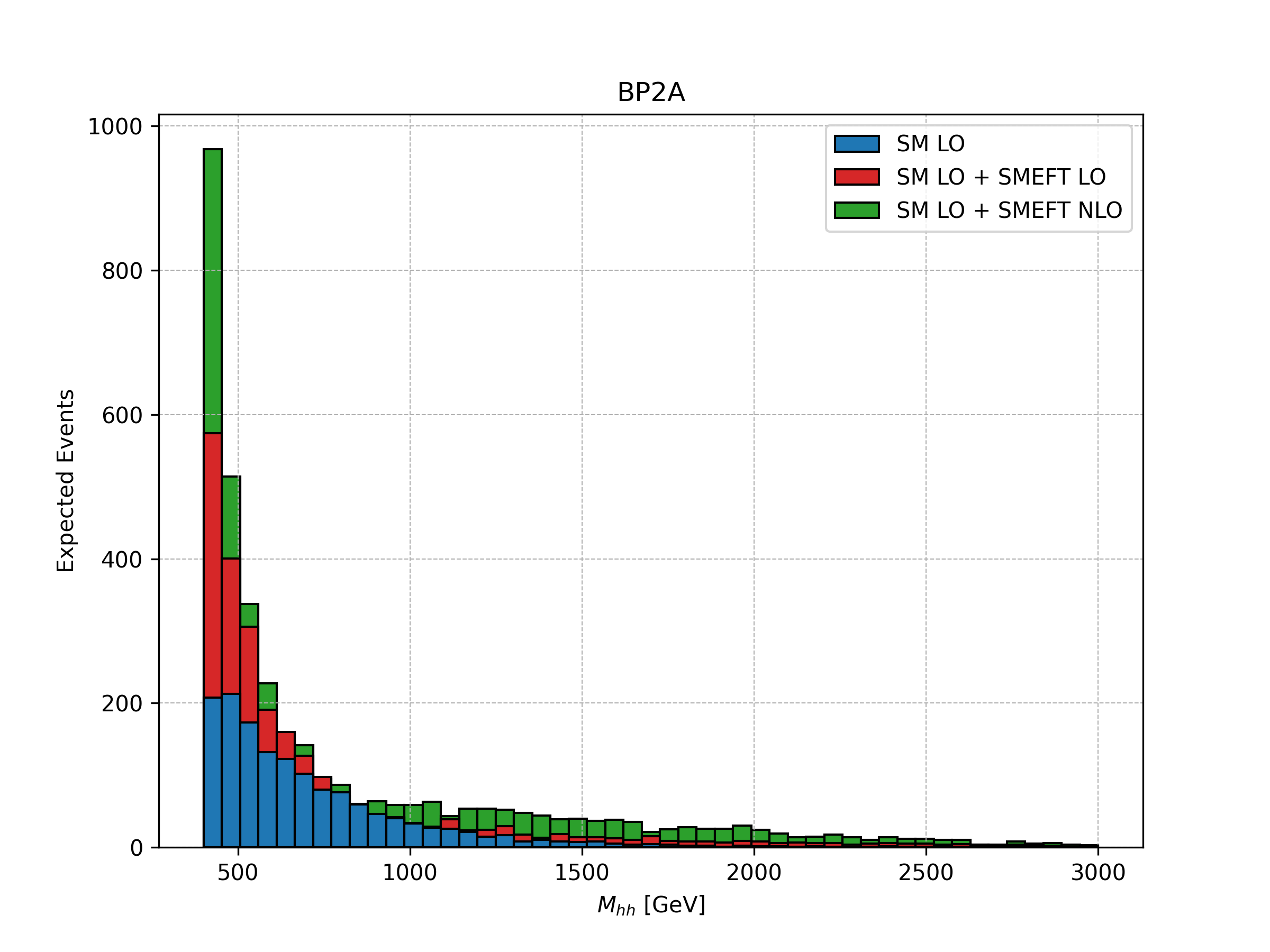}
    \includegraphics[width=0.4\linewidth]{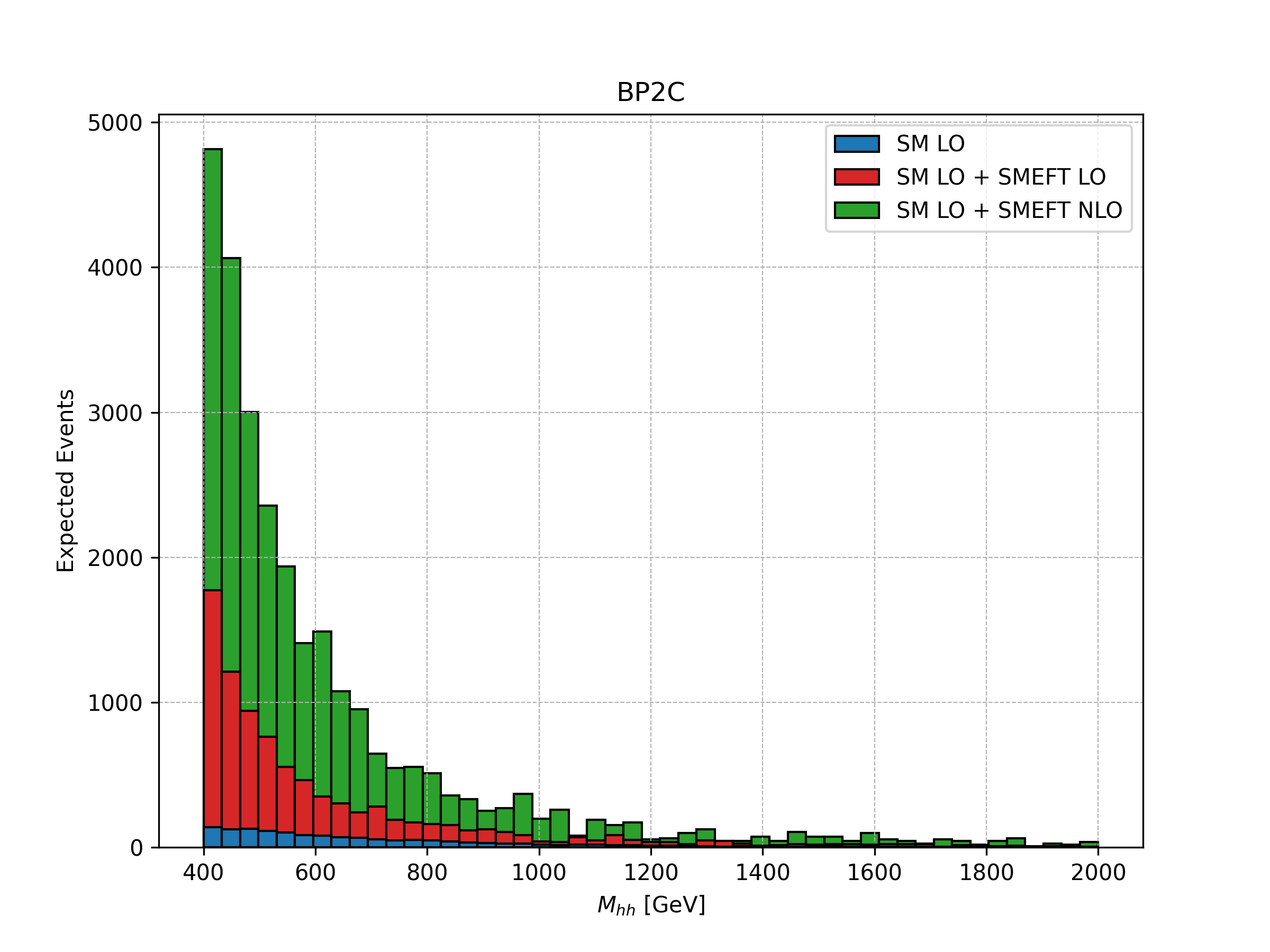}
    \includegraphics[width=0.4\linewidth]{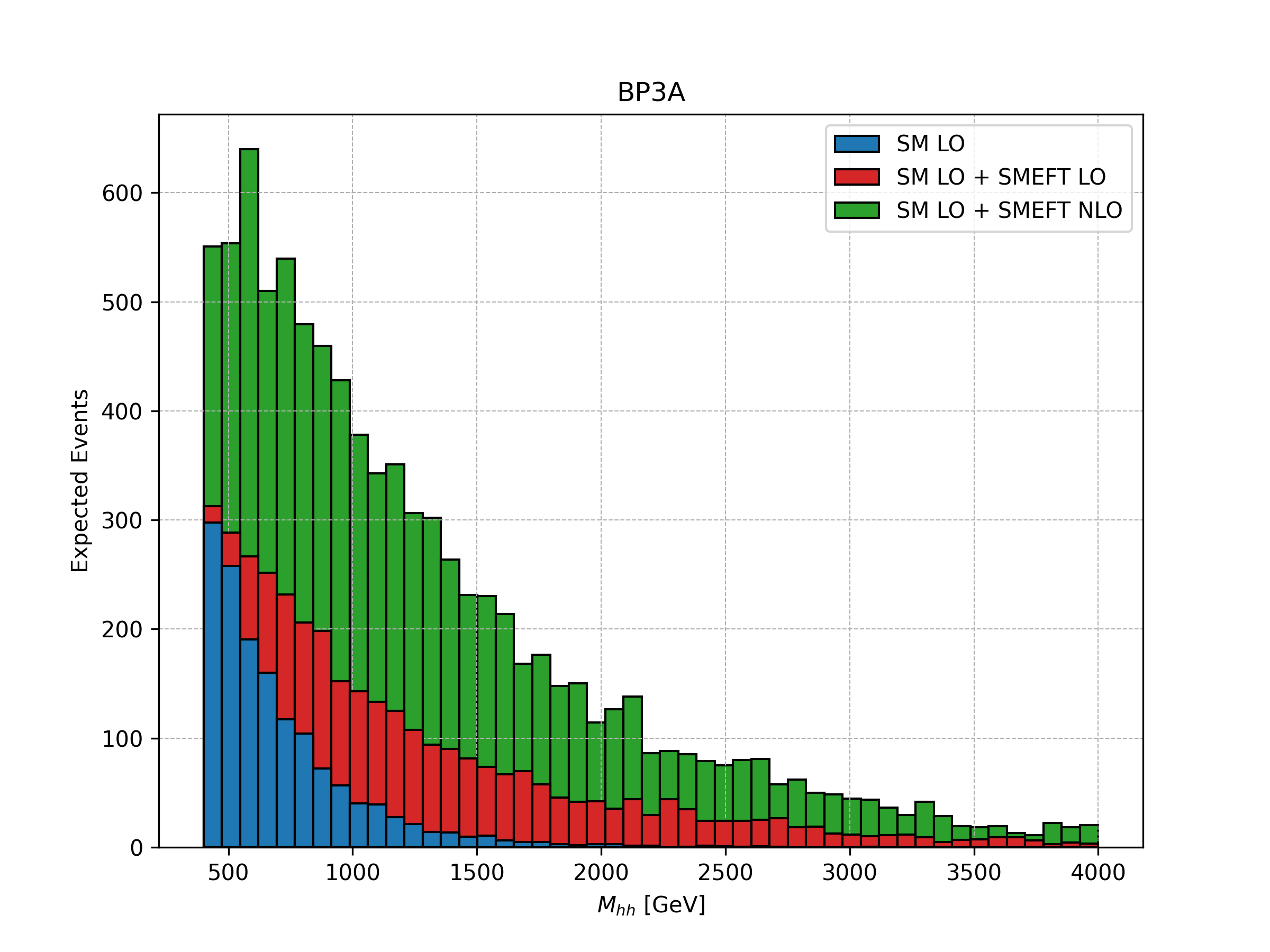}
    \includegraphics[width=0.4\linewidth]{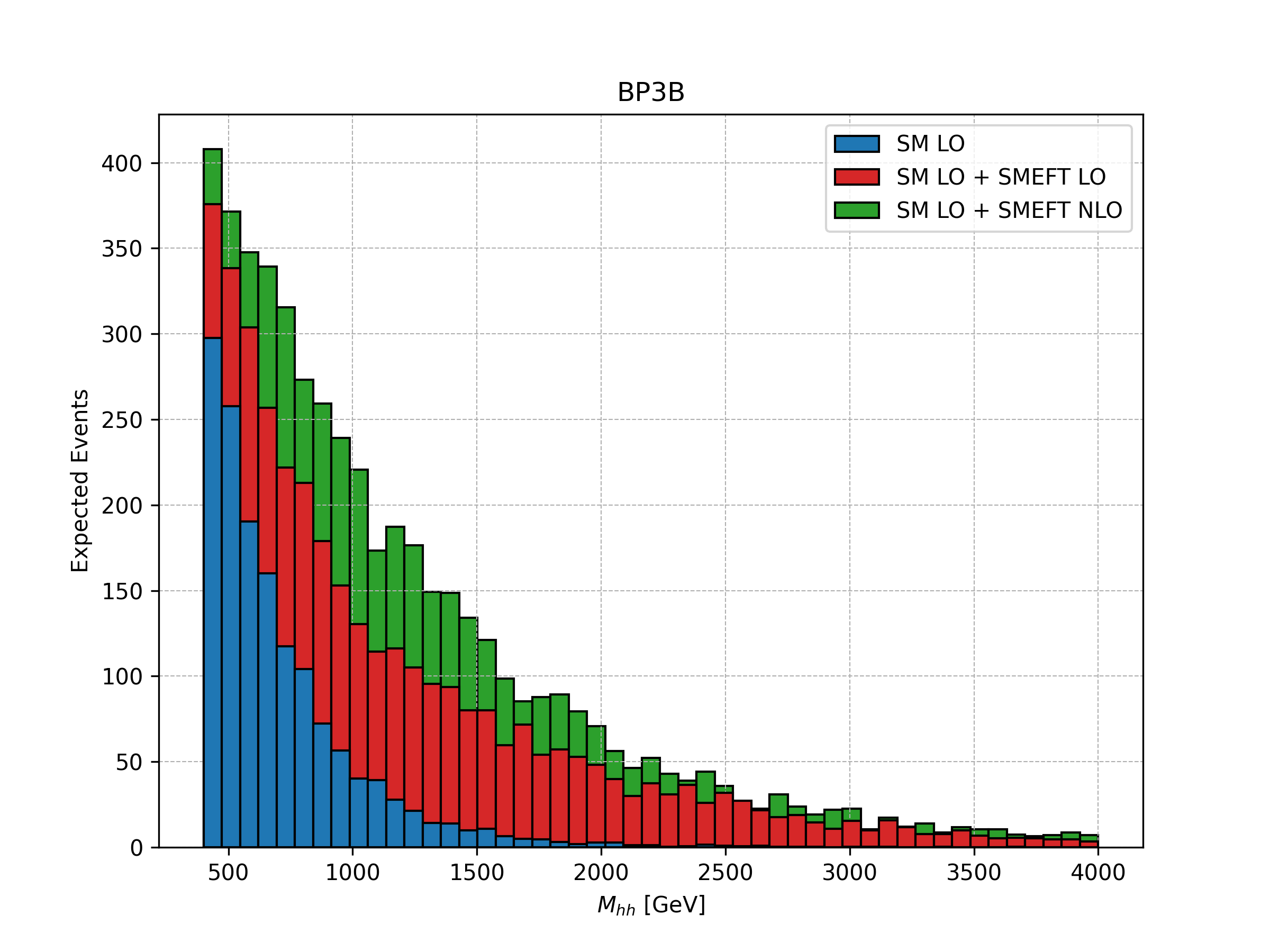}
    \includegraphics[width=0.4\linewidth]{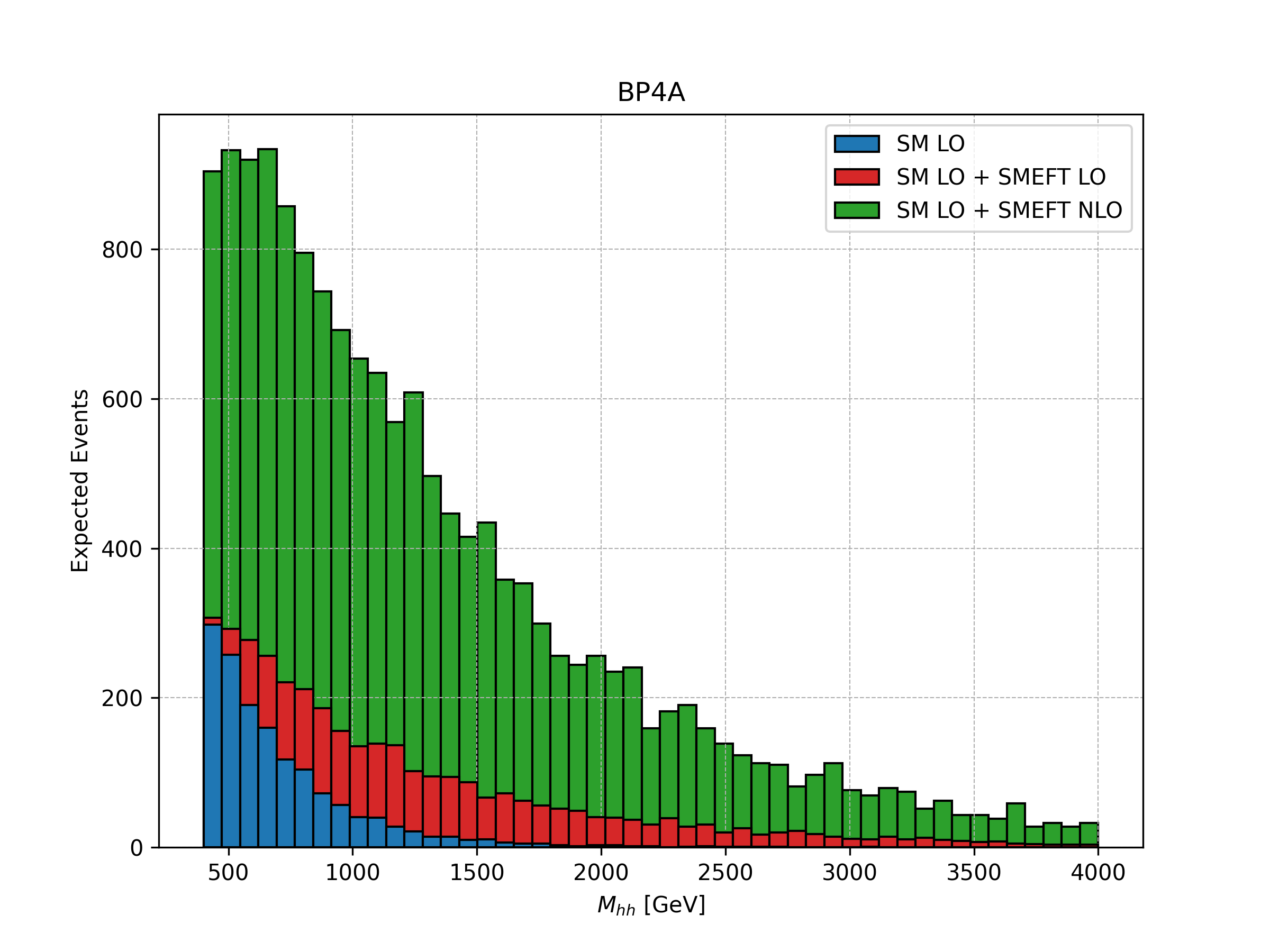}
    \includegraphics[width=0.4\linewidth]{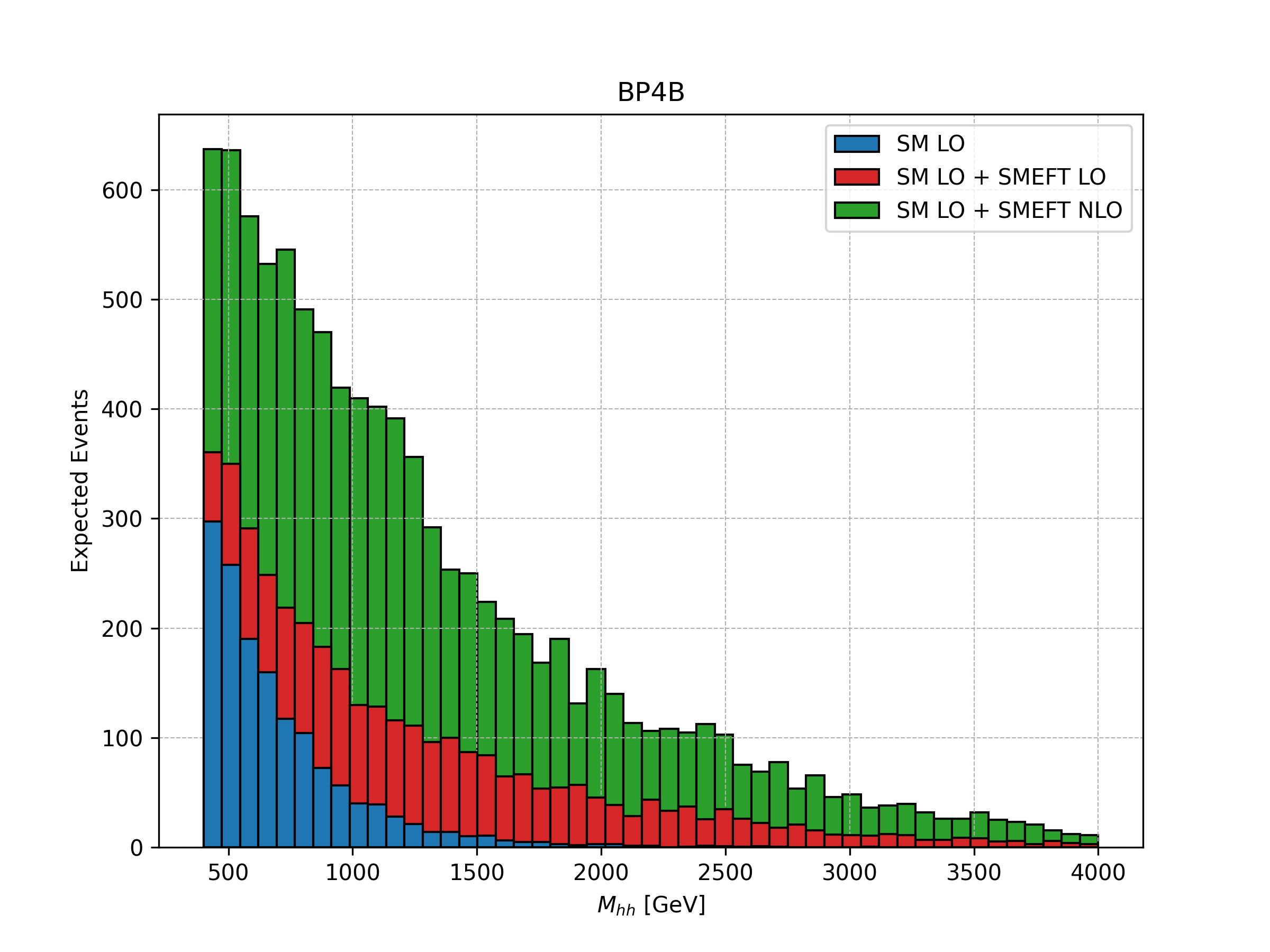}
  \caption{\sl Higgs pair invariant mass $M_{hh}$ distribution for
    chosen BPs from Table~\ref{Tab:pphhjj:Benchmarks} for HL-LHC.
  }
    \label{fig:manyWCs:Histos}
\end{figure}

The results of Monte Carlo simulations are summarised in
Table~\ref{Tab:pphhjj:manyWCs} and in the histograms of
Figure~\ref{fig:manyWCs:Histos}. One can draw the following
conclusions:
\begin{itemize}
\item \textbf{BP1} - The enhancement is primarily driven by
  modifications of $\kappa_{\lambda_3}$, resulting in scenario C in
  maximal increase of $\mathcal{O}(30)$ at linear order and up to
  $\mathcal{O}(65)$ at quadratic order, for both the LHC and
  HL-LHC. With luck, such a significant effect could be observed even
  before the HL-LHC era. Quadratic effects are large and have to be taken
  into account for viable cross-section estimates, especially that at
  the NLO order they can lead to both significant enhancement or
  suppression of the final result, depending on relative WC signs.
\item \textbf{BP2} - The conclusions mirror those for BP1, with the
  main difference arising from the inclusion of $C_{\varphi^8}$ in the
  $\kappa_{\lambda_3}$ modification, along with additional
  contributions from $C_{\varphi^6\Box}$.
\item \textbf{BP3} - In the absence of contributions to
  $\kappa_{\lambda_3}$, the derivative WCs $C_{\varphi\Box}$ and
  $C_{\varphi W}$ can only produce smaller enhancement, reaching
  $\mathcal{O}(5)$ in Scenario A. Again, quadratic effects are large
  and can lead to both constructive or destructive interference
  between WCs.
\item \textbf{BP4} - The conclusions are consistent with those for
  BP3, with the enhancement of the cross-section reaching
  $\mathcal{O}(9)$ after including also WCs of dimension-8 operators
  $C_{\varphi^6\Box}$ and $C_{W^2\varphi^4}^{(1)}$.
\end{itemize}
These results highlight the importance of including quadratic
contributions in the analysis, both when calculating total cross
sections and in finding the shape of invariant mass distribution (as
illustrated in Figure~\ref{fig:manyWCs:Histos}). Although operators
related to $\kappa_{\lambda_3}$ dominate the enhancement, interference
between $C_\varphi$, $C_{\varphi^8}$ and other WCs can lead to large
effects, potentially comparable to those in the gluon fusion
channel. The important role is played here by the relative signs of
various WCs.  Even without tree-level triple Higgs boson coupling
modification, i.e. when $\kappa_{\lambda_3} = 1$, enhancement from
derivative WCs can reach up to $\mathcal{O}(10)$. This conclusion
should be interpreted with caution, as we do not include the full set
of dimension-8 operators; the result may therefore depend on the
choice of operator basis.

We should also comment that our analysis is preliminary and thus does
not specify more refined experimental cuts.  In case of
momentum-dependent Higgs-vector boson couplings, which are
particularly interesting as they are weakly or not at all probed by
the ggF channel, even a simple cut on invariant mass of the Higgs pair
may further improve a ratio of signal from momentum dependent
operators from the SM background.  In Table~\ref{Tab:pphhjj:Cut} we
show the results obtained assuming $M_{hh}\ge 1$ TeV.  As one can see,
comparing to numbers in Tables \ref{Tab:pphhjj:Dim6}
and~\ref{Tab:pphhjj:manyWCs}, such a cut can increase the signal to SM
prediction ratio by additional factors of up to almost 5, while still
leaving the total number of events large enough to be observable and
statistically significant.

\begin{table}[tb!]
  \begin{center}
    \begin{tabular}{|c|c||c|c||c|c|}
    \hline
    \multicolumn{6}{|c|}{\textbf{HL-LHC}}\\
    \hline
    WC & $\frac{C}{1\text{ TeV}^{2,4}}$ & $\Delta\sigma_{D6}$ & $\Delta
    N_{D6}$ & $\Delta\sigma_{D6^2+D8}$ & $\Delta N_{D6^2+D8}$ \\
    \hline
    $C_{\varphi \Box}$ & $2$ & 2.62 & 332 & 4.96 &
    810 \\
    \hline
    $C_{\varphi W}$ & $1$ & 6.35 & 1095 & 11.12 & 2070\\
    \hline
    $C_{\varphi^6\Box}$ & $-4$ & - & - & 1.43 & 88\\
    \hline
    $C_{W^2\varphi^4}^{(1)}$ & $-4$ & - & - & 3.80 & 573\\
    \hline
\multicolumn{2}{|c||}{BP3 scenario A} & 7.92 & 1415 & 23.09 & 4517\\
    \hline
\multicolumn{2}{|c||}{BP4 scenario A} & 7.92 & 1415 & 41.86 & 8357\\ \hline
    \end{tabular}
  \end{center}
  \caption{\sl Maximal enhancement of VBF di-Higgs production for
    dimension-6 Wilson coefficients (WCs) obtained after applying
    $M_{hh}\ge 1$ TeV cut.  }
  \label{Tab:pphhjj:Cut}
\end{table}

The presented results should be put into a context with expected
experimental accuracy. Following the recent ATLAS
report~\cite{ATLAS:2025wdq} on the projected HL-LHC sensitivity to
double Higgs boson production, we quote it for the VBF-induced double
Higgs production as
\begin{equation}
\mu_{\text{VBF}-hh} \equiv \Delta \sigma \gtrsim 10 \;.
\label{eq:sensitivity}
\end{equation}
This value is significantly above the enhancements predicted in the
previous sections, where only a single Wilson coefficient was
considered at a time, leading to at most $\Delta \sigma\lesssim 3$ and
limited prospects for observing VBF di-Higgs boson production at the
HL-LHC within the regime of EFT validity.

However, as demonstrated in this section, in more elaborated multiple
WC scenarios, or using cuts on invariant mass of the Higgs pair, the
enhancement can become large enough to make the VBF channel detectable
in future, in parallel to the ggF channel.  Measuring (or putting
upper limits) on both of them is valuable, as the sets of SMEFT WCs
which they probe overlap only partially.  The results shown highlight
also the importance of considering the quadratic terms in the EFT
expansion and interference-driven effects between multiple Wilson
coefficients.  Such effects can produce sizeable cross-section
enhancement in the VBF channel, even in cases where the ggF channel
remains unchanged.

\section{Conclusions}
\label{sec:summary}

Our study explores the influence of higher-dimensional operators of
the Standard Model Effective Field Theory on the double Higgs boson
production process via Vector Boson Fusion.  Although the VBF channel
is subdominant compared to gluon fusion, it is characterised by a
unique experimental signature and provides sensitivity to a different
spectrum of BSM interactions.

We incorporated all bosonic CP-conserving dimension-6 operators and a
chosen representative subset of dimension-8 operators, arranged in
Table~\ref{tab:DiHiggs:Operators}. These operators may significantly
impact the tree-level Higgs bosons self-interactions, and its
interactions with $W$ and $Z$ gauge bosons.\footnote{In SMEFT there
are also $VV'h$ and $VV'hh$ vertices with $V,V'={\gamma,Z}$ that are
driven by $C_{\varphi W}, C_{\varphi B}, C_{\varphi WB}$ WCs. These
are, of course, included in our numerical results but not shown
analytically in the Appendices.}  For the dimension-8 operators, we
believe that our selection is adequate for capturing the overarching
conclusions regarding the maximum possible enhancement of the VBF
double Higgs production process within SMEFT.  Our choice encompasses
all main classes of dimension-8 operators corresponding to possible
configurations of energy growth in the amplitudes. The complete set of
such operators is very large and would obscure, rather than clarify,
the conclusions and discussion presented here. Moreover, since the
specific set of operators generated depends on the underlying UV
model, a broader selection would not necessarily be more valid or
useful. Our aim in this work is to illustrate representative patterns
of discussed behavior, which can then be straightforwardly
extrapolated to more realistic scenarios.

We determined realistic numerical values for the Wilson coefficients
associated with the chosen SMEFT operators in
Section~\ref{subsec:DiHiggs:Assumptions}.  This involved either
leveraging existing bounds (as detailed in Table~\ref{tab:fits}), that
are derived from numerical fits to experimental data, or extrapolating
such bounds to currently unconstrained WCs by applying Naive Dimension
Analysis assumptions.
To further illustrate these concepts, we employed examples of singlet
and triplet scalar extensions of the SM in
Section~\ref{sec:DiHiggs:UVScalar}.  This enabled us to demonstrate
when and how such Wilson coefficient values could be realised after
the decoupling of complete UV-models.  We utilised a natural matching
technique, deployed directly in the SMEFT mass basis, and leveraged
the Goldstone Boson Equivalence Theorem.  This allowed us to directly
map UV-model parameters to the relevant SMEFT Wilson coefficients, as
evident from the high energy analytical forms of specific amplitudes
(provided in the Appendices~\ref{App:DiHiggs:WWHH:Ampl}
and~\ref{App:DiHiggs:ZZHH:Ampl}).  Concurrently, we also examined the
convergence criteria for the Effective Field Theory expansion using
these same SM scalar extensions.

Subsequently, in Sections~\ref{sec:DiHiggs:Ampls}
and~\ref{sec:DiHiggs:Num}, we conducted an in-depth analysis of the
$VV\to hh$ and the full $pp\to hhjj$ cross-sections, respectively.
Our analysis included the full suite of next-to-leading order bosonic
contributions (in the EFT expansion), specifically dimension-6,
dimension-6$^2$, and dimension-8 terms, which contribute to the
tree-level amplitudes.  Through this analysis, we pinpointed which WCs
exert the most significant impact and how these contributions vary
with the invariant mass of the Higgs boson pair in the final state.
The key findings and implications of our work, arranged in
Tables~\ref{Tab:pphhjj:KappaLambda}-\ref{Tab:pphhjj:Cut} and
Figures~\ref{fig:Dim6:Histos:KappaLambda}-\ref{fig:manyWCs:Histos} are
summarised as follows:
\begin{enumerate}
\item \textit{Non-derivative operators} ($C_\varphi,\,
  C_{\varphi^8}$): these WCs contribute to the trilinear Higgs
  coupling and are by far the most significant in potential
  enhancement of the VBF di-Higgs cross-section, by a factor of at
  most $\mathcal{O}(20)$ (see Table~\ref{Tab:pphhjj:KappaLambda}).
  Nevertheless, we also note that this kind of modification would also
  drastically change the ggF channel, which could result in its much
  earlier detection.
\item \textit{Derivative operators} (e.g. $C_{\varphi\Box},\,
  C_{\varphi W}$): the VBF di-Higgs production cross-section exhibits
  only modest enhancements, typically at most a factor of
  $\mathcal{O}(2-3)$ for dimension-6 WCs, as presented in
  Table~\ref{Tab:pphhjj:Dim6}.  Even smaller improvement can be seen
  in dimension-8 WCs, where the increase is no more than of order
  $\mathcal{O}(1.5)$.  Consequently, these effects are difficult to
  detect at the LHC and continue to be marginal with the
  High-Luminosity LHC (HL-LHC) upgrade.
  Nevertheless, we noticed that the inclusion of dimension-6$^2$ terms
  in the amplitude, may alter the predicted enhancement by $\approx
  50\%$ baring cancellations with the full set of dimension-8
  operators.
\item \textit{In scenarios where multiple WCs contribute
  constructively}, particularly those involving derivative operators
  (e.g., $C_{\varphi\Box}$, $C_{\varphi W}$), the cross-section can be
  enhanced by up to an order of magnitude. Such enhancements are most
  pronounced in kinematic regions with high invariant mass of the
  Higgs pair ($M_{hh} \gtrsim$ 1 TeV), where the energy-dependent
  terms dominate -- see Tables~\ref{Tab:pphhjj:Benchmarks}
  and~\ref{Tab:pphhjj:Cut}. This underscores the significance of
  incorporating contributions from dimension-6$^2$ and dimension-8
  operators, as their interference can substantially alter
  predictions.
\end{enumerate}
In summary, while the VBF di-Higgs channel alone may not provide
definitive evidence of BSM physics at the HL-LHC, it serves as an
important complementary probe, especially for derivative interactions
and more realistic, multi-WC scenarios.  Our results highlight the
importance of including higher-order EFT contributions, and emphasise
the need for a careful assessment of EFT validity, motivating further
experimental and theoretical efforts to explore this process in
greater detail.

\section*{Acknowledgements}
The authors acknowledge support from the COMETA COST Action CA22130.
The work of JR was supported by the Polish National Science Center
under the research grant DEC2019/35/B/ST2/02008.  MR acknowledges
support from the University of Padua under the 2023 STARS Grants@Unipd
programme (Acronym and title of the project: HiggsPairs – Precise
Theoretical Predictions for Higgs pair production at the LHC). We
would like to thank Stefan Pokorski for numerous fruitful discussions.


\begin{appendices}

  \include{AppendixDiHiggs.tex}

\end{appendices}

\bibliography{bibliografia}{}
\bibliographystyle{JHEP}

\end{document}

%% file: AppendixDiHiggs.tex
\setcounter{equation}{0} 
\setcounter{table}{0} 

\section*{Appendices: Dominant contributions to helicity amplitudes}

In the following two sections, we present the helicity amplitudes for
the $WW\rightarrow HH$ and $ZZ\rightarrow HH$ processes, including
SMEFT contributions from the dimension-6 and dimension-8 operators
collected in Table~\ref{tab:DiHiggs:Operators}. For each amplitude, we
display only the leading order terms in the high energy limit
($M_V^2/s\rightarrow 0$), including interference terms between
different Wilson coefficients.  We start with the $WW\rightarrow HH$
process in section~\ref{App:DiHiggs:WWHH:Ampl} and present the
Standard Model contribution in Eq.~\eqref{eq:DiHiggs:WWHH:SM}, the
dimension-6 and dimension-6$^2$ contributions in
Eq.~\eqref{eq:DiHiggs:WWHH:Dim6}, and the dimension-8 contributions in
Eqs.~\eqref{eq:DiHiggs:WWHH:Dim8:1}
and~\eqref{eq:DiHiggs:WWHH:Dim8:2}. Similarly, in
section~\ref{App:DiHiggs:ZZHH:Ampl}, for the $ZZ\rightarrow HH$
process, we present the SM contribution in
Eq.~\eqref{eq:DiHiggs:ZZHH:SM}, the dimension-6 and dimension-6$^2$
contributions in Eq.~\eqref{eq:DiHiggs:ZZHH:Dim6}, and the dimension-8
contributions in Eqs.~\eqref{eq:DiHiggs:ZZHH:Dim8:1}
and~\eqref{eq:DiHiggs:ZZHH:Dim8:2}.

\section{$W^+W^-\rightarrow HH$ helicity amplitude}
\label{App:DiHiggs:WWHH:Ampl}

\subsection{SM}

\begin{equation}
\begin{aligned}
\mathcal{M}^{WW,\, SM}_{00}&=\sqrt{2} G_F
M_H^2\left[1+2\left(1-\frac{4}{\sin
  ^2 \theta}\right) \frac{M_W^2}{M_H^2}\right]\; ,\\[2mm]
\mathcal{M}^{WW,\, SM}_{\pm\pm}&=\mathcal{O}\left(\frac{1}{s}\right)\; , \\[2mm]
\mathcal{M}^{WW,\, SM}_{\pm \mp} & =2\sqrt{2} G_F M_W^2\; ,\\[2mm]
\mathcal{M}^{WW,\, SM}_{0 \pm} & =\mathcal{M}^{WW,\, SM}_{\pm 0} =
\mathcal{O}\left(\frac{1}{\sqrt{s}}\right)\,.
\label{eq:DiHiggs:WWHH:SM}
\end{aligned}
\end{equation}

\subsection{Dimension-6 SMEFT}
\begin{align}
\mathcal{M}^{WW,\, D6}_{00}=&-3\sqrt{2} C_\varphi
\left(\frac{1}{G_F\Lambda^2}\right)-20C_{\varphi W} \left(\frac{M_W^2}{\Lambda^2}\right) + \frac{1}{2}(4 C_{\varphi\Box} - C_{\varphi D})
\left(\frac{s}{\Lambda^2}\right)\hspace{3cm} \textbf{dim-6}\nonumber\\[2mm]
&+ \frac{1}{2\sqrt{2}} (4 C_{\varphi\Box} - C_{\varphi
D})^2\left(\frac{1}{G_F\Lambda^2}\right)\left(\frac{s}{\Lambda^2}\right)
-3 C_\varphi (4 C_{\varphi\Box} - C_{\varphi D})\left(\frac{1}{G_F\Lambda^2}\right)^2\; , 
\hspace{1.3cm} \textbf{(dim-6)}^2\nonumber\\[2mm]
\mathcal{M}^{WW,\, D6}_{\pm\pm} =& -(4 C_{\varphi\Box} - C_{\varphi D})\left(\frac{M_W^2}{\Lambda^2}\right) +2C_{\varphi W}\left(\frac{s}{\Lambda^2}\right)\nonumber \\[2mm]
&+\sqrt{2}C_{\varphi W} (4 C_{\varphi \Box} -C_{\varphi D}+6 C_{\varphi W}) \left(\frac{1}{G_F\Lambda^2}\right)\left(\frac{s}{\Lambda^2}\right)
-6 C_\varphi C_{\varphi W}\left(\frac{1}{G_F\Lambda^2}\right)^2\; ,\label{eq:DiHiggs:WWHH:Dim6}
\\
\mathcal{M}^{WW,\, D6}_{\pm\mp}=& + (4 C_{\varphi\Box} - C_{\varphi D})\left(\frac{M_W^2}{\Lambda^2}\right)\nonumber\\[2mm]
&+2\sqrt{2}C_{\varphi W}^2\left(\frac{1}{G_F\Lambda^2}\right)\left(\frac{s}{\Lambda^2}\right)\; ,  \nonumber\\[2mm]
\mathcal{M}^{WW,\, D6}_{0\pm}=&-4\sqrt{2}\cot\theta C_{\varphi W} \left(\frac{\sqrt{s} M_W}{\Lambda^2}\right)\nonumber\\[2mm]
&-2 C_{\varphi W}\left( 4C_{\varphi W} +  4C_{\varphi \Box} -  C_{\varphi D}\right)\cot\theta\left(\frac{1}{G_F\Lambda^2}\right)\left(\frac{\sqrt{s} M_W}{\Lambda^2}\right)\; .\nonumber
\end{align}

\subsection{Dimension-8 SMEFT}
For clarity, we divide the expressions for the dimension-8 terms into 2 classes. 
\subsubsection{ $\varphi^8$, $\varphi^6 D^2$, $\varphi^4 D^4$ operator classes}

\begin{align}
\mathcal{M}^{WW,\, D8}_{00}=&-6C_{\varphi^8} \left(\frac{1}{G_F^2\Lambda^4}\right)
+\frac{\sqrt{2}}{4}\left(8C_{\varphi^6 \Box}-C_{\varphi^6D^2}\right)\frac{s}{\Lambda^2}\frac{1}{G_F \Lambda^2}\nonumber\\[2mm]
&- \frac{1}{8} \, \left [\left ( C_{\varphi^4 D^4}^{(1)} +
  C_{\varphi^4 D^4}^{(2)} + 4\, C_{\varphi^4 D^4}^{(3)} \right ) +
  \left ( C_{\varphi^4 D^4}^{(1)} + C_{\varphi^4 D^4}^{(2)} \right )
  \cos^2\theta \right ]\, \, \left (\frac{s}{\Lambda^2} \right )^2\; ,\nonumber \\[2mm]
\mathcal{M}^{WW,\, D8}_{\pm\pm}=& 
-\frac{1}{\sqrt{2}}\,\left ( 8\, C_{\varphi^6\Box} - C_{\varphi^6
  D^2} \right )\, \left (\frac{M_W^2}{\Lambda^2} \right)\, \left (
\frac{1}{G_F \Lambda^2} \right ) \nonumber \\[2mm]
&+\frac{1}{4} \left [ 4\, C_{\varphi^4 D^4} ^{(3)} + \left (
  C_{\varphi^4 D^4} ^{(1)} + C_{\varphi^4 D^4} ^{(2)}\right )\,
  \sin^2\theta \right ] \, \left (\frac{M_W^2}{\Lambda^2} \right ) \,
\left (\frac{s}{\Lambda^2} \right )\; ,\label{eq:DiHiggs:WWHH:Dim8:1} \\[2mm]
\mathcal{M}^{WW,\, D8}_{\pm\mp}=&
+\frac{\sqrt{2}}{4}\left(8C_{\varphi^6 \Box}-C_{\varphi^6D^2}\right)\frac{M_Z^2}{\Lambda^2}\frac{1}{G_F \Lambda^2}+
\frac{1}{4} \left( C_{\varphi^4 D^4}^{(1)} + C_{\varphi^4 D^4}^{(2)}\right) \sin^2\theta \left(\frac{M_W^2}{\Lambda^2}\right)\left(\frac{s}{\Lambda^2}\right)\; ,\nonumber\\[2mm]
\mathcal{M}^{WW,\, D8}_{0\pm}=& +\frac{1}{8\sqrt{2}} \left( C_{\varphi^4 D^4}^{(1)}+C_{\varphi^4 D^4}^{(2)}\right) \sin{2\theta}\left(\frac{s^{3/2}M_W}{\Lambda^4}\right)\; . \nonumber
\end{align}

\subsubsection{$X^2\varphi^4$, $X^2 \varphi^2 D^2$, $X \varphi^4 D^2$ operator classes}

\begin{align}
\mathcal{M}^{WW,\, D8}_{00}=&
+\left[\frac{1}{2}C_{W^2\varphi^2 D^2}^{(1)}(\sin^2\theta+1)+ 2 C_{W^2 \varphi^2 D^2}^{(2)}\right]\left (\frac{M_W^2}{\Lambda^2} \right )\left(\frac{s}{\Lambda^2}\right)
\nonumber\\
&
-14\sqrt{2} C_{W^2\varphi^4}^{(1)}\left(\frac{M_W^2}{\Lambda^2}\right)\left(\frac{1}{G_F \Lambda^2}\right)+\frac{5i}{2^{5/4}} C_{W\varphi^4D^2}^{(1)}\left(\frac{M_W}{\sqrt{G_F}\Lambda^2}\right)\left(\frac{s}{ \Lambda^2}\right)\; ,\nonumber\\[2mm]
\mathcal{M}^{WW,\, D8}_{\pm\pm}=&
-\left(\frac{1}{4}C_{W^2\varphi^2 D^2}^{(1)}+C_{W^2\varphi^2 D^2}^{(2)}\right)\left(\frac{s^2}{\Lambda^4}\right)\nonumber\\[2mm]
&+3 \sqrt{2} C_{W^2 \varphi^4}^{(1)}\left(\frac{1}{G_F \Lambda^2}\right)\left(\frac{s}{\Lambda^2}\right)
-\frac{3i}{2^{5/4}} C_{W\varphi^4D^2}^{(1)}\left(\frac{M_W}{\sqrt{G_F}\Lambda^2}\right)\left(\frac{s}{ \Lambda^2}\right)\; , \label{eq:DiHiggs:WWHH:Dim8:2}\\[2mm]
\mathcal{M}^{WW,\, D8}_{\pm\mp}=&
-\frac{1}{8}\sin^2\theta C_{W^2\varphi^2
D^2}^{(1)}\left(\frac{s^2}{\Lambda^4}\right) +2^{3/4}
C_{W\varphi^4D^2}^{(1)}\left(\frac{M_W}{\sqrt{G_F}\Lambda^2}\right)\left(\frac{M_W^2}{ \Lambda^2}\right)\; ,\nonumber\\[2mm]
\mathcal{M}^{WW,\, D8}_{0\pm}=& + 2^{1/4}i C_{W\varphi^4D^2}^{(1)}\cot\theta\left(\frac{M_W}{\sqrt{G_F}\Lambda^2}\right)\left(\frac{M_W \sqrt{s}}{ \Lambda^2}\right)\nonumber\\[2mm]
&- 4
C_{W^2\varphi^4}^{(1)}\cot{\theta}\left(\frac{s^{1/2}M_W}{\Lambda^2}\right)\left(\frac{1}{G_F \Lambda^2}\right)
- \frac{1}{8\sqrt{2}}C_{W^2\varphi^2
D^2}^{(1)}\sin{2\theta}\left(\frac{s^{3/2}M_W}{\Lambda^4}\right)\; . \nonumber
\end{align}

\setcounter{equation}{0} 
\setcounter{table}{0} 

\section{$ZZ\rightarrow HH$ helicity amplitude}
\label{App:DiHiggs:ZZHH:Ampl}

\subsection{SM}

\begin{equation}
\begin{aligned}
\mathcal{M}^{ZZ,\, SM}_{00}&=\sqrt{2} G_F
M_H^2\left[1+2\left(1-\frac{4}{\sin ^2 \theta}\right)
  \frac{M_Z^2}{M_H^2}\right]\; ,\\[2mm]
\mathcal{M}^{ZZ,\, SM}_{\pm\pm}&=\mathcal{O}\left(\frac{1}{s}\right)\; , \\[2mm]
\mathcal{M}^{ZZ,\, SM}_{\pm \mp} & =2\sqrt{2} G_F M_Z^2\; ,\\[2mm]
\mathcal{M}^{ZZ,\, SM}_{0 \pm} & =\mathcal{M}^{ZZ,\, SM}_{\pm 0}=\mathcal{O}\left(\frac{1}{\sqrt{s}}\right)\; .
\label{eq:DiHiggs:ZZHH:SM}
\end{aligned}
\end{equation}

\allowdisplaybreaks

\subsection{Dimension-6 SMEFT}
\begin{align}
\mathcal{M}^{ZZ,\, D6}_{00}=&-3\sqrt{2}C_\varphi\left(\frac{1}{G_F\Lambda^2}\right)
+ (2 C_{\varphi\Box} + C_{\varphi D})\left(\frac{s}{\Lambda^2}\right)\hspace{6.7cm} \textbf{dim-6}\nonumber\\[2mm]
&-20C_{\varphi W}\left(\frac{M_W^2}{\Lambda^2}\right)+20C_{\varphi B}\left(\frac{M_W^2-M_Z^2}{\Lambda^2}\right)-20C_{\varphi WB}\left(\frac{M_W\sqrt{M_Z^2-M_W^2}}{\Lambda^2}\right)\nonumber\\[2mm]
&+\sqrt{2} (4 C_{\varphi\Box}^2 - \frac{3}{4}C_{\varphi D}^2)\left(\frac{1}{G_F\Lambda^2}\right)\left(\frac{s}{\Lambda^2}\right)
-\frac{3}{2}C_\varphi (8 C_{\varphi\Box} - C_{\varphi D})\left(\frac{1}{G_F\Lambda^2}\right)^2\; , \hspace{1.7cm} \textbf{(dim-6)}^2\nonumber\\[2mm]
\mathcal{M}^{ZZ,\, D6}_{\pm\pm}=& -2(2 C_{\varphi\Box} + C_{\varphi D})\left(\frac{M_Z^2}{\Lambda^2}\right)\nonumber\\ &+2C_{\varphi W}\left(\frac{M_W^2}{M_Z^2}\right)\left(\frac{s}{\Lambda^2}\right)+2C_{\varphi B}\left(1-\frac{M_W^2}{M_Z^2}\right)\left(\frac{s}{\Lambda^2}\right) +2C_{\varphi WB}\frac{M_W}{M_Z}\sqrt{1-\frac{M_W^2}{M_Z^2}}\left(\frac{s}{\Lambda^2}\right)  \nonumber\\[2mm]
&+6\sqrt{2}C_{\varphi W}^2\left(\frac{M_W^2}{M_Z^2}\right)\left(\frac{1}{G_F\Lambda^2}\right)\left(\frac{s}{\Lambda^2}\right) \label{eq:DiHiggs:ZZHH:Dim6}\nonumber \\[2mm]
&+6\sqrt{2}C_{\varphi B}^2\left(1-\frac{M_W^2}{M_Z^2}\right)\left(\frac{1}{G_F\Lambda^2}\right)\left(\frac{s}{\Lambda^2}\right) 
+2\sqrt{2}C_{\varphi WB}^2\left(1-\frac{M_W^2}{\blue{2}M_Z^2}\right)\left(\frac{1}{G_F\Lambda^2}\right)\left(\frac{s}{\Lambda^2}\right)\nonumber\\[2mm]
&-6 C_\varphi\left[C_{\varphi W} \left(\frac{M_W^2}{M_Z^2}\right)+ C_{\varphi B} \left(1 - \frac{M_W^2}{M_Z^2}\right)+
C_{\varphi W B} \left(\frac{M_W}{M_Z}\right) \sqrt{1 - \frac{M_W^2}{M_Z^2}}
\right]\left(\frac{1}{\Lambda^2  G_F}\right)^2\nonumber\\[2mm]
 & +\Biggl[ - \frac{1}{\sqrt{2}} C_{\varphi W} C_{\varphi D}
\left ( \frac{M_W^2}{M_Z^2}\right )
+ 4 \sqrt{2}C_{\varphi W}C_{\varphi \Box}
 \left ( \frac{M_W^2}{M_Z^2} \right ) \nonumber \\[2mm]
&- \sqrt{2} C_{\varphi B} C_{\varphi D} \left (1-\frac{M_W^2}{2 M_Z^2}\right )
+ 4 \sqrt{2}C_{\varphi B}C_{\varphi \Box}
\left (1-\frac{M_W^2}{M_Z^2}\right ) \\[2mm]
&- \frac{3\sqrt{2}}{4} C_{\varphi WB} C_{\varphi D}\frac{M_W}{\sqrt{M_Z^2-M_W^2}} \left (1-\frac{2M_W^2}{3 M_Z^2}\right )
+ 4 \sqrt{2}C_{\varphi WB}C_{\varphi \Box} \frac{M_W}{M_Z}
\sqrt{1-\frac{M_W^2}{M_Z^2}}\nonumber \\[2mm]
&+ 5 \sqrt{2} C_{\varphi B} C_{\varphi W B}\left (\frac{M_W}{M_Z}\right )\sqrt{ 1- \frac{M_W^2}{M_Z^2}} + 
 7 \sqrt{2} C_{\varphi B} C_{\varphi W B}\left (\frac{M_W}{M_Z}\right )\sqrt{ 1- \frac{M_W^2}{M_Z^2}} \Biggr ]
 \left(\frac{1}{G_F\Lambda^2}\right)\left(\frac{s}{\Lambda^2}\right)\; ,
 \nonumber \\[2mm]
\mathcal{M}^{ZZ,\, D6}_{\pm\mp}=& +(4 C_{\varphi\Box} + C_{\varphi D})\left(\frac{M_Z^2}{\Lambda^2}\right)\nonumber\\[2mm]
 &+\Biggl [4\sqrt{2}C_{\varphi W}^2\left(\frac{M_W^2}{M_Z^2}\right)
+4\sqrt{2}C_{\varphi B}^2\left(\frac{M_W^2}{M_Z^2}\right)\left(1-\frac{M_W^2}{M_Z^2}\right)+\sqrt{2}C_{\varphi WB}^2\nonumber\\[2mm]
&+ 2 \sqrt{2} C_{\varphi W B} (C_{\varphi W}+ C_{\varphi B)}  \left(\frac{M_W}{M_Z}\right)\sqrt{1-\frac{M_W^2}{M_Z^2}}\Biggr ] \left(\frac{1}{G_F\Lambda^2}\right)\left(\frac{s}{\Lambda^2}\right) \; , \nonumber\\[2mm]
\mathcal{M}^{ZZ,\, D6}_{0\pm}=&-4\sqrt{2}\cot\theta
  C_{\varphi W} \left(\frac{M_W^2}{M_Z^2}\right)
  \left(\frac{\sqrt{s} M_Z}{\Lambda^2}\right)\nonumber\\[2mm]
  &-4\sqrt{2}\cot\theta
  C_{\varphi B} \left(1-\frac{M_W^2}{M_Z^2}\right)
  \left(\frac{\sqrt{s} M_Z}{\Lambda^2}\right)
  -4\sqrt{2}\cot\theta
  C_{\varphi WB}\sqrt{1-\frac{M_W^2}{M_Z^2}}\left(\frac{\sqrt{s} M_W}{\Lambda^2}\right)\nonumber\\[2mm]
  & + \Biggl [ -16\cot\theta
  C_{\varphi W}^2 \left(\frac{M_W^2}{M_Z^2}\right)
  \nonumber\\[2mm]
  &-16\cot\theta
  C_{\varphi B}^2 \left(1-\frac{M_W^2}{M_Z^2}\right)\left(\frac{1}{G_F\Lambda^2}\right)\left(\frac{\sqrt{s} M_Z}{\Lambda^2}\right)
  -8\cot\theta
  C_{\varphi WB}^2 \left(1-\frac{M_W^2}{M_Z^2}\right)\nonumber\\[2mm]
  &-2\cot\theta
  C_{\varphi W}(4 C_{\varphi \Box} + C_{\varphi D}) \left(\frac{M_W^2}{M_Z^2}\right)-2 \cot\theta
  C_{\varphi B}\left( 4C_{\varphi\Box} \left(1-\frac{M_W^2}{M_Z^2}\right)
  +C_{\varphi D} \left(\frac{M_W^2}{M_Z^2}\right)\right)\nonumber\\[2mm]
  &-8 \cot\theta
  C_{\varphi WB}C_{\varphi\Box} 
 \left(\frac{M_W}{M_Z}\right)\sqrt{1-\frac{M_W^2}{M_Z^2}}-\cot\theta
  C_{\varphi WB}C_{\varphi D} 
 \left(\frac{M_W}{M_Z}\right)\left(1-2\frac{M_W^2}{M_Z^2}\right)\left(1-\frac{M_W^2}{M_Z^2}\right)^{-1/2}\nonumber\\[2mm]
 &- 4 C_{\varphi W B} (3C_{\varphi W}+ C_{\varphi B)}  \left(\frac{M_W}{M_Z}\right)\sqrt{1-\frac{M_W^2}{M_Z^2}}\Biggr ] \left(\frac{1}{G_F\Lambda^2}\right)\left(\frac{\sqrt{s} M_Z}{\Lambda^2}\right) \; . \nonumber
    \end{align}

\subsection{Dimension-8 SMEFT}

\subsubsection{ $\varphi^8$, $\varphi^6 D^2$, $\varphi^4 D^4$ operator classes}
\begin{align}
\mathcal{M}^{ZZ,\, D8}_{00}=&-6C_{\varphi^8} \left(\frac{1}{G_F^2\Lambda^4}\right)
+\sqrt{2}\left(2C_{\varphi^6 \Box}+C_{\varphi^6D^2}\right)\frac{s}{\Lambda^2}\frac{1}{G_F \Lambda^2}\nonumber\\[2mm]
&- \frac{1}{2} \,\left ( C_{\varphi^4 D^4}^{(1)}
 - \frac{1}{2} C_{\varphi^4 D^4}^{(2)}\sin^2\theta + \, C_{\varphi^4 D^4}^{(3)} \right )\, \, \left (\frac{s}{\Lambda^2} \right )^2\; , \nonumber \\[2mm]
\mathcal{M}^{ZZ,\, D8}_{\pm\pm}=&
-2\sqrt{2}\left(2C_{\varphi^6 \Box}+
 C_{\varphi^6D^2}\right)\frac{M_Z^2}{\Lambda^2}\frac{1}{G_F \Lambda^2}\nonumber\\[2mm]
&+ \left[\left(C_{\varphi^4 D^4}^{(1)}-\frac{1}{2}C_{\varphi^4
D^4}^{(2)}+C_{\varphi^4 D^4}^{(3)}\right)-\frac{1}{2}C_{\varphi^4
D^4}^{(2)}\cos^2\theta\right]\left(\frac{M_Z^2}{\Lambda^2}\right)\left(\frac{s}{\Lambda^2}\right)\; ,\label{eq:DiHiggs:ZZHH:Dim8:1}
\\[2mm]
\mathcal{M}^{ZZ,\, D8}_{\pm\mp}=&
+\sqrt{2}\left(2C_{\varphi^6 \Box}+\frac{3}{4}C_{\varphi^6D^2}\right)\frac{M_Z^2}{\Lambda^2}\frac{1}{G_F \Lambda^2}+\frac{1}{2}C_{\varphi^4 D^4}^{(2)}\sin^2\theta\left(\frac{M_Z^2}{\Lambda^2}\right)\left(\frac{s}{\Lambda^2}\right)\; ,\nonumber\\[2mm]
\mathcal{M}^{ZZ,\, D8}_{0\pm}=& +\frac{1}{4\sqrt{2}}C_{\varphi^4 D^4}^{(2)}\sin{2\theta}\left(\frac{s^{3/2}M_Z}{\Lambda^4}\right)\; .\nonumber
\end{align}

\subsubsection{$X^2\varphi^4$, $X^2 \varphi^2 D^2$, $X \varphi^4 D^2$ operator classes}

\begin{align}
\mathcal{M}^{ZZ,\, D8}_{00}=&+14\sqrt{2} C_{B^2\varphi^4}^{(1)}\left(\frac{M_W^2-M_Z^2}{\Lambda^2}\right)\left(\frac{1}{G_F \Lambda^2}\right)
+14\sqrt{2}
C_{W^2\varphi^4}^{(1)}\left(\frac{M_W^2}{\Lambda^2}\right)\left(\frac{1}{G_F \Lambda^2}\right)\nonumber\\[2mm]
&+\left[\frac{1}{2}C_{B^2\varphi^2 D^2}^{(1)}(\sin^2\theta+1)
+2C_{B^2\varphi^2 D^2}^{(2)}\right]\left(1-\frac{M_W^2}{M_Z^2}\right)\left(\frac{M_Z^2}{\Lambda^2}\right)\left(\frac{s}{\Lambda^2}\right)\nonumber\\[2mm]
&+\left[\frac{1}{2}C_{W^2\varphi^2 D^2}^{(1)}(\cos^2\theta+1)
+2C_{W^2\varphi^2
D^2}^{(2)}\right]\frac{M_W^2}{M_Z^2}\left(\frac{M_Z^2}{\Lambda^2}\right)\left(\frac{s}{\Lambda^2}\right)\nonumber\\[2mm]
& -\frac{5i}{2^{5/4}}
C_{B\varphi^4D^2}^{(1)}\left(\frac{\sqrt{M_Z^2-M_W^2}}{\sqrt{G_F}\Lambda^2}\right)\left(\frac{s}{ \Lambda^2}\right)
+\frac{5i}{2^{5/4}}
C_{W\varphi^4D^2}^{(1)}\left(\frac{M_W}{\sqrt{G_F}\Lambda^2}\right)\left(\frac{s}{ \Lambda^2}\right)\; ,\nonumber\\[2mm]
\mathcal{M}^{ZZ,\, D8}_{\pm\pm}=&
+3\sqrt{2}C_{B^2\varphi^4}^{(1)}\left(1-\frac{M_W^2}{M_Z^2}\right)\left(\frac{1}{G_F\Lambda^2}\right)\left(\frac{s}{\Lambda^2}\right)+3\sqrt{2}C_{W^2\varphi^4}^{(1)}\frac{M_W^2}{M_Z^2}\left(\frac{1}{G_F\Lambda^2}\right)\left(\frac{s}{\Lambda^2}\right)\nonumber\\[2mm]
&-\left[\frac{1}{4}C_{B^2\varphi^2 D^2}^{(1)}+C_{B^2\varphi^2
D^2}^{(2)}\right]\left(1-\frac{M_W^2}{M_Z^2}\right)\left(\frac{s^2}{\Lambda^4}\right)-\left[\frac{1}{4}C_{W^2\varphi^2
D^2}^{(1)}+C_{W^2\varphi^2
D^2}^{(2)}\right]\frac{M_W^2}{M_Z^2}\left(\frac{s^2}{\Lambda^4}\right)\nonumber\\[2mm]
&-\frac{3i}{2^{5/4}}
C_{B\varphi^4D^2}^{(1)}\left(\frac{\sqrt{M_Z^2-M_W^2}}{\sqrt{G_F}\Lambda^2}\right)\left(\frac{s}{ \Lambda^2}\right)
-\frac{3i}{2^{5/4}}
C_{W\varphi^4D^2}^{(1)}\left(\frac{M_W}{\sqrt{G_F}\Lambda^2}\right)\left(\frac{s}{ \Lambda^2}\right)
\; ,\label{eq:DiHiggs:ZZHH:Dim8:2}
 \\[2mm]
\mathcal{M}^{ZZ,\, D8}_{\pm\mp}=&
-\frac{1}{8}C_{B^2\varphi^2
 D^2}^{(1)}\left(1-\frac{M_W^2}{M_Z^2}\right)\sin^2\theta\left(\frac{s^2}{\Lambda^4}\right)-\frac{1}{8}C_{W^2\varphi^2
 D^2}^{(1)}\frac{M_W^2}{M_Z^2}\sin^2\theta\left(\frac{s^2}{\Lambda^4}\right)\nonumber\\[2mm]
&+2^{3/4}i
C_{B\varphi^4D^2}^{(1)}\left(\frac{\sqrt{M_Z^2-M_W^2}}{\sqrt{G_F}\Lambda^2}\right)\left(\frac{M_Z^2}{ \Lambda^2}\right)
+2^{3/4}i
C_{W\varphi^4D^2}^{(1)}\left(\frac{M_W}{\sqrt{G_F}\Lambda^2}\right)\left(\frac{M_Z^2}{ \Lambda^2}\right)\; ,\nonumber
\\[2mm]
\mathcal{M}^{ZZ,\, D8}_{0\pm}=& - 4 C_{B^2\varphi^4}^{(1)}\left(1-\frac{M_W^2}{M_Z^2}\right)\cot{\theta}\left(\frac{s^{1/2}M_Z}{\Lambda^2}\right)\left(\frac{1}{G_F \Lambda^2}\right)- 4 C_{W^2\varphi^4}^{(1)}\frac{M_W^2}{M_Z^2}\cot{\theta}\left(\frac{s^{1/2}M_Z}{\Lambda^2}\right)\left(\frac{1}{G_F \Lambda^2}\right)\nonumber\\[2mm]
&- \frac{1}{8\sqrt{2}}C_{B^2\varphi^2
D^2}^{(1)}\left(1-\frac{M_W^2}{M_Z^2}\right)\sin{2\theta}\left(\frac{s^{3/2}M_Z}{\Lambda^4}\right)
- \frac{1}{8\sqrt{2}}C_{W^2\varphi^2
D^2}^{(1)}\frac{M_W^2}{M_Z^2}\sin{2\theta}\left(\frac{s^{3/2}M_Z}{\Lambda^4}\right)\nonumber\\[2mm]
&+2^{1/4} i
C_{B\varphi^4D^2}^{(1)}\cot\theta \left(\frac{\sqrt{M_Z^2-M_W^2}}{\sqrt{G_F}\Lambda^2}\right) \left(\frac{M_Z \sqrt{s}}{ \Lambda^2}\right)
+2^{1/4} i
C_{W\varphi^4D^2}^{(1)} \cot\theta\left(\frac{M_W}{\sqrt{G_F}\Lambda^2}\right)\left(\frac{M_Z \sqrt{s}}{ \Lambda^2}\right)\; .\nonumber
\end{align}